# Quantum information processing at the cellular level. Euclidean approach.

Vasily Ogryzko[1]

Institute Gustave Roussy

Villejuif, France

## Summary

Application of quantum principles to living cells requires a new approximation of the full quantum mechanical description of intracellular dynamics. We discuss what principal elements any such good approximation should contain. As one such element, the notion of 'Catalytic force' ***Cf*** is introduced. ***Cf*** is the effect of the molecular target of catalysis on the catalytic microenvironment that adjusts the microenvironment towards a state that facilitates the catalytic act. This phenomenon is experimentally testable and has an intriguing implication for biological organization and evolution, as it amounts to "optimization without natural selection of replicators". Unlike the statistical-mechanical approaches to self-organization, the ***Cf*** principle does not encounter the problem of 'tradeoff between stability and complexity' at the level of individual cell. Physically, the ***Cf*** is considered as a harmonic-like force of reaction, which keeps the state of the cell close to the ground state, defined here as a state where enzymatic acts work most efficiently. Ground state is subject to unitary evolution, and serves as a starting point in a general strategy of quantum description of intracellular processes, termed here the 'Euclidean approach'. The next step of this strategy is transition from the description of ground state to that one of growing state, and we suggest how it can be accomplished using arguments from the fluctuation-dissipation theorem. Finally, given that the most reliable and informative observable of an individual cell is the sequence of its genome, we propose that the non-classical correlations between individual molecular events at the single cell level could be easiest to detect using high throughput DNA sequencing.

---

[1] vogryzko@gmail.com; http://sites.google.com/site/vasilyogryzko/

# Content

## Part 1. Catalytic force

## Part 2. Ground state

## Part 3. Self-reproduction

# Part 1. Catalytic Force

**Abstract 1**

A new approximation to the description of intracellular dynamics is considered. For every catalytic event taking place in the cell, we propose to consider the reciprocal effects of the target molecule on the catalytic microenvironment, which we term catalytic force *Cf*. In full quantum mechanical account, cell is an aggregate of nuclei and electrons described with the density operator formalism. To describe the origin of the *Cf*, the cell is bipartitioned into two physical parts – a particular molecule (the subject of catalytic activity) and the rest of the cell (the catalytic microenvironment). In analogy to the exchange force mechanism, the catalytic microenvironment experiences a physical force adjusting it to a state where the catalytic act happens more efficiently. One of the most striking biological consequences of this proposal is a complementary to the natural selection way to optimize a biological structure, simply from the physical principle of minimum energy. This mechanism could serve as a new physical justification for the self-organization phenomenon, alternative to the approaches based on statistical mechanics of open systems far from equilibrium.

## *1. The need for a new approximation.*

I thank the organizers of this workshop Vladko Vedral and Elisabeth Rieper for giving me an opportunity to share with you my view of the potential role of quantum theory in explaining Life.

I am biologist, and although I will use some formulae, I am not going to use them for calculations, but mostly for conceptual purposes, to clarify what I want to say.

In any case, Life is too diverse and complex, and we also seem to be lacking some unifying principles needed to understand it. Therefore, at this point, I do not believe that direct calculations on any particular model would give us an immediate deep insight – especially if our main interest is in how quantum mechanics could be used *at the level of the whole cell*, and not at the level of an individual protein. Accordingly, my main goal here is finding the most appropriate language to describe intracellular dynamics that could naturally incorporate quantum principles. And developing new conceptual approach will require a level of abstraction that would allow us to see the most important aspects, and not to be lost in the details.

Essentially, my main point is that we need a new approximation to the description of intracellular dynamics. This means that we start from the full quantum mechanical description – considering a cell as a system of electrons and nuclei, governed mostly by the laws of electromagnetism. We do not take for granted any approximation, usually implied when describing intracellular dynamics – and we look how can we simplify such 'from the first principles' description. Note that, far from proposing a specific mathematical formalism, my goal here is rather to discuss what principal elements any such 'good' approximation should contain.

But first, what could be wrong with the approximations commonly used in chemistry or systems biology? The main problem is that most of them were developed based on the *in vitro* systems as a model – and thus cannot account for the important difference between *in vivo* and *in vitro* cases. A good approximation has to match theory with experiment – more specifically, theoretical description should distinguish between *dynamic variables* and *external parameters*, the latter corresponding to the aspect of the studied system that we can control experimentally (i.e., fix with arbitrary precision). However, this is where the difference between the *in vitro* and *in vivo* situations transpires. When modeling an individual enzymatic reaction *in vitro*, we can control its input in many respects, including the structure and the amount of substrate molecules, as well as the amounts and the state of the enzyme. However, to exercise such a control over an individual enzymatic event becomes impossible when we are working with the whole cell – hence all this information should enter into our description as dynamic variables when we are considering the *in vivo* situation.

Another (arguably related) problem with the commonly used approximations is that, from the outset, they get rid of *quantum entanglement*. For example, the Born-Oppenheimer approximation (BOA), widely used in quantum chemistry to describe molecular structure, starts with separating the nuclear and electronic degrees of freedom, i.e., with representing the state of the molecule mathematically as a product state – whereas taking the electronic-nuclear coupling into account would require a more general 'entangled' description. Given that transformations of molecular structure are an essential part of intracellular dynamics, the applicability of BOA to the description of the *in vivo* case is very limited. On the other hand, as I have argued previously (Ogryzko, 2008a), taking quantum entanglement into account might be important in explaining stability of intracellular dynamics (the 'tradeoff problem', see *24* and *35*) – which underscores the need for a principally new approximation that does not get rid of entanglement at the outset.

Another feature of a good approximation is in providing novel insights, preferably leading to testable predictions. As you will see, the proposed ideas do have intriguing biological consequences – most importantly, a new way to understand the nature of *the forces that shape intracellular order*. Thus, for me as a biologist, the stakes are incredibly high – essentially we are proposing here a new paradigm for the origin of meaning of biological organization, alternative to the Darwinian paradigm of natural selection. Although I do foresee many technical challenges in developing the proper mathematical formulation and in finding the best ways to test these ideas experimentally, the stakes certainly justify the endeavor.

Finally, this might not only be about biology. Living nature provides us with the first clear example of natural objects that meaningfully process and store information (Ogryzko, 1994) – that is, the first example of IGUSes (Gell-Mann and Hartle, 1990; Saunders, 1993). Given that the concepts of information and computation acquire growing importance in the foundations of physics itself ('it from bit' (Zurek, 1990)), the marriage between quantum mechanics and biology could have an impact on the physics foundations as well (see *32, 35a, 45*).

## *2. Summary in terms of Quantum Information Theory.*

Here, I would like to summarize my talk from the perspective of the main question that this workshop aims to ask: 'Is living nature engaged in any kind of quantum information processing?'

I am going to advocate the following provocative idea: 'If quantum information processing in biology is possible, it can only be done in the form of *molecular interconversions*'.

The reason for this statement is the following. All approaches to fight decoherence use redundancy in one way or another – which means that they have to use *aggregates of elementary particles* to encode a qubit. In a sense, because of the destructive effects of decoherence, the Moore's law has to stop one step before the level of electrons and protons. In the physicist's laboratory, we can think of various ingenious contraptions to generate redundancy and fight decoherence: quantum dots, ion traps, etc – all of which are ultimately aggregates of elementary particles. However, in biological systems the only aggregates of elementary particles that we ever encounter are molecules. Hence the above provocative statement.

On the other hand, aren't all molecules already classical entities, which do not exhibit quantum behavior?

The three main points of my talk will be related to clarification of the above provocative statement. I will discuss:

1a. How the intracellular environment could help to revive the quantum properties of molecules,

And vice versa, a more important part of the same question,

1b. How the quantum aspects of molecular behavior could contribute to the organization of the intracellular environment that facilitates this very revival.

I will discuss some very intriguing features of this proposal that appeal to me as a biologist – in particular, its relation to the self-organization phenomenon, i.e., a way to optimize biological order, which is alternative to the mechanism of natural selection. In short, I will suggest that enzymatic activities, when considered in the *in vivo* context, generate a new kind of physical force that adjusts the state of the cell in such a way that the enzymes function more efficiently.

Incidentally, this proposal also has intriguing implications for quantum information processing, for the following reason. As long as we use isolated molecules, our attempts to take advantage of nontrivial quantum effects will be challenged by the need to precisely control their environment 'by hand' in order to fight decoherence. To the contrary, as I will argue below, in the context of the intracellular environment, the cells will be taking care of this problem for us, so to speak.

2. I will also discuss what it would take to make this idea work.

In particular, I will suggest that intracellular dynamics will have to be described as a unitary process – and I will clarify, in the context of what I call an euclidean approach, the implications of this suggestion.

3. Finally, I will also talk about the potential use of these ideas for practical purposes – in particular, about the phenomenon of adaptive mutations, with which I was involved.

# *3. Quantum vs classical properties.*

What do I mean by 'the revival of quantum properties'?

First, I take quantum description as fundamental (i.e., the 'first principles') description of nature, and the classical description as its derivative – resulting from the so called 'quantum to classical transition'. Hence the term 'revival'. Second, in the same context, the dichotomy between 'quantum versus classical *properties*' is more appropriate than dividing the world between quantum and classical *objects*. The latter dichotomy is misleading – because it implies that all macroscopic objects are classical and thus cannot have any nontrivial quantum-mechanical behavior.

I suggest we take a different perspective, where any object (no matter how large or small) can have some classical properties, but also some quantum properties.

It is well established, for example, that even the electron, an obviously 'quantum object', has some classical properties – such as its mass or charge. Therefore, for the discussion of the 'quantum versus classical divide' in this case, the language of 'properties' is clearly more appropriate than the language of 'objects'. But then, reciprocally, a seemingly classical macroscopic system might have features that have to be described by non-commuting operators – and thus have some 'quantum properties'.

Overall, whether some observables are classical or not is related to existence of superselection rules, forbidding superpositions of some states of the system. Interestingly, the superselection rules were introduced first axiomatically to explain the classical nature of charge and mass (Wick et al., 1952), but more recently have been proposed to emerge as a result of environmentally induced decoherence (Giulini, 2000 ; Giulini et al., 1995).

Accordingly, for the sake of this presentation, I will keep to the following view on the issue of 'classical versus quantum properties'. It is a *practical question* of whether we can have an object in a superposition of different eigenstates of a given operator or not – and this depends on the practical availability of reference frames permissive for such superpositions (Bartlett et al., 2007).

Compare it to how statistical physics explains the second law of thermodynamics and the origin of physical irreversibility. In principle, fundamental physics is perfectly consistent with us designing a mechanical system of many degrees of freedom in such initial conditions that it would evolve toward a more ordered state. However, such a feat is impossible for all practical purposes – except for the nanosystems, where the fluctuation theorem describes the situation more accurately (Evans and Searles, 2002).

Now back to the quantum vs classical properties. Similarly, an electron cannot be in the state of superposition of different charges because it is very difficult to arrange for a reference frame (environment) that would allow us to achieve this – although Aharonov and Susskind demonstrated a long time ago how it can be done in principle (Aharonov and Susskind, 1967). By the same token, in the right environment, Schroedinger's cat can be in the state of being alive and dead – but it is *practically impossible* to arrange for such an environment (although, again, for nano-objects – such as fullerens – it has been done (Arndt et al., 1999)).

What encourages me to seek the place of quantum principles in explanation of Life is the following observation. While we all agree that it is impossible for a cat to be dead and alive at the same time – many biological problems are, in fact, in a ' *grey zone'* in respect to the practical availability of reference frames. In biology, reference frames (environment) change all the time. Therefore, some properties of a particular object that appear to be classical and subject to superselection rules in one environment – might exhibit superposition behavior in another. Conversely, a state einselected in one environment – could be destroyed by decoherence in another. Moreover, all these situations could be equally realistic and happening in the life-time of the same biological system. In particular, this nontrivial quantum behavior could manifest itself at the molecular level in living cells – which is most relevant to the further discussion.

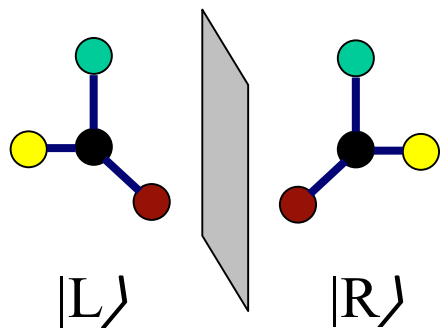

$$|1\rangle = (|L\rangle + |R\rangle)/\sqrt{2}$$

$$|2\rangle = (|L\rangle - |R\rangle)/\sqrt{2}$$

*Real states*     *Energy eigenstates*

$$\rho^A(t_0) = Tr_E|\Psi_{AE}\rangle\langle\Psi_{AE}| = \Sigma\alpha_i\alpha^*_j\langle e_i|e_j\rangle|a_i\rangle\langle a_j| \rightarrow \rho^a(t) = \Sigma\alpha^2_i|a_i\rangle\langle a_i|$$

## *4. Quantum properties at molecular level*

So what about quantum properties of molecules?

Let's first consider the following example. Hund, in the early days of quantum mechanics brought up a question: 'why molecules have a definite shape?' (Hund, 1927) As a simple illustration to this problem, he used an example of molecular chirality. According to the principles of quantum mechanics, all molecules are supposed to be in the eigenstates of the energy operator. However, for chiral molecules it is obviously not the case – in reality we observe only left or right enantiomers, whereas the energy eigenstates would be the symmetric or asymmetric superpositions of the enantiomer states.

How to explain this discrepancy between the predictions of quantum mechanics and reality?

Different attempts to reconcile this phenomenon with quantum mechanics followed (Amann, 1991; Pfeifer, 1980 ). They eventually converged on the idea that it is environmentally induced decoherence that is responsible for the stability of alternative enantiomers (Zeh, 1970; Zurek, 2003). In more technical language it means that due to the entanglement of the molecule (**A**) with its environment (**E**), its state should be described as a reduced density matrix:

$$\rho^A(t) = Tr_E|\Psi_{AE}\rangle\langle\Psi_{AE}|,$$     [4.1]

and the values of the off-diagonal terms in this matrix $|a_i\rangle\langle a_j|$, when described in the 'Left' versus 'Right' basis ($|L\rangle$, $|R\rangle$), vanish due to the orthogonality between the corresponding states of the environment

$$\langle e_i|e_j\rangle \rightarrow 0$$     [4.2]

In a more pedestrian language it means that the environment serves as an observer that can *distinguish* between the 'Left' $|L\rangle$ and 'Right' $|R\rangle$ states of the molecule – and this observation destroys the superposition of these states.

The two main implications of this slide are: first – as a default, the molecules can be in a superposition of different molecular shapes (i.e., unitary transitions between them are allowed in quantum theory); and second – it is the environment that is responsible for forbidding these transitions, i.e., for the emergent classicality of the molecular shape.

But on the other hand, what do enzymes do? Nothing but convert one molecular shape to another. Which leads me to the next slide.

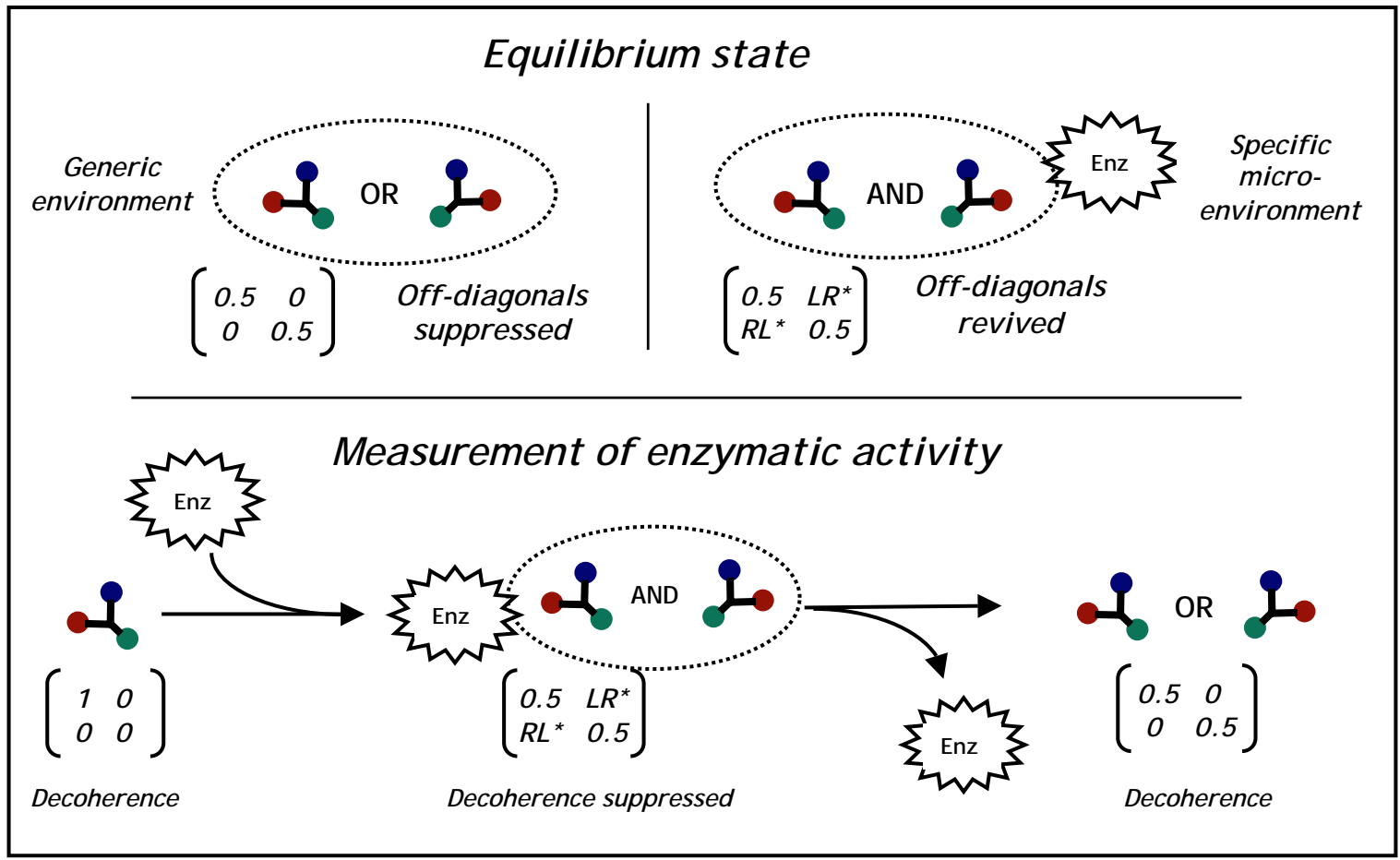

## 5. General description of enzymatic act as 'decoherence suppression'

For the sake of the argument, we will keep to the conventional view on enzymes, suggested by Pauling – that they help their substrate molecules to cross the kinetic barriers (separating their alternative shapes) by stabilizing the transition states better than the substrate/product states (Garcia-Viloca et al., 2004; Pauling, 1946). Given that 'kinetic barrier' is a classical notion, we will revisit the Pauling's idea from the fundamental 'first principles' perspective. This will require formulating the notion of kinetic barrier in the language of off-diagonal terms of a reduced density matrix (see also *section 29*). We will not discuss the role of tunneling.

For simplicity, let us focus on the molecular chirality as a toy model, and worry about the more general case later. Also, we will consider the equilibrium situation (i.e., with both enantiomer forms present in equal amounts) – which is sufficient to capture the Pauling's idea, and is easier to describe with the reduced density matrix approach (Figure, top left part).

The most important is to recognize the crucial fact that enzymes represent very irregular and 'non-generic', but at the same time *highly reproducible microenvironments*. Since it is not a thermal bath, all bets are off – as far as decoherence is concerned. In general, one should expect that a particular microenvironment could significantly affect the outcome of decoherence process – otherwise responsible for the classical behavior of our molecule.

Now suppose that we have a protein **Enz**, generated during billions of years of biological evolution, that, when in contact with our chiral molecule **T**, could suppress the ability of the surrounding environment to distinguish between the *'Left'* and *'Right'* states. If we put **Enz** in contact with **T**, this will lead to revival of the off-diagonal terms (depicted here as *LR\** and *RL\**) in the reduced density matrix of the molecule **T**, expanded in the (*'L'*, *'R'*) basis (Figure, top right part). Essentially, the transition between the two enantiomers, forbidden in the case of generic environment due to decoherence, will be allowed when in contact with **Enz**.

By this criterion, our protein should qualify as an enzyme (more specifically, racemize) – as this is what enzymes do – they do nothing else but facilitate transitions between alternative molecular configurations. Importantly, no full revival of the off-diagonal terms is necessary – any increase in the value of the off-diagonals will eventually have the effect of accelerating the transition – and thus has to qualify as catalysis.

Using this simple model as an example, could we then understand the way enzymes work, in general terms, as decoherence suppression? Could their catalytic mechanism be in providing a specific microenvironment that protects the target molecule from the effects of a generic environment – the effects that are otherwise responsible for the classical shape of the molecule?

Several comments are in order:

1. Admittedly, as presented, this suggestion looks like a gross oversimplification. For example, an enzymatic mechanism could involve more than one transition step (and thus have intermediate stable states $I_1, I_2, \ldots$), and several alternative pathways could be involved as well. There is, however, a general mathematical result, based on the Schmidt decomposition theorem, that claims that for any density matrix $\rho_A$ describing a mixed state of a system $A$, we can always add an ancillary system $B$ such that our density matrix $\rho_A$ will appear as a reduced density matrix of a system $(A+B)$ in a particular pure state $|\psi\rangle$:

$$\forall \rho_A, \text{mixed}, \exists \text{ auxiliary system B, such that } \rho_A = Tr_B(|\psi\rangle\langle\psi|) \text{ for some } |\psi\rangle \in H_A \otimes H_B \qquad [5.1]$$

In other words, any desired evolution of your system (generally involving more than two 'classical' shapes – which in this description corresponds to the density matrix having more than two basis states: $L, I_1, I_2, \ldots, R$, plus off-diagonal terms), can be modeled as a result of its interaction with a particular microenvironment. And an enzymatic molecule could be understood as responsible for such microenvironment (see the appendix 1 (**46**) for the clarification of this statement).

2. Another way to talk about enzymes is to say that they help a particular molecular system to reach the equilibrium between different molecular configurations (in the density matrix language, they accelerate its approach to the mixed state – which could otherwise take millions of years). This formulation is more experiment-friendly, as one typically *measures* enzymatic activity by generating a nonequilibrium situation – adding the substrate (e.g., *L* state) and enzyme together and detecting the appearance of alternative molecular shape (*R* state, respectively) afterwards. However, as the *measurements* go, this is an irreversible process (see also slide *17*), and its description from the first principles would be beset with technical and foundational problems. As noted above, we do not need it to capture the Pauling's idea.

Nevertheless, we can depict the process as consisting of two steps (Figure, bottom). At the first step, the molecule *T* in its *L* configuration (i.e., having only one diagonal term in its density matrix) is put in contact with the *Enz*, and their interaction will lead first to entanglement and then to dynamic evolution of the composite system '*T + Enz*' towards a state where the *'R'* state has an equal amplitude to be detected (Figure, bottom, middle). Importantly, the complex '*T + Enz*' is not expected to be stable and will eventually fall apart. After dissociation of the enzyme from *T*, fast decoherence will ensue (because now the molecule is surrounded by the generic outside environment **E**, which can distinguish between the alternative forms of *T*) resulting in our molecule acquiring a mixed state of being either in the *'Left'* or *'Right'* configuration (on the right[2]). The *'R'* state can be thus considered a stable (molecular) record of enzymatic activity. Alternatively, if we now start with *T* in its *'Right'* state and add enzyme, we will again end up with *T* in a mixed state of either *'Left'* or *'Right'* configurations. Thus, either *'L'* or *'R'* state could serve as a blank state of the target molecule, which plays a role of 'measurement device' and correspondingly produces either *'R'* or *'L'* as the record of the measurement of enzyme's activity.

3. Finally, one might object to this proposal as too abstract and actually telling us nothing about the mechanism of any particular enzyme – we simply state, using the density matrix formalism, something that we knew all along – that enzyme, when added to a target molecule, helps to transform it from one configuration to another [3]. However, this is exactly my aim, which I wish to strongly emphasize – that it is not the goal of this presentation to propose a new mechanism of enzymatic catalysis based on some exotic quantum effect. I suggest something more trivial – how, starting from the first principles, one can describe an enzymatic process as a revival of off-diagonal terms in the reduced density matrix of the target molecule represented in the molecular shape basis. (And if we want to see the vanishing off-diagonals as result of superselection rules and decoherence, then we can formulate the principle of enzymatic mechanism as 'decoherence suppression'). In other words, I am looking for a *general description of an enzymatic act* – a proper level of abstraction that would not depend on any particular molecular mechanism and would allow us to integrate it into the global picture of intracellular dynamics. If, following Zeh (Zeh, 2007), we consider decoherence as a general *'dequantization'* procedure of the fundamental quantum description of the world – then the suppression of decoherence would naturally qualify as the way to *'quantize'* the description of enzymatic act.

---

[2] shown is the density matrix of the system in equilibrium, although the shorter times of interaction between the target molecule and enzyme could lead to other than 50% contributions of the alternative states in the resulting mixture.

[3] more accurately, helps to accelerate the equilibration between its two alternative shapes, usually termed substrate and product

*Mandelate racemaze:* 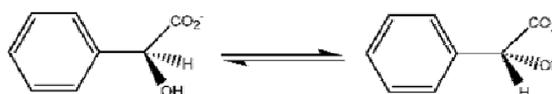

## 6. New experimental systems to study decoherence, reference frames, superselection rules etc?

Before proceeding further, I would digress. Regardless of whether the description proposed above is of any use for enzymology in understanding particular mechanisms of enzymatic activity, it might help fundamental physics in search of new experimental models to study decoherence, measurement, superselection rules and related concepts.

Here is an example of mandelate racemaze – one of the most studied cases from the family of enzymes catalyzing transformation between molecular enantiomers. This protein can be prepared in large amounts, and much information about this enzyme is available – such as its many substrates and inhibitors, as well as the crystal structure of wild type form and several mutants. Thus, mandelate racemaze (and related enzymes) might provide a new convenient model, prepared for us by living Nature, for theoretical and experimental studies of decoherence, measurement, superselection rules and related phenomena.

In this respect, it is also relevant to note the work done more than a decade ago. An application of a particular series of laser impulses has lead to the generation of superpositions between the *'Left'* and *'Right'* states of an optically active molecule (Cina and Harris, 1995; Shao and Hanggi, 1997). This result illustrates that to suppress decoherence by arranging an environment oblivious to the difference between the *'Left'* and *'Right'* states is not dramatically difficult and has been accomplished in laboratory conditions (although not using enzymes as microenvironments).

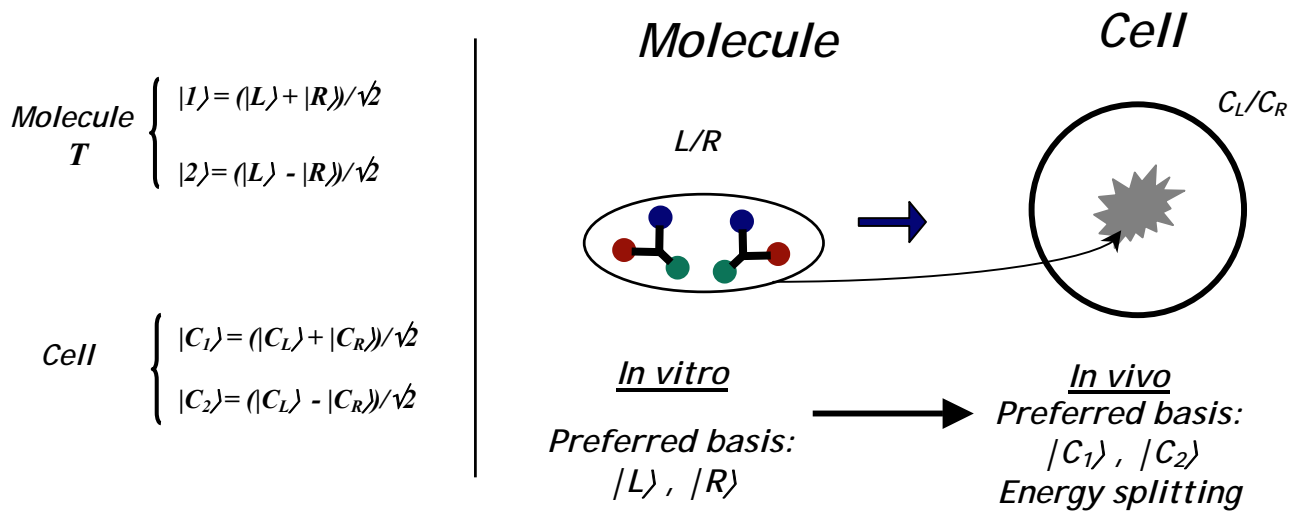

## 7. Reciprocal effect of catalyzed system on the catalytic microenvironment

I am now coming to the first main thrust of my talk. As pointed out before, the proposed general description of enzymatic act using density matrix language seems too abstract and not providing much of mechanistic insights. Moreover – even if we had a specific enzyme in mind, this formal description would not allow us to do any useful calculations. So what is it good for?

These concerns, however, are beside the point. Our main interest will be not in the effects of enzyme (or more generally, microenvironment) on the target molecule $T$ – but rather in the reverse, reciprocal effect of the catalyzed molecule $T$ on its microenvironment (e.g., enzyme). So far we were treating our microenvironment as a black box, with only one essential difference from the other black boxes – upon its interaction with the target molecule it could make its alternative states (e.g., $|L\rangle$ and $|R\rangle$) look nonorthogonal. But now let us analyze the other side of this interaction. Consider that our black box has a structure – e.g., at least two states ($|A\rangle$ and $|B\rangle$) that differ in the ability to 'distinguish' between the $|L\rangle$ and $|R\rangle$ states of the target molecule $T$. In what follows, I will explore the following crucial novel idea – due to the effect of the catalyzed molecule on its microenvironment, we should expect existence of physical forces of a new kind that will adjust the intracellular environment to optimize the catalytic transitions (e.g., drive the black box from $|A\rangle$ state towards $|B\rangle$ state).

Here is a brief description of how this idea works (see the Figure). I again use molecular chirality as an example.

As discussed previously, due to strong coupling to its environment **E**, the preferred basis for the optically active molecule $T$ is *'Left'* and *'Right'* configurations ($|L\rangle$ and $|R\rangle$ states) – instead of their linear superpositions ($|1\rangle$ and $|2\rangle$), which correspond to different energy eigenstates of the molecule's Hamiltonian. However, this pertains to the generic environment, commonly encountered *in vitro* (see the middle part of the Figure). Now let's change the setting and consider the same molecule in the context of a specific (i.e., not generic) microenvironment, able to suppress decoherence. It could be an enzyme, but for good measure, let us take the whole cell as this microenvironment (*in vivo*, the right part of the Figure). For simplicity, I will consider a case where there is complete protection from decoherence. Then the preferred states (now we have to consider the whole cell as our system, instead of $T$) will correspond to different energy eigenstates ($|C_1\rangle$ and $|C_2\rangle$). It is exactly the energy splitting between the preferred states that will be responsible for the existence of the physical force proposed above.

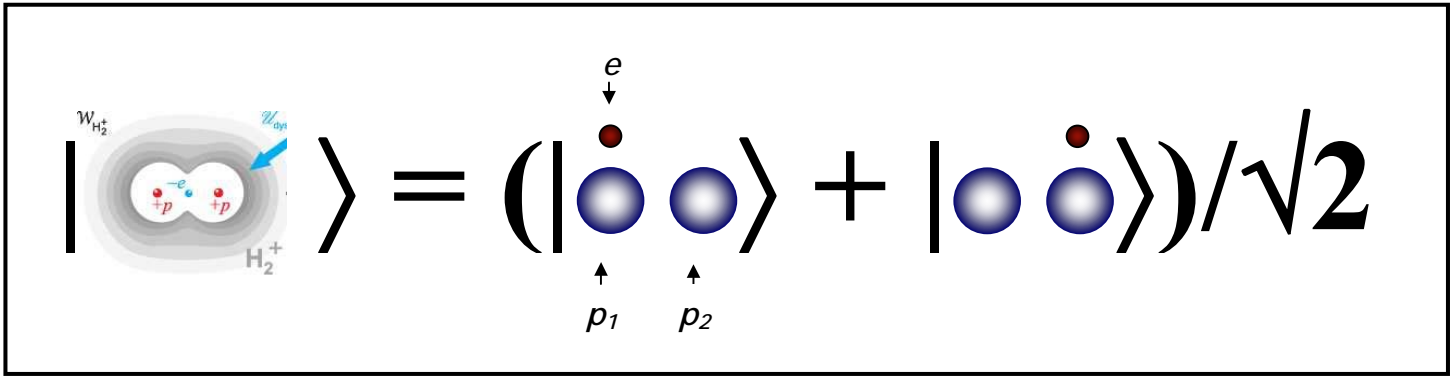

## 8. Example – molecular hydrogen ion

Let me illustrate the idea using the example of another simple system – a molecular hydrogene ion. This system consists of two protons and one electron, the two protons being attracted to each other due to the exchange of the electron between them. Formally, the ground state of this system is described as a superposition of two states:

$|\psi\rangle = (|1\rangle + |2\rangle)/\sqrt{2},$  [8.1]

where $|1\rangle$ and $|2\rangle$ correspond to the electron being located close to either one or the other proton.

To explain how the force of attraction between the two protons emerges, all we need is to have two degrees of freedom and a particular relation between their dynamics. One degree of freedom (we call it *'e'*) assumes only two values, corresponding to the location of the electron near one or the other proton. The second degree of freedom (we call it *'p'*) is continuous and corresponds to the distance between two protons. The 'force' that holds the two protons together results from the fact that when they are close to each other (the *'p'* degree of freedom has a small value), the electron exchange (the dynamics along the *'e'* degree of freedom) is more efficient compared to the case when the two protons are far away. More efficient exchange corresponds to the total energy of the system being lower – and as a physical system favors the states with lower energy, this explains the 'attraction force' between the two protons.

This is rather general mechanism of physical attraction, which works for many other physical forces that hold material things together. One might even argue that this is the main reason *why we need quantum superpositions* in the first place – to ensure that the world does not fall apart. True as it may, the above example does not do the full justice to the role of superposition principle in stability of physical objects. The superposition does not always have to be presented as an exchange of a 'particle' between two objects in physical space. As I will argue below, such a representation conceals more fundamental mechanism at work – with a potentially much broader scope. In particular, the second degree of freedom (i.e., *'p'*), forced to adjust to the state that facilitates the exchange along the first degree of freedom (i.e., *'e'*), does not have to correspond to a physical distance between two objects. It could be a more abstract property of one part of the system (*'p'*), interacting with the second part (responsible for the *'e'* degree of freedom). I will explore the potential implications of this more general mechanism in generation of a new type of physical force operating in the cell.

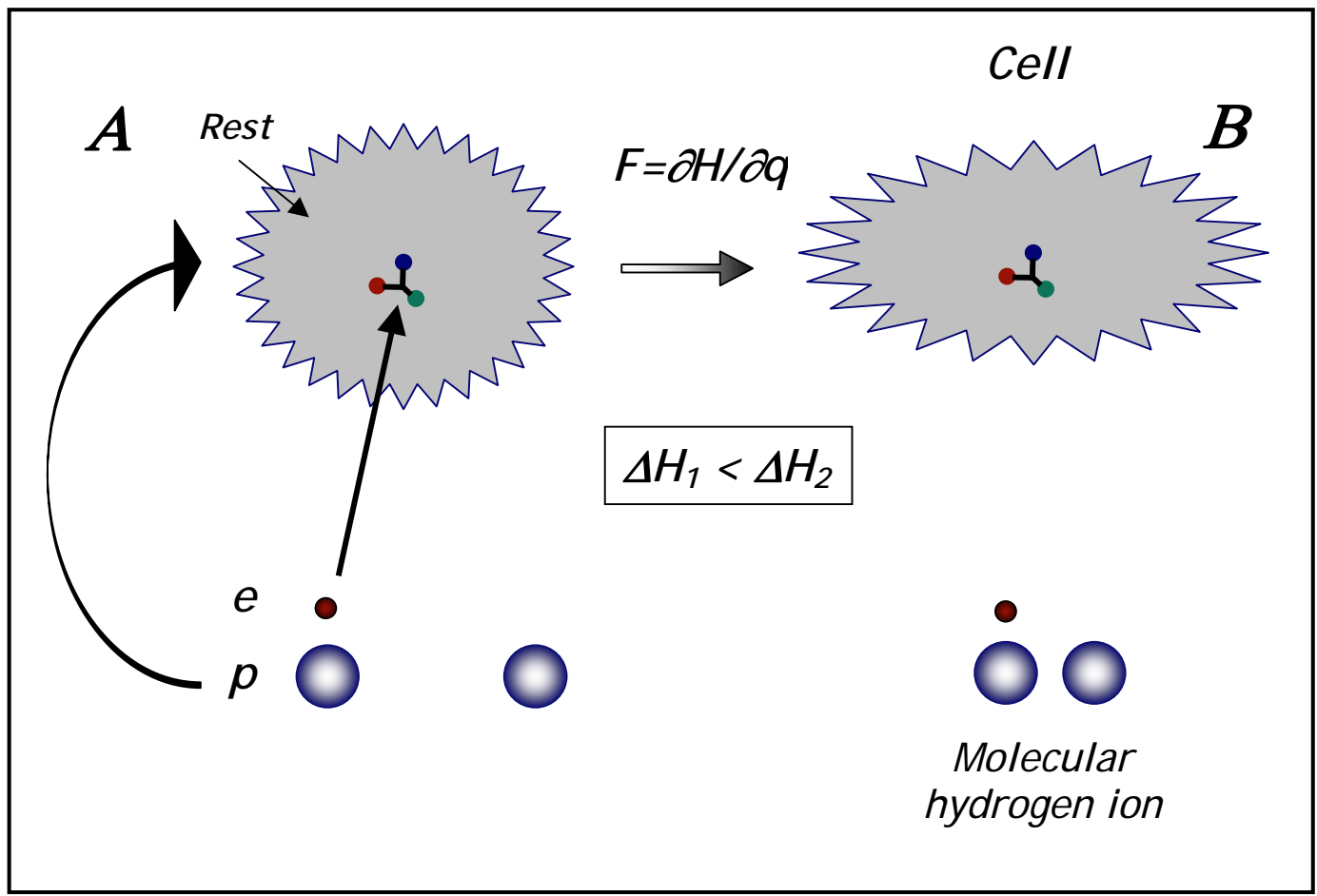

## 9. Catalytic act in vivo as an 'exchange force'

Here, using the chiral molecule in the context of intracellular environment as a toy model, I am going to provide a formal one-to-one correspondence between this system and the molecular hydrogene ion, considered above. I will map the electron degree of freedom *'e'* (its location close to one or another proton, $|1\rangle$ or $|2\rangle$) onto the state of the chiral molecule *T* (being in the $|L\rangle$ or $|R\rangle$ configuration), and I will map the second degree of freedom *'p'* (the distance between the two protons) onto the state of the rest of the cell *Rest* (the microenvironment of the chiral molecule).

Now we will compare two situations: **A.** on the left, a microenvironment where the transition between the $|L\rangle$ and $|R\rangle$ states of *T* is less efficient (formally equivalent to the large distance between the protons), and **B.** on the right, a microenvironment where the same transition is more efficient (i.e., short distance between the protons). According to the same logic as presented above, the more efficient exchange along the first degree of freedom ($|L\rangle$ versus $|R\rangle$) in the latter case leads to greater energy splitting. This lowers the total energy of our system to a greater extent – thus generating a physical force that drives the second degree of freedom *Rest* (the rest of the cell) towards the second configuration. In other words, the catalytic act forces the microenvironment to adjust towards the state where this very act occurs more efficiently!

Below, I will discuss in greater detail this intriguing prediction, which undoubtedly could provide many novel insights into intracellular dynamics. For the absence of a better term, I will refer to it as 'catalytic force' – *Cf*. As promised, to see this reciprocal action of one part of a composite system on the other one in the most transparent way, we had to use the proper level of abstraction (strip our description down to two interacting degrees of freedom) that did not get us buried in inessential details.

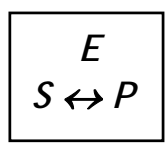
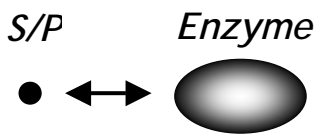

Attraction between enzyme E and its substrate/product S/P

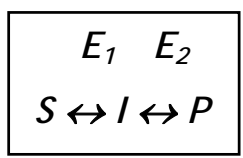
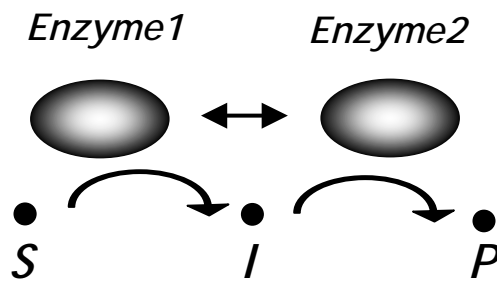

Attraction between enzymes $E_1$ and $E_2$ via intermediate I

## *10. Physical and biological implications*

First, even if there could be a quantum-mechanical effect of the sorts just described, one can wonder what difference could it make in realistic situations (what is its energy scale, does it work in anything other than chiral molecule cases, etc). Before we go any further in this direction, I would like to make two comments. The first is a clarification of the ontological status of the proposed phenomenon, and the second one concerns its intriguing biological implications, which should make this investigation worthwhile.

1). First, I do not propose the 'catalytic force' ***Cf*** to be in the same league with the four known fundamental physical fields (electromagnetism, weak, strong and gravitation). Its ontological status is rather that of an effective field, analogous to van der Waals forces, hydrogene bonds, ion-ion interactions, covalent bonds and other interactions that hold the cell together. They are all *approximations* to the fundamental description of a living cell as a complex system of electrons and nuclei – ultimately governed by nothing but the laws of electrodynamics and quantum mechanics. The concept of effective force is used in order to circumvent the practically impossible task of accurately calculating dynamics of a system composed of billions of degrees of freedom – such as a single living cell. Similarly, I propose to see ***Cf*** as another convenient approximation having analogous quantum-mechanical origin. Like the other effective forces, it is formally based on the 'exchange principle': we represent the considered system as two degrees of freedom (*'e'* and *'p'*), then assume that the 'efficient exchange' along one degree of freedom (*'e'*) depends on the value of the second degree of freedom (*'p'*) – and as a result, obtain the reverse effect of *'e'* on *'p'*.

There is, however, one important difference of ***Cf*** from the regular inter- and intra-molecular forces operating in the cell. Most of them (with some notable exceptions, e.g., forces responsible for aromatic structure) can be represented as an interaction between two physical entities (molecules or atoms, described by the *'p'* variable) – via an exchange of a third entity (a particle, described by the *'e'* variable) in the physical three-dimensional space. This makes sense because these forces were deduced from *in vitro* experiments on isolated molecules (or atoms) interacting with each other. They are convenient for conceptualizing the processes *in vivo* in the terms that have their easily identifiable counterparts *in vitro* – which has an advantage when we want to model individual steps of intracellular processes in cell-free systems. In this respect, ***Cf*** is different – generally, it cannot be represented as a binary interaction, as both its origin and operation belong to the configuration space of the described system. This is why it would be difficult to observe and deduce its existence from experiments *in vitro* (however, see ***Appendix 2, 47***). On the other hand, as will be argued below, taking it into account might be crucial for understanding what is going on *in vivo*.

2). That said, there should be ways to observe at least some effects of ***Cf*** as a binary interaction in the physical three-dimensional space. This is important for making the idea testable, and also for recognizing its biological implications. Since we are looking for examples of how the configuration of the intracellular environment could facilitate an enzymatic transition, *in vivo* situation would be the most relevant in detecting possible manifestations of ***Cf***. Here are two examples (among potentially many others): 1. Top of the Figure. We can predict the existence of an attraction between a molecule of enzyme ***E*** and a molecule of the corresponding substrate ***S/P*** *in vivo* (we can extend this prediction by including adjustment of molecular orientations that facilitate the reaction, i.e., attraction in 'orientation space'). 2. Bottom of the Figure. We can also predict an attraction between molecules of enzymes ***$E_1$***, ***$E_2$*** that catalyze two consecutive reactions $S \leftrightarrow I \leftrightarrow P$ (including the orientation degrees of freedom as well).

| Mantra of <u>Molecular Biology</u>: | Mantra of <u>Self-organization</u>: |
|---|---|
| *"Study structure first!"* | *"Dynamics comes first!"* |
| *Strong and weak interactions are encoded in genome, hold the cell together and determine dynamics* | *Dynamic flow organizes system to ensure 'optimal performance'* |
| ***Function is a consequence of structure*** | ***Structure is explained by function*** |
| ***Optimization mostly via <u>natural selection</u>*** | ***<u>Alternative to natural selection</u> way of optimization. Can help to address the problem of 'irreducible complexity'*** |

## 11. 'Optimization without natural selection of replicators'. New physical justification of self-organization principle.

For a molecular biologist, the above predictions look very counterintuitive – we are accustomed to thinking of enzymatic processes in the cell as occurring due to collisions between randomly moving molecules. To the contrary, here we claim to uncover an aspect of intracellular dynamics that endows molecular motions with direct functional meaning. Indeed, this is the most interesting feature of the *Cf* – its effects on the intracellular dynamics are reminiscent of the phenomenon of self-organization (Eigen and Schuster, 1979; Goodwin, 2001; Haken, 1983; Kauffman, 1993).

The phenomenon of self-organization challenges the dominant mode of thinking in molecular biology – which always considers *function* as a consequence of *structure*. Speaking of intracellular organization in particular, the supramolecular order (the way how different components of the cell are located and oriented in respect to each other) is a result of weak interactions between the cellular components – as exemplified by the role of macromolecular docking in formation of multiprotein complexes. The docking surfaces, in turn, are determined by domain folding – also due to the weak interactions, all eventually determined by the primary aminoacid sequence. This sequence is encoded in the genome – an ordered sequence of nucleotides maintained due to the covalent sugar-phosphate backbone of DNA.

Thus, in molecular biology the explanatory arrow always goes from the structure of genome to the structure of the protein and only then to function. A tacit ideal of molecular biological knowledge (a limit to which all of it should ultimately converge) is in having detailed information of position of every atomic nucleus and electron in the cell – and then deducing the dynamics of the system based on the knowledge of the structure. Accordingly, molecular biology should study structure first – as the progress in understanding intracellular function can result only from obtaining ever more detailed structural information. In this view, structure has clear priority over function; therefore the only effect that function can have on the cell structure in this paradigm is indirect – via selection at the population level among structural variations caused by random changes in the genome sequence. In other words, the only way of optimizing a biological structure to fit a particular function (literally, *to give it a meaning*) is by natural selection.

To the contrary, the concept of self-organization gives dynamics a more independent role in the inner workings of Life. A popular model here is the convection (Bénard) cell phenomenon (Karsenti, 2008). Due to space limitations, I cannot give full justice to this very interesting phenomenon. Its import is in illustrating that it is physically possible for the flow of energy (i.e., dynamics of the system) to play a critical role in determining the spatial order of the system. This is the principal feature of self-organization – with dynamics having a role on its own and directly contributing to the emergence and stability of order. Accordingly, in the purest case of self-organization, function comes first, and the structure second.

For biology, the biggest impact of this idea is in providing a way to optimize biological order that is alternative to the mechanism of natural selection. One dramatic difference is that we do not need a *population of replicators* for this kind of optimization to work – as it can occur at the level of individual non-reproducing system. Among the reasons why the idea of *optimization without natural selection* is so attractive is that it can provide a solution to the problem of 'irreducible complexity', promoted by Intelligent Design as a challenge to the evolutionary paradigm in modern biology. Simply put, if self-organization does play a role in Life, an ordered structure with a particular function can emerge and be maintained in the cell without necessarily having all genes ready and in place.

But if the idea of self-organization is so widely known[4], what is new in my proposal? It should help to get around the following problem. Despite all the buzz around this concept, the current attempts to ground self-organization in physical principles do not apply to many biological systems – as these attempts rely on statistical mechanics of open systems far from equilibrium (Glansdorff and Prigogine, 1971). As discussed later, this approach is not satisfactory if individual living cells are concerned. Briefly, if we want to apply the laws of statistical mechanics to ensure stable dynamics of the system (which has to unravel in an enormously high-dimensional space of relevant variables), the numbers of participating components have to be unrealistically large. In this regard, what the *Cf* notion brings to the table is an *alternative physical justification* for self-organization phenomenon – its main appeal being the fact that large numbers are not required for this idea to work (see *19, 34, 35* for the clarification of this statement).

---

[4] to be sure, this is true mainly for the physically-minded researchers, whereas most of the biologists are still under the spell of the Darwnian paradigm

# Part 2. Ground state.

*Abstract 2*

To put the idea of 'catalytic force' ***Cf*** in accord with the notions of the physics of condensed matter, we propose to consider ***Cf*** as a harmonic-like force of reaction, which keeps the physical state of the cell close to the ground state, defined as a state where all enzymatic acts work most efficiently. We propose an estimate of how significant the energetic contribution of the ***Cf*** could be. Given that ground state is subject to unitary evolution, this notion is proposed as a starting element in a more general strategy of quantum description of intracellular processes, termed here 'Euclidean approach'. In addition, we argue why quantum principles will be necessary for the understanding the physics of Life – namely, for addressing the problem of stability of intracellular dynamics ('tradeoff between complexity and stability') at the level of individual cell. This issue, although not sufficiently appreciated yet, will be brought to the fore with the increased role of 'nano-' and '-omics' approaches in biology.

## *12. How to make this idea work?*

The above arguments render the idea of a new physical force operating in the cell sufficiently intriguing. However, to warrant its further analysis, we need to address several questions first:

1) So far, we used molecular chirality as a toy model. This is not very interesting – given that not much racemization is going on in cells, and it does not play important role in cell functioning. The effect would acquire truly fundamental significance if we could apply it to every enzymatic act happening in the cell. However, in the more general case of catalytic transition, the alternative molecular configurations (substrate and product states of a target molecule *T*) correspond to different energy values – complicating the treatment of molecular interconversions as a unitary process.

2) We do not know yet how significant our effect could be energy-wise. Is the associated energy gain comparable to thermal energy – so it could withstand the interaction with external environment? Only then can we consider our force as playing a legitimate role in intracellular dynamics.

3) Finally, we described the system: 'molecule + the rest of the cell' as evolving in a unitary way. How could this description be relevant to any biological system? Aren't they all supposed to be open systems – far from equilibrium and dissipating energy for survival?

We could take our effect more seriously only if we could address these concerns. They will be discussed in the next several slides.

# Confinement of activation energy to the intracellular microenvironment

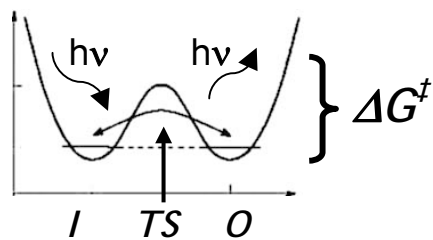

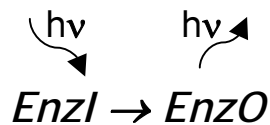

$$k = \frac{k_B T}{h} e^{-\frac{\Delta G^\ddagger}{RT}}$$

(Eyring-Polanyi)

Activation energy
≅ thermal energy

In vitro

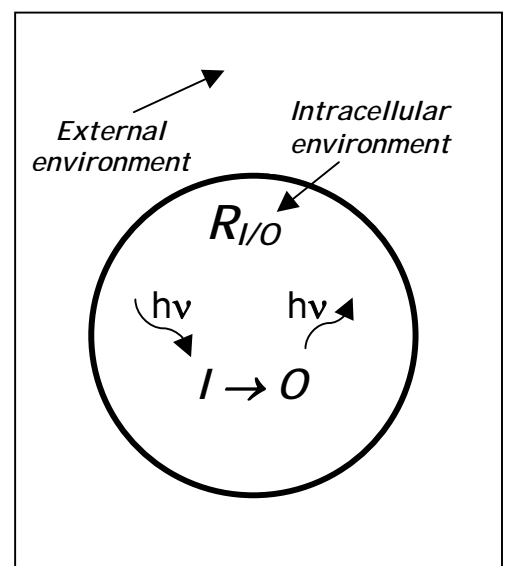

Activation energy
≠ thermal energy

In vivo

## 13. Requirements for unitary evolution

We will leave the third question for later, but will note that, in fact, the requirement of unitary evolution of the system 'molecule + the rest of the cell' might help us to deal with the first question. To avoid terminological confusion with the mathematical concept of 'product state', we will refer to the substrate and product states of a target molecule $T$ as the 'input' $|I\rangle$ and 'output' $|O\rangle$ states, correspondingly. Let us consider description of a particular intermolecular transitions at the level of the whole cell, as a change between two states of the cell: $|C_I\rangle = |I\rangle|R_I\rangle$ and $|C_O\rangle = |O\rangle|R_O\rangle$, where $|I\rangle$ and $|O\rangle$ are the two alternative states of the catalyzed molecule $T$ (input and output, correspondingly) and $|R_I\rangle$ and $|R_O\rangle$ correspond to the state of the rest of the cell. Given the condition of unitary dynamics, the energies of the two alternative states of the cell $E_{CI}$ and $E_{CO}$ have to be equal – which implies an exact compensation of the energy differences between substrate and product states of $T$ by reciprocal differences in the energies of the rest of the cell:

$\Delta E_{I/O} = -\Delta E_{Ri/Ro}$                                                                              [13.1]

Thus, we are able to extend our description to enzymatic transitions other than a change in molecular chirality. This is important, as it supports the general applicability of our idea and thus potentially fundamental role of the catalytic forces $Cf$ that we want to study.

However, we got a little more than we bargained for. There is an additional side to the requirement that enzymatic transition can be described as a unitary process at the level of the whole cell. It concerns the origin and fate of the *activation energy* ($\Delta G^\ddagger$) – we will have to require that the activation energy has to be confined to the intracellular environment.

Usually, when we consider a molecular transition at the level of an isolated target molecule, or as a part of enzymatic process *in vitro*, we presume that the energy necessary to overcome the kinetic barrier[5] is thermal energy, originating from outside of the described system (i.e., complex between enzyme and the target molecule, depicted here in two states: *EnzI* and *EnzO*). At least, the Eyring-Polanyi equation in chemical kinetics (Evans and Polanyi, 1935; Eyring, 1935) is based on calculating the probability of transition state *TS* on the basis of the energy differences between this state and the ground state – and this approach is most easily justified if we assume that these two states are in equilibrium due to interaction with a thermal environment

---

[5] activation energy, graphically depicted here as quantum of energy hν delivered from outside of the system

(middle of the Figure).

However, in order to describe enzymatic process as unitary dynamics, we cannot use such a simple idea. The reason is the following – if our system (cell) needs to interact with its environment for the enzymatic transition to happen, it will imply its entanglement with the environment – and thereby decoherence and loss of unitary character of its dynamics. Therefore, the activation energy for an enzymatic transition has to be confined to the cell, if we want the unitary description to apply. If we divide all the environment of our molecule **T** into two parts (right part of the Figure): the intracellular environment (the rest of the cell, $R_{I/O}$) and the external environment of the cell (the rest of the Universe), we can formulate the condition imposed by unitary evolution as the requirement for the activation energy h𝜈 for any enzymatic act to come from and be lost back to the intracellular environment ($R_{I/O}$) – with no information about the transition leaking outside the cell.

…

Incidentally, this second requirement might help us to derive the proposed self-adjusting effect of **Cf** on cellular microenvironment in the following economical way:

1) If the activation energy is confined to the cell, it has to be included in the physical description of our system as part of its internal energy.

2) Then the *principle of minimum energy* has to apply – seeing to it that the cell will favor the state where this energy is minimized.

3) However, a decrease in the activation energy necessary for an enzymatic transition to happen corresponds to a more efficient catalytic process – that is, the cell will be forced to assume the state where the enzymatic transition occurs most efficiently.

Even more briefly, the physics itself, via the principle of minimum energy, tells the cell to assume an 'optimal' configuration.

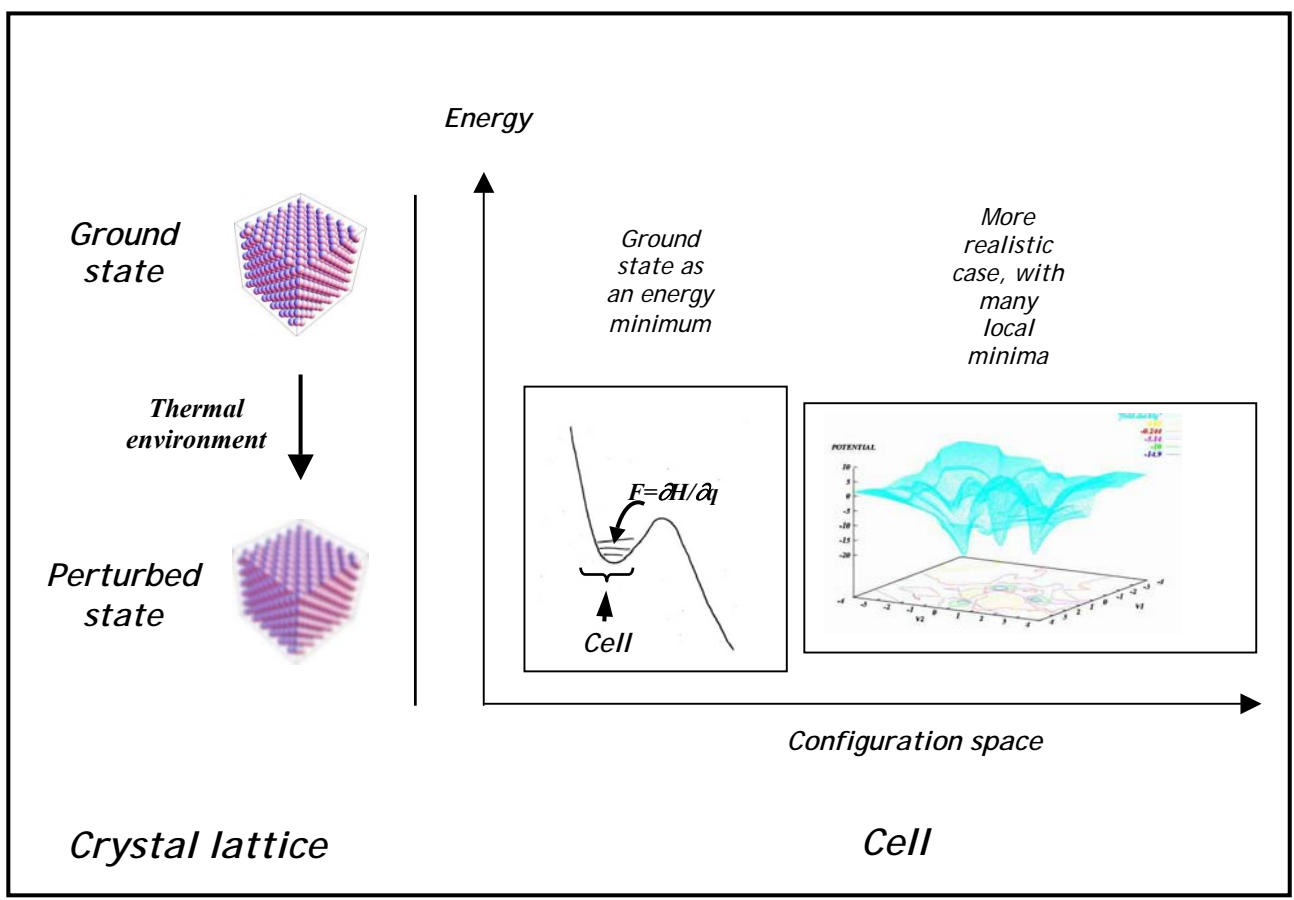

## 14. Ground state

It might seem too much to ask – if we want to describe intracellular dynamics as a unitary process, every enzymatic act occurring in the cell has to satisfy two extraordinarily strong conditions: 1). The energy difference between substrate $|I\rangle$ state and product state $|O\rangle$ of the target molecule $T$ has to be *exactly* compensated by the differences in the energies of corresponding states of the rest of the cell ($|R_I\rangle$ and $|R_O\rangle$), and 2). The activation energy has to be confined to the intracellular environment (i.e., the target molecule $T^6$ should receive it from the rest of the cell $R_{I/O}$ to cross the barrier and then release it back to $R_{I/O}$ after the enzymatic transition occurred).

Needless to say, unitary description has its appeal – it is simpler, and also implies the intriguing self-organizing effect of catalysis *in vivo*. On the other hand, how realistic such perfect fine-tuning of intracellular microenvironment could be? The cell is a 'wet and warm system' – and the interaction of the cell with the external environment is expected to destroy any delicate arrangement of this sort.

I will argue now that, in order to benefit from such a description, we do not need the real cell to be in a unitary evolving state. It is enough for us if it was kept sufficiently close to it. Another side of the same argument will be that it is exactly the *Cf* that can ensure this. Namely, we should look at the unitary evolving state as an ideal situation akin to the notion of a *ground state* used in the physics of condensed matter – whereas the *Cf* force will assume the role of a reactive harmonic-like force keeping cell close to this ground state.

Consider an example of a crystal lattice (left part of the Figure). In mathematical physics, the way to describe this object is to start with the most symmetric state – where all atoms (or molecules) are perfectly aligned relative to each other (top). This state of the lattice (ground, or vacuum state) has lowest energy, but it can only exist on paper, i.e. at the zero temperature. To have a more realistic case, we consider interaction of the crystal with external environment, which has an effect of disturbing the perfect alignment in the lattice (bottom). The fact that real crystals exist at regular temperatures implies that they can withstand this interaction. It is convenient to explain the resistance of the lattice to a thermal perturbation via action of harmonic forces that pull the system back to its ground state. This implies that a disturbance will generally increase the energy of the lattice – and the energy difference between the disturbed and the ideal (ground) states will determine the strength of the reaction force.

---

[6] or the complex between enzyme and $T$

It is precisely from the same perspective of 'ground state' that I suggest to consider the unitary evolving state of the cell. Consider first an ideal case – a hypothetical state of our cell characterized by enzymatic activities happening with most efficiency. Consider now a change in the cell state that decreases the enzymatic efficiency (e.g., **B → A**). According to the scheme presented previously (*slide 9*), this change will not be favored energetically − and thus will lead to generation of a physical force (*Cf*) pulling our system back towards the ground state after the perturbation (right part of the Figure, top). Thus, from the ground state perspective, we should interpret our catalytic forces *Cf* as the harmonic-like forces of reaction, keeping the state of the cell close to the state where all enzymatic acts work perfectly well. Such a state would be characterized by the constraints imposed by unitary evolution, and exist in the same ideal sense as the ground state of a crystal lattice[7]. Although highly unlikely and unphysical at the first glance, the 'unitary state' concept can be valuable in helping us to understand the physics behind the organization of real cells.

Two comments are in order.

1) First, regular condensed states of matter can relatively easily emerge via phase transition from the less disordered states. However, no biologist would seriously consider a possibility that if we take a cell apart and then add all intracellular components (proteins, nucleic acids, etc) back together, an alive cell would emerge from such a mix. Evidently, the similarity between 'living matter' and more regular states of condensed matter breaks down at this point.

However, the notion of a ground state preserved by the reactive enzymatic forces *Cf* is helpful regardless of the differences. We need to discriminate between two questions: 1). One of stability - how the functionally organized state of the cell is preserved in spite of its interactions with thermal environment, and 2) one of history - how this ordered state has came into existence in the first place. Unlike in the case of regular condensed matter, these two questions are clearly different – due to the huge number of possible forms that Life can take. Importantly, only the first question (of stability of particular organized state of cell) is addressed by the ground state concept – in a similar to the condensed matter physics way. As for the second separate issue of why and how such a state came about – we have a difference between physics and biology. Given unique character of the ordered phase in many cases of inanimate condensed matter, it usually suffices to appeal to the shift in balance between the forces of disorder and forces of order (e.g., at a critical temperature) to explain its appearance. Biology, on the other hand, has to implicate *history* in the origin of the ground state and involve mechanisms of inheritance and billions of years of evolution.

2) Second, since DNA contains information about all enzymatic activities responsible for the reaction forces, one could be tempted to think of a unique ground state of the cell encoded by its genetic sequence[8]. However, as implied above, a single ground state of the cell uniquely determined by genome is a gross oversimplification. There are two additional reasons to expect that there should be multiple local minima instead of a global one (right part of the Figure, bottom). First, epigenetic information introduces an hierarchy of stabilities and time scales in the state and dynamics of living cell (Ogryzko, 2008a). Second, the requirement of all enzymatic acts being optimized at once might never be satisfied even at the zero temperature – the logic of organization of all intracellular processes in the four-dimensional space-time might be internally inconsistent. Encouragingly though, condensed matter physics is familiar with this situation. For example, several theoretical approaches has been developed to deal with the presence of frustrations in spin glasses (and other cases of soft matter), typically associated with existence of multiple local energy minima. Importantly, each local minimum will still be maintained via the harmonic-like reaction force mechanism outlined above – thus the ground state concept is still useful.

---

[7] What is ideal and what is real is relative – for those subscribing to the idealistic philosophy, the ground state would be the most real state of the cell, its platonic 'form'

[8] again, for philosophically inclined, this would mean that genetic information literally encodes the platonic form of the cell

## *14A. Decoherence suppression and molecular stability.*

I started with suggestion to describe enzymatic activity as 'decoherence suppression by a specific microenvironment' (section *5*), and then proposed the existence of a *Cf* force effect that should act back on this microenvironment and adjust it to a more efficient 'decoherence suppressing' state (section *9*). To illustrate the origin of the *Cf* force, I used as a model the molecular hydrogen ion, held together by the electron exchange between two protons (section *8*).

But, reciprocally, this parallel also implies that we can treat the issue of molecular stability according to the same 'decoherence protection scheme'. In the case of our molecular hydrogen ion, we should be able to consider the two protons *'p'* as a decoherence-suppressing microenvironment, 'catalyzing'[9] the transition of the electron *'e'* between two alternative locations. From such a perspective, the molecular structure is stable due to the *Cf*-like force that acts to optimize the 'catalytic action' of *'p'* on *'e'* – i.e., adjusts this microenvironment towards better protection of the delocalized electron state from decoherence due to its interaction with the external environment.

The above view is not common – entanglement, let alone 'decoherence protection', usually do not enter the description of the forces that shape molecular structure. However, given that the problem of molecular shape is related to the issue of 'quantum to classical transition', and due to the increasing role of the notion of entanglement in physics, it would be interesting to see if we can bring entanglement in the picture. Let us explore further the possibility to consider molecular structure from the 'decoherence suppression' point of view.

First, a clarification. As long as we limit our description to the ground state, neither the entanglement between the '*p*' and '*e*' degrees of freedom, nor decoherence (as entanglement of the molecule with the external environment) appear to be essential to explain molecular structure. Indeed, the notion of 'ground state' implies the absence of external environment – and hence no need for the concept of decoherence. As for the entanglement between the '*p*' and '*e*' degrees of freedom – the ground state is fairly well described with the Born-Oppenheimer approximation (BOA). The very idea of the BOA simplification scheme is in separating the electron and nuclei degrees of freedom, by representing the state of the molecule as a product: $|\Psi\rangle = |\Psi\rangle_e |\Psi\rangle_n$. Thus, as long as we focus on the ground state (a molecule with the nuclei in the equilibrium position), we have no particular reason to include entanglement (either between the parts '*p*', '*e*' of the system or between the system and the external environment) in consideration.

Nevertheless, given the strong coupling between the nuclei and electrons in a molecule, one still needs to appreciate the fact that the proper description of ground state requires describing these two parts of the molecule as entangled. In the case of molecular hydrogen ion this corresponds to:

$$|\Psi\rangle = (|\Psi_1\rangle_e |\Psi_1\rangle_n + |\Psi_2\rangle_e |\Psi_2\rangle_n)/\sqrt{2}, \qquad [14a.1]$$

where the two components of the superposition correspond to the states of the molecule with the electron close to the 'left' or the 'right' proton. Thus, we need rationale why we are allowed to disregard this entanglement and represent $|\Psi\rangle$ simply as a product $|\Psi\rangle = |\Psi\rangle_e |\Psi\rangle_n$. I want to propose now that the notion of 'decoherence protection' (i.e., avoiding entanglement of *'e'* with environment) could provide such a justification.

To be able to talk about decoherence, we now want to consider the molecule in a more realistic setting – i.e., with outside environment implied. We are interested in obtaining conditions when the information about the electron location in the molecule cannot leak to the outside environment. Theoretically, the information about electron location could be accessed either *a)* directly by measuring the position of electron itself, or *b)* indirectly via observation of the proton state (i.e., accessing the information about the *'e'* state that is encoded in the *'p'* degree of freedom via entanglement, as depicted in **[14a.1]**). The notion of 'decoherence protection' implies that neither way to access the state of the electron by the outside environment is possible in the case of molecular structure.

*a)* The analysis of the first possibility (direct measurement of electron position) leads to another clarification. In fact, the electron position can be observed – but only if the molecule is irradiated with light of a sufficiently short wave-length. However, this will have a price of destroying the molecular structure – as the energy of photon will be high enough to excite the electron and break the molecule apart. Therefore, we have to clarify that we are limiting our discussion to the bound state – and do not consider high temperatures or other laboratory conditions that are not compatible with the existence of the bound state.

---

[9] 'Facilitating' could be a more common way to describe it, with no change in meaning.

*b)* Let us now focus on the nuclei degree of freedom *'p'*. 'Decoherence protection' means in this case that the environment cannot tell the difference between the $|\Psi_1\rangle_n$ and $|\Psi_2\rangle_n$ states of the protons. But then, for most intents and purposes we can consider $|\Psi_1\rangle_n = |\Psi_2\rangle_n$. Therefore, we can take it out of the brackets in the expression **[14a.1]**, obtaining

$$|\Psi\rangle = (|\Psi_1 + \Psi_2\rangle_e)|\Psi\rangle_n/\sqrt{2}. \qquad [14a.2]$$

In other words, we can represent the state of the molecule as a product – thus essentially using the decoherence protection argument to arrive at BOA.

Now, starting with the same entangled description of the ground state **[14a.1]**, we will explore a slightly different route to BOA, which will automatically include 'corrections' to it. Let us consider an important property of entanglement – it's invariance to a change in basis. For example, if we introduce a different basis for description of both *'e'* and *'p'* parts of the molecule:

$$|\Psi_+\rangle = (|\Psi_1\rangle + |\Psi_2\rangle)/\sqrt{2}$$

$$|\Psi_-\rangle = (|\Psi_1\rangle - |\Psi_2\rangle)/\sqrt{2} \qquad [14a.3]$$

, in this new basis, the same state of our molecule will again look entangled:

$$|\Psi\rangle = (|\Psi_+\rangle_e|\Psi_+\rangle_n + |\Psi_-\rangle_e|\Psi_-\rangle_n)/\sqrt{2}, \qquad [14a.4]$$

But what is the meaning of the two components in this new representation? Given the above decoherence protection argument,

$$|\Psi_1\rangle_n \sim |\Psi_2\rangle_n, \text{ then } |\Psi_+\rangle_n \sim |\Psi_1\rangle_n \sim |\Psi_2\rangle_n{}^{10} \qquad [14a.5]$$

i.e., the 'symmetric' term $|\Psi_+\rangle_e|\Psi_+\rangle_n = (|\Psi_1 + \Psi_2\rangle_e)|\Psi_+\rangle_n$ could more or less correspond to the BOA description of the state of the molecule. But what about the second term $|\Psi_-\rangle_e|\Psi_-\rangle_n$, reflecting the asymmetric combination of the 'left' and 'right' states? Since we assumed that $|\Psi_1\rangle_n \sim |\Psi_2\rangle_n$, the contribution of the second term at first seems negligible, because it appears that

$$(|\Psi_1\rangle_n - |\Psi_2\rangle_n) \sim 0, \text{ thus } |\Psi_-\rangle_e|\Psi_-\rangle_n \sim 0 \qquad [14a.6]$$

Following this route takes us back to the BOA, as we obtain $|\Psi\rangle = |\Psi_+\rangle_e|\Psi_+\rangle_n + 0.$

Now the crucial point. The two states of microenvironment: $|\Psi_1\rangle_n$ and $|\Psi_2\rangle_n$ do not have to be identical for the decoherence protection to work. Consider, for example, that information about the state of *'e'*: $|\Psi_1\rangle_e, |\Psi_2\rangle_e$ is encoded in the phase difference between the corresponding states of *'p'*: $|\Psi_1\rangle_n, |\Psi_2\rangle_n$. Then no information about the state of *'e'* will leak into the outside environment of the molecule – i.e., the state will be protected from decoherence. But strictly speaking, the asymmetric term $|\Psi_-\rangle_e|\Psi_-\rangle_n$ will not be equal zero. In other words, the exact description of the ground state **[14a.1, 14a.4]** differs from the BOA description (which is an approximation) – by the correction term $|\Psi_-\rangle_e|\Psi_-\rangle_n$.

Does this correction term matter? As long as we are working with an isolated molecule in statistical ensemble, it is reasonable to disregard the phase difference between the $|\Psi_1\rangle_n$ and $|\Psi_2\rangle_n$ states. For all intents and purposes, the exact presentation is not necessary – this is perhaps why the BOA works so well in such situations. However, if we now consider a larger system *A* composed of many interacting molecules, the relative phases of their correction terms could come into play in the description of the total system. As the phases can add both positively and negatively, the resulting interference can affect observable characteristics of the system *A* by changing probabilities of some events. In this respect, it is tempting to relate the $|\Psi_-\rangle_e|\Psi_-\rangle_n$ term with another famous correction to BOA – Berry phase and related concepts (Berry, 1984; Mead and Truhlar, 1979), but the exact link between these notions is beyond the scope of this presentation.

So far, we were discussing molecular structure. But similar corrections should equally apply for our BOA-like deviation of the *Cf* force, based on a bipartition of the cell to a target molecule and the catalytic microenvironment. For every enzymatic act (*I* ↔ *O*), we can represent the state of the cell close to the ground state as an entangled state, in different ways:

---

[10] Modulo the normalization coefficient

$$|C\rangle = |I\rangle|R_I\rangle + |O\rangle|R_O\rangle \quad , \text{or} \quad |C\rangle = |+\rangle|R_+\rangle + |-\rangle|R_-\rangle \qquad [14a.7]$$

, where $|I\rangle$ and $|O\rangle$ are the two alternative molecular shapes of the catalyzed molecule (input and output of a catalytic act, correspondingly), $|R_I\rangle$ and $|R_O\rangle$ correspond to the state of the rest of the cell; and $|+\rangle$, $|-\rangle$, $|R_+\rangle$, $|R_-\rangle$ are their symmetric and asymmetric combinations. (For simplicity of the argument, we presume equal contribution to the superposition).

Incidentally, this is an entanglement between an element of the structure and the rest of the structure, something that has been termed 'global entanglement' (Chandran et al., 2007) – as opposed to regional entanglement, which links separate (and more or less localized) elements of a complex structure.

There is an important difference from the molecular BOA discussed above. For every individual bipartition of cell to a target molecule and the corresponding microenvironment, the 'nuclei-like degree of freedom' (microenvironment $R$) will be composed of elements (molecules) – each of which being a target part for some other bipartition. One could hypothesize that the asymmetric correction terms $|-\rangle|R_-\rangle$ could play a role in the integration of the energy contributions of different bipartitions into the effective cell Hamiltonian – a nontrivial task, which will be briefly discussed again in section *27.*

*15. Force to recon with*

Next we address the question of how strong can the 'catalytic forces' be – as compared to the regular weak and strong inter- and intra- molecular forces holding the cell together. I will suggest arguments supporting their potential to play appreciable role in the stability of the cell's ordered state. I admit that these arguments might not completely satisfy every reader. However, given the potentially high biological significance of the ***Cf*** (i.e., its 'self-organization' role), the stakes justify even a 'back of the envelope' estimates – which will hopefully lead to more rigorous treatments of the ***Cf*** force energy.

Following the arguments at the end of the section *13*, we consider the activation energy as a part of the internal energy of the cell[11] – and then apply the principle of minimum energy that should force the cell towards the state where its energy is minimized. Given that the decrease in activation energy corresponds to how much the kinetic barrier (i.e., the energy of the transition state TS) has been lowered, my estimate is based on the identification of the energy contribution of ***Cf*** with how much the energy of the transition state of the catalyzed molecular transformation is lowered by this enzymatic activity, as compared to the non-catalyzed transition.

Typical acceleration in enzymatic reactions is $10^{10}$ to $10^{15}$ times (Wolfenden, 2003) – which translates to the decrease in the energy of transition state (TS) around 24 kT to 34.5 kT. Now consider two cell states: $|A\rangle$ and $|B\rangle$, such that a particular transition $|C_I\rangle \leftrightarrow |C_O\rangle$ is facilitated 100 times in the $|B\rangle$ state compared to the $|A\rangle$ state. Accordingly, we can expect energy gain around 4.6 kT. Compared to the typical energies of the conventional physical forces contributing to the cell organization (4 kT for dipole-dipole interactions, 5-19 kT for hydrogen bonds, 100-150 kT for covalent bonds), this contribution is comparable to that of weak interactions and thus might indeed have a considerable role in shaping intracellular structure and dynamic.

Moreover, we have estimated a contribution of only one enzymatic act (i.e., $|I\rangle \leftrightarrow |O\rangle$ in the response to a particular perturbation of the cell state (i.e., $|A\rangle \rightarrow |B\rangle$). However, in general, a single perturbation could affect several enzymatic activities at once – each individually contributing to the total force of reaction to this perturbation. Accordingly, the total force could be significantly stronger than the above estimate made for a single enzymatic act. On the other hand, if we have a frustrated system (see the slide *14*), we will need to consider a local minimum instead of a global one – together with a possibility that some enzymatic transitions could actually favor the perturbation. It is a nontrivial issue how to calculate and then integrate the contributions of different enzymatic activities in the effective Hamiltonian of the cell. Although it is beyond the scope of this presentation, this question will be briefly touched upon later (***27, 33***).

*Slide 16. How to make this idea work?*

We still have one question unresolved. What relevance this could have to real biological systems – which are all supposed to be open systems, far from equilibrium and dissipating energy for survival? The requirement of unitary description is not consistent with any of these properties.

---

[11] Strictly speaking, the notion of activation energy is meaningless, when the catalytic transition is considered in the context of whole cell in ground state, which is evolving in unitary way. However, if we focus on an individual enzymatic act *in vivo,* we can trace out the state of the rest of the cell and consider the catalyzed transition $\rho_{es} \leftrightarrow \rho_{ep}$ (between the enzyme-substrate complex ***ES*** and enzyme-product complex ***EP***, correspondingly) 'as if' it was thermally activated barrier crossing (modulo tunneling effects). See the Appendix 1 (Section 46) for the discussion of the similarities between the description of an individual enzymatic act *in vivo* and *in vitro,* that should justify the use of the notion of activation energy.

## Slide 17. Role of non-equilibrium and irreversibility in Life is overrated

The above objection is not as damaging as it might seem. In fact, the role of irreversibility and non-equilibrium in Life is significantly overrated.

To clarify this claim, I propose first to distinguish between two situations when physical irreversibility appears in dealing with a biological system. One is an experimental situation of measuring a particular property of the system − typically, an enzymatic activity (either *in vivo* or *in vitro*). As has been discussed in section **5**, in this case the irreversibility is unavoidable. The experiments could be only performed by addition of a substrate (input state of the molecule *T*) to the enzyme[12] – and observing the appearance of the output state. Clearly, if both input and output states are present in equilibrium quantities, there could be no increase in the output state even if the enzyme was active – i.e., no activity could be *detected*. In the language of quantum theory, to measure enzymatic activity, we have to prepare a 'product state' (in the mathematical sense) between the enzyme $|E\rangle$ and substrate $|I\rangle$ – and dynamic evolution of this system will eventually lead to detectable presence of the output state $|O\rangle$:

$$|E\rangle|I\rangle \rightarrow \alpha_1|E\rangle|I\rangle + \alpha_2|E\rangle|O\rangle \qquad [17.1]$$

, which can be described as a quantum operation performed on the state of the target molecule *T*, leading to an irreversible process of transition of pure state $|I\rangle\langle I|$ ('input only') to the mixed state $\alpha^2_1|I\rangle\langle I| + \alpha^2_2|O\rangle\langle O|$ ('input or output').

A very different situation arises when we do not perturb a biological system to measure its property, but leave it to its own devices. Granted, in most of the cases and scales that we observe in living nature (growth, adaptation, evolution etc), we will have to describe a biological system as an open system that dissipates energy. However, this is by no means unavoidable. There are many cases of '*suspended animation*' or dormant states − such as anabiosis, cryptobiosis (exemplified by such polyextremophiles as tardigrades), and sporulation. In these states a living object is not dissipating any energy − but its functional order is preserved enough so in the right conditions it could rise and shine again. Overall, there seems to be a confusion of whether it is necessary for Life to be physically irreversible and nonequilibrium phenomenon.

---

[12] or to the cell, if the experiments are performed *in vivo*

# *The notion of non-equilibrium state is too vague*

## *Quasiequilibrium (metastable) state* or *Disspative structure*

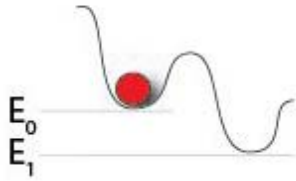

*Double-well potential*

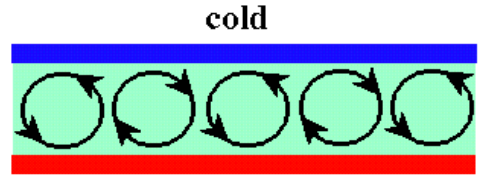

*Bénard convection cells*

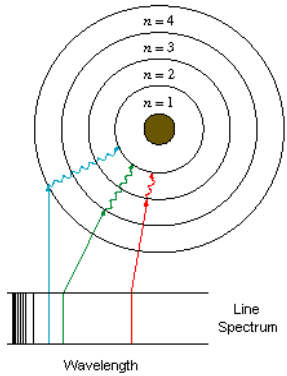

*Borh's atom*

*No need for constant 'work of maintenance'*

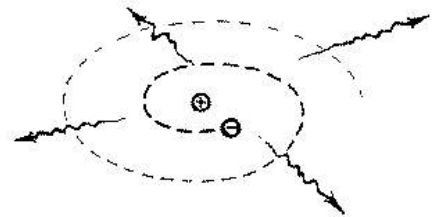

*Rutherford's atom*

*Would require constant pumping of energy*

---

## *18. The notion of non-equilibrium state is too vague*

My solution to the problem is to avoid the notion of non-equilibrium state altogether. This term is too vague and misleading – because it does not allow us to discriminate between two very different situations.

The first one can be called *quasi-equilibrium* or *metastable* state (left part of the Figure) – this is a kind of state that the majority of stable physical objects on Earth (rocks, tables, cars, etc) are in. It corresponds to a local minimum of the energy potential, protected from falling into the lower energy states by kinetic barriers. At a sufficiently long time scale, this state will either tunnel through to the lower energy state, or cross the barrier because of a thermal fluctuation – thus it will be changing with time and has to be considered as out of equilibrium. Most importantly, however, at sufficiently short time scale it will not be changing – thus it can be also considered as a stable state in equilibrium with its environment.

An alternative situation (right part of the Figure) corresponds to the more exotic so called *dissipative structures* (Prigogine, 1969), exemplified by the phenomenon of Bénard instability. An important difference of dissipative structures from the metastable states is in the requirement of constant energy flow needed to keep the dissipative structures. There is practically no time scale at which these structures could persist on their own and thus be considered as in equilibrium with environment – switching off the energy flow will immediately start the process of their relaxation to the equilibrium state.

Intriguingly, the classical and quantum models of atomic structure could also serve as examples of the two alternative meaning of the term 'non-equilibrium'.

The Bohr's model of atom (Bottom, left) could be considered as a metastable state. At sufficiently short time scale any bound state (corresponding to electron 'orbiting' around proton) is stable – although the excited states will eventually decay into bound states with lower energy, or into the ionized atom state, consisting of proton and electron moving separately. On sufficiently long time scale, even the ground state is unstable and will decay into the ionized atom state, provided that the atom is in vacuum.

On the other hand, the Rutherford's model of atom (Bottom, right) is intrinsically unstable. Classical physics does not allow the electron to complete a single orbit around the nucleus without dissipating energy. If we were to keep the state of electron in the Rutherford atom as a stationary orbital, we would require constant 'work of maintenance' and supply of energy to the system. Thus, although usually not considered from this perspective, the Rutherford's atom would correspond to a dissipative structure.

A typical cell contains billions of nuclei and electrons, instead of one proton and one electron. Nevertheless, as one could have guessed from the previous discussion, I favor considering cell as a physically bound state, held together by the laws of quantum theory – much closer in the spirit to the Bohr's atomic model than to the Rutherford's atom.

### *19. Is life a dissipative structure?*

Can the physics of Life be properly described by the theory of dissipative structures? There are many arguments against this idea. I refer you to one influential opinion (Anderson and Stein, 1987) proposing that, physically, living matter rather corresponds to so called 'state of generalized rigidity', not far from equilibrium.

Here I will suggest an independent argument against dissipative structures as a proper physical theory for biology. It is based on what Prigogine, the originator of the theory of dissipative structures, admits himself (Prigogine, 1980) – the mathematical theory of dissipative structures requires *large numbers* of participating components (to satisfy the criterium of local equilibrium, essential for his theory).

However, to expect large numbers of all essential components is not realistic in many cases of biological systems. The physics of Life is *nanophysics*, operating with few copies of many participating molecules. The cells of bacteria provide the most vivid example. If we calculate the number of free protons in an individual cell of *E. coli* − we come to a ridiculously low number of 5 free protons per cell. Taking a more exotic nanocell of micoplasma *Acholeplasma laidlawii,* having the internal volume 100 times smaller that that of *E. coli* − we obtain an absurd number of 0.05 free protons in a single cell. I discuss this problem in more detail later (***24 and 35).***

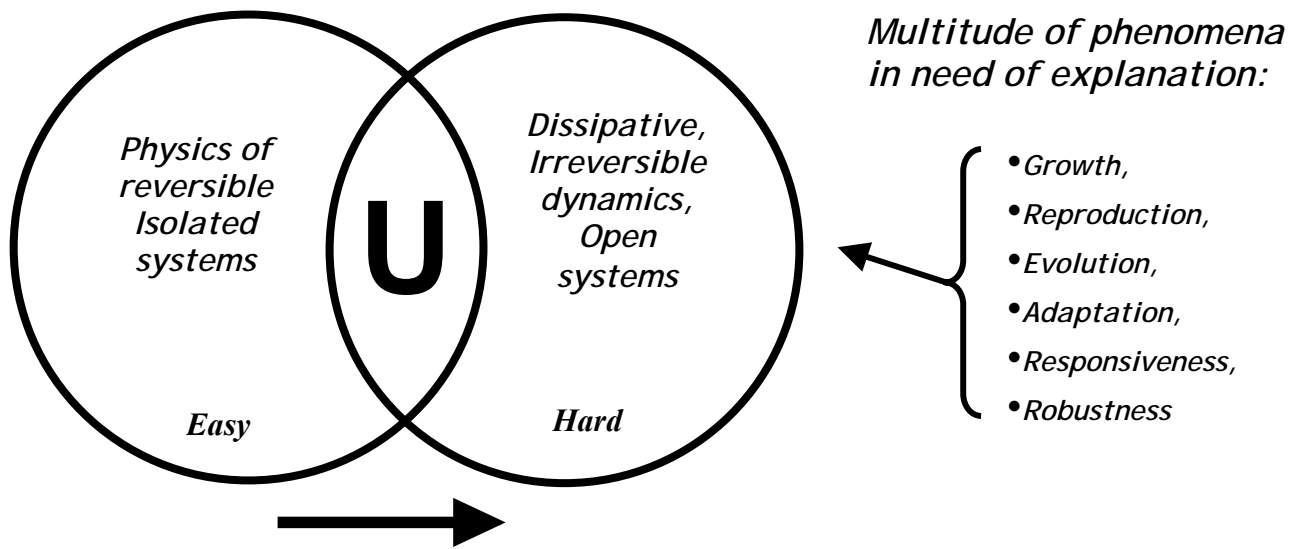

*Start with a problem that is most tractable physically, but at the same time meaningful biologically - 'U' problem*

*Multitude of phenomena in need of explanation:*
- *Growth,*
- *Reproduction,*
- *Evolution,*
- *Adaptation,*
- *Responsiveness,*
- *Robustness*

*After solving the 'U' problem, expand the solution by 'analytic continuation' to other biological phenomena*

## *20. Euclidean strategy. Motivation.*

The previous discussion suggests a new crucial element that any serious approach to the physics of intracellular organization should contain. As already discussed before, we start 'from the first principles' (i.e., from the full quantum mechanical description) – without taking for granted any approximation, usually implied when describing intracellular dynamics. We consider a cell as a system of electrons and nuclei, governed by the laws of electromagnetism – and seek a new approximation that would allow us to write the 'effective equation of motion of the cell'.

The crucial novel requirement that any such approximation should satisfy is the following. For every catalytic act happening in the cell, it has to take into account the reciprocal action of the catalyzed molecule on its microenvironment. This effect could be considered as a novel kind of effective force *Cf* operating in a living cell and contributing to its organization. We emphasized an important difference of *Cf* from the other inter- and intra- molecular weak and strong forces – it has a "self-organizing" effect on the cell state. Also, we argued that for this idea to work, we should use the notion of a ground state of a cell described by unitary equation – so that the actual state of cell is kept in close proximity to it via the action of these harmonic-like 'catalytic forces'. To justify the plausibility of the ground state approach, we also argued that physical irreversibility and non-equilibrium are not unavoidable physical features of the 'living matter'.

As appealing as this approach might be, now we need to address two questions: 1). All that we can say so far is that quantum mechanics suggests an intriguing twist on the physics of biological organization. However, we have not sufficiently explored whether quantum principles are really required to explain Life as we know it. Could we do just fine with classical concepts? 2). Even if unitary description is convenient in some cases, it is evident that eventually we will have to deal with physical irreversibility. Plentitude of biological processes, such as growth, reproduction, adaptation and evolution are associated with energy dissipation (not conservative physical evolution) – and thus will require treatment of biological objects as open systems out of equilibrium. Then, even if useful – should the concept of a cell in the ground state be only a part of a larger picture?

The next part of my presentation will be concerned with these two questions. First, I will argue why quantum theory will be needed for understanding Life. I will propose that taking quantum principles into account will be necessary for addressing the problem of *stability of intracellular dynamics at the level of individual cell* (Ogryzko, 2008a). This issue, although not sufficiently appreciated yet, will be brought to the fore with the increased role of single-molecule methodology and systems methods in biology.

On the other hand, regardless of the issue of stability, there are independent reasons to resort to the quantum theory formalism for description of intracellular dynamics. In fact, it is related to the technical problems of describing physically irreversible processes. All fundamental laws of physics are unitary − i.e., evolving an initially pure quantum state into another pure state. Change to a non-unitary description is beset with technical and foundational problems (Zeh, 2007). Therefore, although we are bound to deal with physical irreversibility, it is a good strategy to keep as long as possible to the unitary description – and give it up as a last resort only.

Accordingly, I propose to see the idea of the cell in a ground state as a part of a broader strategy of description of intracellular processes. Previously, I termed this strategy the 'Euclidean approach' (Ogryzko, 2008b) – in analogy to the related method from Quantum Field Theory. Its main motivation is the following. Mathematical physics has been quite successful in dealing with physically conservative processes – such as unitary evolution in quantum theory. Thus, it would be most useful to find a problem that lies at the intersection between cellular biology and the physics of conservative, unitarily evolving systems. This problem would be most tractable for a mathematical physicist – and on the other hand, still would be meaningful for a biologist. Starting with such a problem would have many technical benefits and should be relatively easy. Only after we have studied it well enough, have found nontrivial solutions and translated their meaning to molecular biology language – would we go one step further and expand these solutions to the more difficult problems of growth, reproduction, adaptation, evolution etc.

We label this hypothetical problem a *'U'* problem, after the term 'Unitary'[13]. The idea of a ground state supported by the harmonic-like reactive enzymatic forces, proposed previously (slides *13, 14*), would be relevant exactly for this part of our approach – and our discussion of the role of irreversibility in biology (slides *17- 19*) suggests that it should be possible to find a problem that fits the description. As the term 'Euclidean' suggests, expanding the solutions of the *'U'* problem to the more difficult cases will correspond to the mathematical procedure of *analytic continuation*, to be discussed later. Admittedly, 'Euclidean' is somewhat an abuse of the term here. However, I hope to demonstrate that it is sufficiently close to its original meaning.

Finally, 'learning to stop worrying and love the unitary description' has an additional advantage. Among many nice properties of the dynamics of conservative systems are symmetries, which we hope to uncover in the analysis of the *'U'* problem[14]. Importantly, in quantum theory symmetries are directly related to the observables of the system. Therefore, if we understand well enough the *'U'* problem, we might be able to mathematically infer, which of its properties can be treated as quantum-mechanical observables. Later, we will see how this helps in justification of our approach to adaptive mutation phenomenon. In fact, this will serve as an example of a more general idea – the extension of our description to more difficult problems that involve irreversibility, growth or adaptation might be interpreted as measurements of some observables of the biological system by its environment.

---

[13] In the hindsight, this is not the best label to use, as one might confuse it with the logical sign 'Or' (union) on a Venn diagram, very suggestive on this slide. Our hypothetical problem lives in the intersection, not in the union of the two sets. Thanks to Bruno Sanguinetti and Nathan Babcock for pointing this out

[14] Most of them will not be obvious, since we are not working with a periodic system like a crystal lattice

## 21. What could be the U state?

What such a (hypothetical) **'U'** problem could be?

To make things easier, we should focus on as simple object as possible. Given our interest in the role of 'enzymatic forces' *Cf* in intracellular organization, this object should be a cellular life-form, i.e., a virus will not do. Procaryotes are simpler than eucaryotes and better studied. Vast amount of information about bacteria has been accumulated in 50 years of molecular biology – including delineation of metabolic pathways, systematic analysis of essentiality of all genes in the model organism *E.coli* and complete sequencing of its several strains and many close and distant relatives. Thus, it could very well be a bacterial cell.

On the other hand, since we want to deal first with a physically conserved system (**'U'** problem), we have to exclude any energy or matter flow through it. We do not want this cell to grow or even consume any substrate. In other words, the object of our study should be a '*starving bacterial cell*'. This state is simple to prepare experimentally – we grow cells in a nutrient medium first, then remove all nutrients by extensive washing, and finally place them in a solution (such as phosphate buffer) that contains only inorganic salts to preserve physiological osmotic pressure. Notably, the cells in this state can survive for several days at room temperature.

One might object that, even when the cell is deprived of any external substrate, it will still dissipate energy derived from internal resources to preserve itself. The answer to this objection depends crucially on the distinction between the two physical views on cellular organization – a dissipative structure versus a metastable (quasi-equilibrium) state. As argued previously (sections *17-19*), the most important difference between these two views is in the role of time scales. Whereas in the case of dissipative structure there is no time scale at which it could be considered as stable on its own – a metastable state will not require energy supply at sufficiently short time scales. Given that we favor the latter view on cellular organization, we can now clarify what we propose to study – it is the intracellular dynamics at the time scales that are sufficiently short, i.e. the time scale when the energy dissipation can be neglected. We wish to explore the idea that enzymatic processes inside the cell are happening on the time scale shorter than the energy dissipation of the starving cell.

Regrettably, although absence of substrate (outside resources, food, nutrient etc) is normal in many biological situations, not much is known about the biology of a starving cell. Most of the data on bacteria have been obtained by studying cells growing exponentially (or in a related so called 'steady state' in industrial fermenters) – i.e., in artificially created conditions of nutrient excess. This is not a physiological situation. The closest to our case would be the studies of stationary phase of cell culture growth. Although extensive literature exists on this subject (Potrykus and Cashel, 2008), it is not immediately applicable to the object that we are considering.

Now, after we pinpointed the object of study, what would be the nontrivial problem that we want to address? As I will be discussing in some length further, it is the problem of stability of intracellular dynamics in the context of 'small numbers and fluctuations'.

## 'Classical' molecular biology

*Studying **one** property using **many** objects*

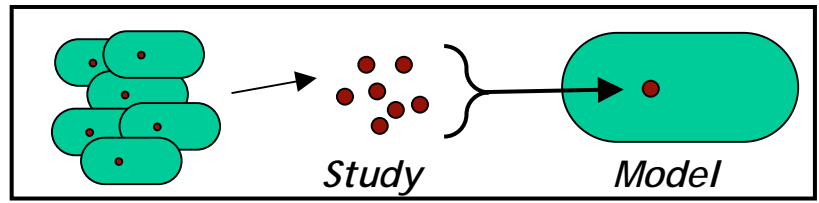

*Study*      *Model*

### Systems biology

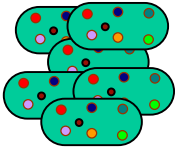

*Study **all** properties using **many** objects*

### Nanobiology

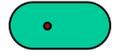

*Study **one** property focusing on a **single** object*

### Systems nanobiology?

*Study of **all** properties of a **single** object*

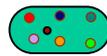

| Can all relevant properties of an 'elementary living object' be known'? | ?? | Fluctuations in context of high dimensionality of intracellular dynamics |

### 22. Why do we need quantum biology on cellular level?

The main reason why I expect the problem of stability of intracellular dynamics to become increasingly important is the progress of *'nano-'* and *'-omics'* approaches in biology. The development of these technologies will bring about new standards of rigor – and will require critical reassessment of the assumptions used for analysis of physics behind Life.

A crucial step in establishment of molecular biology was the recognition of importance of controlling the genetic background of one's experimental system. In the case of bacteria, keeping to this standard of rigor is relatively easy – by isolating an individual bacterial colony from a Petri dish and establishing the so called pure culture. The aim of this cloning procedure is to control effects of genetic variations and eventually tease out the roles of different genes in the phenomenon under question. Accordingly, the pure culture is routinely analyzed by propagating to generate billions of cells – which can be either studied alive or broken apart to isolate many copies of one molecular part of the cell (typically an enzyme) and to analyze these large homogenous ensembles of molecules *in vitro*.

Setting this level of rigor required forsaking the issue of *cell individuality*, manifested in fluctuations of many parameters in the single cells with identical genome[15]. Otherwise, it proved very beneficial for the success of molecular biology as a science. Nevertheless, from the modern perspective, two limitations of the molecular biological approach are becoming increasingly evident:

1) The use of an ensemble of objects instead of an individual object,

---

[15] as well as neglecting the role of epigenetic factors

and

2) Focusing on only one (or few at best) properties of the studied object.

Recent technological advances open a possibility to overcome these limitations. On one hand, development of high throughput methodologies has led to the 'systems biology' (-omics) approaches. Their ultimate aim is to measure *all relevant properties* of the studied object. However, the technology is not sensitive enough yet to accomplish this analysis at the level of an individual cell.

In parallel, development of ultra-sensitive technologies has led to 'nanobiology', which aims to analyze behavior of *single* molecules – ultimately inside the living cells. On the other hand, we cannot do '-omics' with these methods – i.e., we are still limited to the study of only few of different molecules (or other properties of an individual cell) at once.

Sooner or later, the technological development will merge these two trends into the 'Systems nanobiology' – aiming to study all relevant properties of a single individual cell. The analysis of individual cells, instead of pure cultures, will eventually become a new standard requirement for the study of intracellular processes. My point, however, is that upholding these new standards of rigor will also end molecular biology as we know it – as it will lead to critical revisiting of the assumptions taken for granted in analysis of physics of Life. Namely, it will bring to the fore two fundamental and related questions:

1. Can all relevant properties of an individual cell be known at once?

2. How the intracellular dynamics can be stable?

Addressing these questions will require taking quantum theory into account.

## 23. Can all relevant properties of an individual cell be measured at once?

As for the first question, here is an illustration of the problem that we will face analyzing an individual cell. Consider a cellular process controlled by an individual enzyme E, interconverting its target molecule between states $|I\rangle$ and $|O\rangle$ (i.e., substrate and product)[16]. Suppose that we want to know two different properties of our single cell: 1) the state of a particular copy of this molecule $T$ in the cell at a particular moment and 2) the ability of the cell to interconvert between $|I\rangle$ and $|O\rangle$ states of the same very molecule at the same very moment. Further suppose that with the help of an ultra-sensitive nanotechnology we determined that the cell is in one of the states $|C_I\rangle$ or $|C_O\rangle$ (since we do not want to miss possible role of intracellular microenvironment, we have to use the states of the whole cell in our description). My claim is that the above procedure will provide enough disturbance in the state of the cell to interfere with the process of transition between the cell states $|C_I\rangle$ and $|C_O\rangle$. This is because an enzymatic transition requires interaction with the microenvironment – and placing the cell in the measurement situation alters this environment. As a result, measurement of one observable (cell in state $|C_I\rangle$ or $|C_O\rangle$) will affect another observable – catalytic activity, i.e., the rate of transition between the $|C_I\rangle$ and $|C_O\rangle$ states.

This is an example of the fundamental limitations on our abilities to observe simultaneously several different properties of a single cell. To take these limits into account will require use of an appropriate language to describe the experiments with individual cells. It is natural to expect that the language we are looking for would be the formalism of quantum theory. This would allow us to describe the above experiments as *bona-fide* quantum measurements of non-commuting observables of the whole cells – considered here as enormous conglomerates of billions of nuclei and electrons, but as finite and bounded physical systems nonetheless. As far as the previous example is concerned, the quantum Zeno effect could be a simple way to describe the effect of measurement of the molecular state (either $|I\rangle$ or $|O\rangle$) on catalytic activity (i.e., rate of transition between $|I\rangle$ and $|O\rangle$).

Needless to say, cells also have plenty of classical properties. A typical example are the position and impulse of their center of mass. But biology is hardly interested in the motion of cells as physical bodies in the three-dimensional space. The most interesting things about the cells are the individual enzymatic processes taking place at the molecular level – and as argued above, they will require non-commuting operators for description. Later, I will provide an illustration of how similar consideration could be used to approach the phenomenon of adaptive mutations.

In quantum theory, the limitation on what can be observed simultaneously results from the fact that the procedure of measurement generally perturbs the object under study. On the other hand, it is also common these days to consider the environment of an object as a kind of observer. Thus, the second question, that of stability of intracellular dynamics, is ultimately related to the first one (i.e., 'can everything be known about an individual living cell?') – as it could be formulated as a problem of resistance to perturbations due to the 'measurement' by environment. I will address it in more detail below.

---

[16] for an example that shows that these concerns are relevant, see Choi, P. J., Cai, L., Frieda, K., and Xie, X. S. (2008). A stochastic single-molecule event triggers phenotype switching of a bacterial cell. Science *322*, 442-446.

## 24. Problem of fluctuations and stability of intracellular dynamics

Erwin Schroedinger discussed fundamental difference between the processes studied by the physics of his days and the processes in the living organisms (Schroedinger, 1944). Whereas most of the physical laws have a statistical nature (i.e., physical phenomena are reproducible because of the laws of large numbers) – biological phenomena involve 'incredibly small groups of atoms, much too small to display exact statistical laws'. This observation has led him to support the idea that genetic inheritance, one of the most obvious manifestations of stability of biological organization, is based on molecular structure. Structure of molecules is explained by quantum chemistry – which does not require large numbers to account for the stability of molecules. This is in contrast with statistical physics, where the resistance of a typical macroscopic state to perturbations is enforced by le Chatelier principle ('any change in status quo prompts an opposing reaction in the responding system') – and owes to the fact that average fluctuation size of the number of molecules ($\sqrt{N}$) is extremely small compared to the number of molecules itself ($\sqrt{N} \ll N$).

Translated to a more modern language, the Schroedinger's point was that *physics of life is nanophysics*. Success of molecular genetics confirmed the insight that molecular code-script (based on the stability of the phosphodiester bond, explained by quantum mechanics) serves as the basis for genetic information. On the other hand, another major question remains unresolved – that of the fundamental physics behind the rest of molecular phenomena in the cell. Those are represented by many information-processing activities – such as decoding of genetic information and homeostatic regulation, which supports the rest of cellular metabolism and includes signal transduction, negative and positive feedback loops, check-points etc. Owing to the success and high reproducibility of biochemical approaches in dissecting and modeling the individual steps of these processes, up to this day the rest of the cell is treated as a physically macroscopic system. As a result, we are left with an unsatisfactory 'hybrid-like' view of the fundamental physics behind intracellular organization – whereas the stability of genotype (DNA structure) is enforced by the laws of quantum theory, the stability of phenotype (most of the other molecular processes in the cell) is explained by classical statistical mechanics.

As alluded to previously, the advent of the '-omics' and 'nano-' technologies should bring to the fore the main difficulty of this centaur-like picture of the cell, which is rooted in the *uncritical reliance on coarse graining* procedure implicitly used to account for the stability of intracellular processes. Cell, as the object of 'systems nanobiology' is a 'supra-macromolecular machine' with a tremendous number of variables, on the one hand – and operating at the level of individual molecular interconversions, i.e., in the regime of thermal fluctuations, on the other hand. Starting from the first principles, cell should be considered as a hierarchically organized complex of macromolecules – and its state should be described by a density matrix operating on the high-dimensional Hilbert space specifying positions of every electron and nucleus in it. This is a default description, and any coarse graining approximation should first be thoroughly justified. This is where challenge for the 'nano-' and '-omics' arises – whereas 'nano-' trend limits our ability to average over an ensemble of identical copies of a particular molecular species, the '-omics' approach, taken to the logical extreme, will not leave any other non-relevant degrees of freedom to average over.

Coarse graining is a procedure of replacing the detailed 'microscopic' description of a complex system with more convenient 'macroscopic' variables – usually by averaging out non-relevant degrees of freedom (DOF). A typical example of a coarse grained variable is the notion of 'concentration' – which is based on the assumption that the location of each individual copy of a particular molecular species is not relevant and thus can be averaged out. Coarse graining simplifies the problem at hand and also provides a way to make the variables used for description smooth and well behaved. For example, if we model an enzymatic process *in vitro* using a large homogenous ensemble of molecules that contains many billions of identical components – an average fluctuation is extremely small compared to the number of molecules itself. Equally, when we work with a cell population that contains millions of cells, averaging also allows us to represent the state of the molecule of interest as a concentration correctly to within the limits of an experimental error.

Importantly, however, the notion of concentration is hardly applicable when we are dealing with single cells containing only few copies of a particular molecule – especially if a local context of an individual molecular copy could make a difference. The fluctuations in the 'concentrations' of each individual molecule will be comparable to the concentrations themselves ($\sqrt{N} \sim N$) – i.e. too great to describe intracellular dynamics with a system of partial differential equations (PDE) using concentrations as variables. This 'nano-' perspective severely limits the classical, biochemistry-oriented view on intracellular dynamics.

To overcome this difficulty, stochastic differential equations have been used with some success (Turner et al., 2004). Notably, most of such attempts have been limited to modeling the behavior of *few variables* only. This is where the trouble with the '-omics' approach lies – as it is concerned with describing simultaneously *many* *thousands* of relevant degrees of freedom, which all behave stochastically at the single cell level. The problem is rarely recognized, perhaps because of a tacit assumption that even if we work at the nanoscale and describe an individual molecule, there's got to be an alternative way to smooth our variables and ensure stability of dynamics – for example, by averaging out many other DOFs that are also present in the cell, but can be considered non-relevant for the process that we wish to describe. As long as we limit ourselves to modeling few of cellular properties at the time, this trick appears to work. However, unlike in classical molecular biology, the intent of systems molecular biology is to know *everything* about the cell – and eventually to model the structure and dynamics of the whole cell by integrating all knowledge about conformations, locations, orientations, interactions etc, of every molecular species in it. Then, if in the spirit of the '-omics' perspective, we want all relevant DOFs in our system to be taken into account – it is not evident that we would have enough of truly non-relevant DOFs – in sufficient numbers for every relevant DOF that we want to describe (see also the ***slide 35*** for a formulation of this problem in terms of '*tradeoff between complexity and stability*'). Thus, other ways to explain stability of intracellular dynamics have to be pursued.

To briefly recapitulate the problem, today we know that stability of biological organization cannot be accounted for by genetic information only (Ogryzko, 2008a), which brings us back to where Schroedinger started – to the problem of small numbers. This time, however, it is not only about the few copies of DNA molecules per cell, but about thousands of its other components that are also present in the small numbers of copies – and also about how the compounded effect of their fluctuations in the context of high dimensionality of the state space impacts the stability of intracellular dynamics (i.e., increases the potential for an 'error catastrophe'). This implies that the familiar explanations of molecular structure by quantum chemistry will not be sufficient to understand the stability of biological organization. Just like Schroedinger prescribed, quantum theory will remain involved – but the focus on molecular structure will not suffice, and the dynamics of whole cell will have to be considered from the quantum-mechanical perspective. Among other things, description-wise, this will also require that the notion of *concentration* of a particular molecular species in cell have to be replaced by a more adequate notion. As previously discussed, it should be the concept of *'observables'* – hermitian operators that act on the state space of the cell and assign to every such state the probabilities of finding an individual molecule in a particular location (also orientation, conformation, etc) of the cell.

## 25. Why Quantum Mechanics?

One can bring up several arguments of how quantum theory can help with the problem of stability of intracellular dynamics. It is known, for example, that unlike in classical physics, the linearity of quantum mechanical description makes chaos more difficult to obtain. A similar argument from linearity also suggests that quantum control, somewhat counter-intuitively, is easier to implement than classical one (Chakrabarti and Rabitz, 2007).

Here, introducing terminology that will be useful later, I will provide some additional arguments of how the appealing to quantum mechanical formalism can help with the problem of fluctuations.

The remarkable resilience of living systems to environmentally induced perturbations (including phenomena of homeostatic regulation, repair, regeneration etc...) is commonly attributed to their ability to monitor their own state, recognize a particular perturbation and then develop a response that keeps the value of the controlled variable within the acceptable range. For convenience, let us introduce the notion of cost of maintenance $M_x$ – as the energy needed to be dissipated in order to keep the value of a particular relevant variable $X$ in a range acceptable for the proper functioning of the cell. The work of maintenance will be required to counteract the effects of fluctuations in many relevant variables (DOFs) of the cell. For a particular DOF $X$, the fluctuation scale will be described by its variance: $Var(x) = E[(X-\mu)]^2$ or simply $\sigma_x^2$, where $E$ – operation of taking an expected value, $X$ – value of a variable and $\mu = E(X)$. Accordingly, the cost of maintaining this DOF should be proportional to the variance; $M_x \sim \sigma_x^2$.

Considering now all variables $j$ that we need to describe the state of the cell with, it is natural to assume that the total cost of maintenance has to be $M_T \sim \sum \sigma_j^2$.[17] It is clear that, since for each $\sigma_j^2 \geq 0$, then always $M_T \geq 0$. This property reflects natural assumption about the work of maintenance – no matter in what direction our variable $j$ deviates from the acceptable range, the cost of bringing it back should always have a positive value. Given that all living systems are constantly perturbed by interaction with their environment, this naturally leads to the claim that every living system is necessarily an open system that requires flow of external resources in order to perform 'work of maintenance' $M_T \sim \sum \sigma_j^2$ to sustain its order.

This also means that $M_T = 0$ only in the case of a system devoid of any fluctuations. However, this claim is only true if we limit our description exclusively to real numbers. In fact, if we use complex numbers, the equation $\sum \sigma_j^2 = 0$ will have many nontrivial solutions, according to the fundamental theorem of algebra. For a simple illustration, consider an equation describing circle with a radius $r$: $x^2 + y^2 = r^2$. If we set $r = 0$, the only solution in real numbers will be a trivial point $x,y = (0, 0)$. However, in complex numbers we have many solutions – as long as the following constraint is respected: $x = \pm iy$. Thus, although associated with interpretational challenges, the use of complex numbers helps to support an interesting and rich state space – even in the absence of work of maintenance.

How one could justify appealing to such an exotic device as complex numbers? Putting interpretation issues aside, their use is not an *ad hoc* step – but is entirely expected from the first principles of the quantum mechanical formalism. According to our previous arguments, more fundamental description of intracellular dynamics will have to replace the notion of 'concentration of the substance $X$' with the more fundamental notion – of 'probability to observe the cell in a state with a particular molecule $x$ in a specified location', represented by a projection operator $\mathcal{X}$. Notice that, instead of different molecules as separate objects $X,Y,...$, and the fluctuations in their numbers – the focus now is shifted to many properties $\mathcal{X}, \mathcal{Y}, ...$ of one composite object (i.e., cell) and correlations between these properties. The use of complex numbers is more natural in this setting – their main role is in taking into account the phase relationships between different states of the cell; this implies that, although the fluctuations in different properties $\mathcal{X}, \mathcal{Y},...$ are allowed, they are not independent from each other, thus allowing for much less cost of maintenance.

An alternative way to arrive to the same point is via the notion of entanglement. This idea was discussed previously (Ogryzko, 2008a), therefore I will not go into much detail. I only mention that taking entanglement into account in description of intracellular dynamics can lead to effective lowering of the numbers of dimensions necessary for description of the state of the system (i.e., acknowledging in this way correlations between fluctuations in different properties) – similarly keeping the cost of maintenance under control.

---

[17] we consider for simplicity of the argument the fluctuations as independent.

# Part 3. Self-reproduction.

*Abstract 3*

We use arguments from the fluctuation-dissipation theorem to justify transition from the description of ground state to that one of growth. For every type of a substrate that a given cell can ever consume, the cell in a close to ground state is proposed to be reversibly generating, as a part of fluctuation behavior, molecules of this substrate. Comparison of the process of relaxation of such fluctuations with the actual process of substrate consumption shows formal similarity, which should allow description of cell growth starting from the analysis of ground state. Also, we revisit the problem of tradeoff between complexity and stability, discussed in the previous part (slide 24) – and suggest how quantum entanglement could help a complex dynamic system to both enjoy high number of essential degrees of freedom and, at the same time, have a comparatively low total number of elements.

## *26. Euclidean strategy. Next step.*

We started our presentation with a bold proposal – to consider, for every catalytic act happening in the cell, the reciprocal action of the catalyzed molecule on its microenvironment (e.g., the cell), and to analyze this effect from the first (i.e., quantum-mechanical) principles. Let's now take a stock of where we are.

We have got quite a lot of mileage out of this simple idea. First, it has lead us to propose a novel approximation to the physical description of intracellular dynamics *in vivo* based on considering a ground state of a cell supported by the reactive harmonic-like *Cf* forces, with their intriguing 'self-organizing' properties. Second, we contrasted this idea with the more established approach to biological organization based on the theory of dissipative structures. We emphasized the advantages of our approach – both of a technical (unitary description of the ground state) and a physical (less dissipation, no need for large numbers) nature. Finally, we acknowledged that the idea of ground state has to be a part of a bigger picture – a first step in the strategy to study the physics of intracellular processes, termed here 'euclidean approach'. According to this strategy, we first explain stability of a ground state (which could be several local minima) as supported via the reactive *Cf* forces – and only afterwards consider transitions between these states, which would correspond to irreversible dissipative processes.

There are certainly many open questions remaining for a better understanding of the concept of a ground state of the starving cell. Given that we usually consider molecular processes in the cell as physically irreversible, it is also an interesting challenge to find the counterparts of the familiar molecular biological notions of regulation, feedback control, signal transduction, coding etc, in the description of *unitary* physical evolution of the cell state – in other words, to recover these notions from the unitary formalism. The opportunity to find how these biological notions can be translated in the mathematical language of geometric and topological properties of the state space describing the ground state of the cell (including its symmetries and (co)homological or homotopic invariants) promises a thrilling intellectual adventure.

I will have no time to pursue these questions much further. Most of the remaining talk will be spent exploring the next step of the euclidean strategy. Assuming that we understand enough the ground state of the cell – how we could move on to the description of the open systems and irreversible processes in the proposed framework? In the further discussion, a little bit more emphasis will be put on the formalism and its interpretation.

*For every enzymatic step I ↔ O*

$|+\rangle = (|Cell_I\rangle + |Cell_O\rangle)/\sqrt{2}$

$|-\rangle = (|Cell_I\rangle - |Cell_O\rangle)/\sqrt{2}$

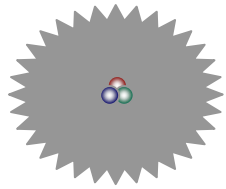

*energy gain $\Delta H_{SP}$*

Molecule     Cell

I/O

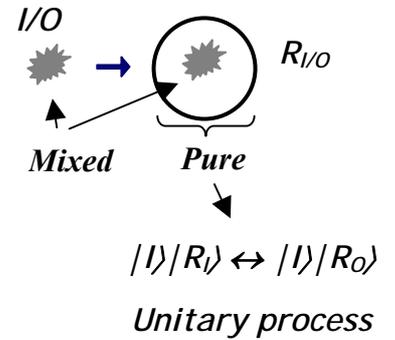

$R_{I/O}$

*Mixed*   *Pure*

$|I\rangle|R_I\rangle \leftrightarrow |I\rangle|R_O\rangle$

*Unitary process*

---

*How to integrate contributions of different enzymatic steps?*

| *Independent* | *consecutive* | *regulation* |
|---|---|---|
| 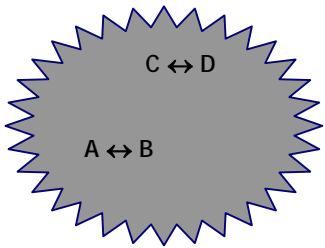 | 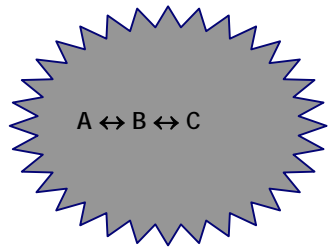 | 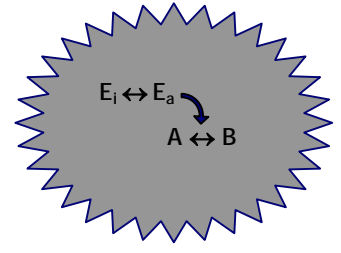 |
| $\Delta H_{AB} + \Delta H_{CD}$ | $\Delta H_{AB} + \Delta H_{BC}$ ? | $\Delta H_{EiEa} + \Delta H_{AB}$ ? |

### 27. How to describe the intracellular dynamics in U state?

Recall now that the *Cf* was proposed as an effective force, i.e., only as an *approximation* to the fundamental description of intracellular dynamics based on the laws of electrodynamics and quantum mechanics. Although admittedly gaining insights into the physics of biological organization, we will need now to consider the limitations of this approximation – if for nothing else but to have a consistent and formalized description of the idea of the ground state. As we will see later, it will also help us in the second stage of the euclidean approach – namely, this description will have to take into account the relation of the cell with its environment, and thus naturally lead to description of irreversible and nonequilibrium processes.

There is something unsettling about the way the existence of the catalytic force *Cf* was derived.

Although we acknowledged that there are many enzymes in the cell all contributing into the lowering of total energy of the ground state, we have not used this fact in any significant way. So far, we limited our analysis solely to the role of a single enzymatic act – by bipartitioning the cell into two subsystems (target molecule *T* that undergoes the enzymatic conversion $|I\rangle \leftrightarrow |O\rangle$ and the rest of the cell $R_T$) and representing the total state of the cell as $|C\rangle = \alpha_1|I\rangle|R_I\rangle + \alpha_2|O\rangle|R_O\rangle$. This was sufficient to illustrate the principle of the *Cf* force (sections *8, 9, 14a*).

However, for a more consistent description of the ground state, we will need to understand how to integrate the contributions $Cf_i$ of all different enzymatic activities $E_i$ operating in the cell – how to add them all together? This is not a simple question. The Figure illustrates three examples, out of many other possibilities. If two enzymatic acts ($E_1: A \leftrightarrow B$, $E_2: C \leftrightarrow D$) are independent from each other and are in different parts of the cell (left), they are more likely to have additive *Cf* contributions to the effective total Hamiltonian $\mathcal{H}_c$. However, there could be more difficult situations, as depicted in the center and on the right: 1) two enzymes ($E_1: A \leftrightarrow B$, $E_2: B \leftrightarrow C$) working in sequence, via an intermediate B; or 2) One enzyme ($E_1: E_{2i} \leftrightarrow E_{2a}$) leading to activation of another ($E_{2a}: A \leftrightarrow B$). Unlike in the first case, the question how to add these contributions is far from trivial. I admit that it constitutes a mathematical challenge that is beyond the scope of this presentation (see *14a*).

Now, after I have briefly touched upon difficulties, a positive note. In fact, acknowledging that there are many enzymes in the cell can serve a constructive purpose – it can provide us with an alternative route to estimate the *Cf* energy contribution for the ground state, as described below.

The operation of many enzymatic acts in the cell imply that taking a particular target molecule *T* in the cell, we cannot limit our kinematic description of the state of *T* to the effect of one enzyme only – i.e., we cannot consider *T* as a binary degree of freedom, represented only by two states: $|I\rangle$ and $|O\rangle$[18]. Every target molecule *T* will be participating in many intracellular processes, which include not only enzymatically driven transformations, but also active transport, passive diffusion, macromolecular assembly etc. All these processes have to be reflected in the description of the ground state as unitary transitions between different (supra)molecular configurations of the cell. Therefore, in the consistent description of the ground state of the cell, our molecule *T* will have to be 'delocalized' (spread in the configuration space) over all possible states that the components of this molecule (ultimately, every nucleus and electron) can assume given all various processes that constitute intracellular dynamics in the ground state. In fact, given that molecule *T* can be decomposed to its elementary parts, it would be more consistent to start our analysis with the nuclei $n_k$ and electrons *e* and consider all their possible locations in the cell and every molecule *T* that $n_k$ or *e* can be a part of. Accordingly, we will be referring to the elements $n_k$ or *e* as a 'target *T*' as well.

That we have to expand the configuration space $H_x$ for each element *T* in cell in this more consistent description, does not change the fact of its entanglement with the rest of the cell *R*. However, instead of representing the ground state of the cell $|C\rangle = \alpha_1|I\rangle|R_I\rangle + \alpha_2|O\rangle|R_O\rangle$, we will have to start with more general description $|C\rangle = \Sigma\alpha_i|T_i\rangle|R_i\rangle$ – where *i* spans, instead of only two states, all possible situations that our element *T* can find itself in the cell. Another way to put it is to say that the ground state of the cell lives in a Hilbert space $H_c$ that is a tensor product $H_c = H_T \otimes H_r$, where $H_T$ and $H_r$ are the state spaces of the molecule and the rest of the cell, respectively. We do not need to stop there. Dividing the rest of the cell *R* further to parts, one obtains a more general representation of $H_c$ as $H_c = \otimes H_j$ where *j* labels all electrons and nuclei in the cell.

We turn now to the estimation of the *Cf* force. Just as this more consistent description does not cancel the fact of entanglement between an element of the cell $T^j$ and its complement $R^j$ (the rest of the cell), neither does it cancel the respective *Cf* force effect of one part of the cell $(T^j)$ on another $(R^j)$. We can now describe all contributions into the total *Cf* force by considering different bipartitions *j* of the cell: $|C\rangle = \Sigma\alpha^j_i|T^j_i\rangle|R^j_i\rangle$, where *j* labels the electrons *e* and nuclei $n_k$ in the cell and *i* spans all possible states of a particular element $T^j$ in the cell. For each individual bipartition *j*, we can estimate the contribution $Cf^j$ to the effective cell Hamiltonian $\mathcal{H}_c$, which could roughly correspond to the delocalization energy – difference in energy between the delocalized and localized states of the *j* element. This is a crucial point - we are appealing here essentially to the same argument as that of Feynman – when he used the uncertainty relation between position and impulse of electron to explain the energy gain in the ground state of the molecular hydrogene ion (Feynman et al., 1964). For an example of such estimate of a higher bound of this contribution, the energy of electron delocalization should be less than the ionization energy of a molecule that the electron *e* is part of[19]. In the case of a hydrogen atom it is about 400 kT, or still 800 higher than thermal energy.

Again, it is a highly nontrivial task to integrate the contributions of different bipartitions *j* into the effective cell Hamiltonian. There are hundreds of billions of electrons and nuclei in a single cell, and the phase factors for every bipartition, alluded to in the section *14A*, could make a big difference, providing both positive and negative interferences. In addition, time hierarchy has to be taken into account somehow. Despite these issues, the above estimate shows that the contribution of the total *Cf* force could be quite significant.

---

[18] the input and the output of this enzymatic act. Also, a reminder – we are not talking about a concentration of *X*, but about a change of the state of the cell, characterized by the presence of our molecule *X* being in a particular location of the cell in one or another state. This description is not limited to regular enzymes. For example, if our protein were a transporter, the input and output would correspond to 'inside' and 'outside' states, respectively.

[19] Otherwise the electron will be driven to be lost spontaneously from the molecule.

## 28. How enzymatic activities can contribute to the stability of the U state?

The previous slide discussed mathematical challenges that we will have to face when describing how the cooperative actions of all enzymes contribute to the stabilization of the state of starving cell. Before proceeding further, I would like to clarify one other problem. It is of rather conceptual nature, but the one that mathematical formalism can help to understand better.

From a naive point of view, the very idea that enzymatic activity in a starving cell could be supporting its order contradicts the elementary knowledge of what enzymes do. We suggested that the ground state of the cell is stable because it is protected by kinetic barriers (it is a metastable or quasiequilibrium state). Moreover, we also suggested that it is the coordinated actions of enzymes in the cell that are responsible for the existence of these barriers.

On the other hand, how does it square with the overwhelming biochemical evidence that *the job of enzymes is to lower kinetic barriers – not to raise them*? Should not, to the contrary, the presence of enzymatic activity lead to an accelerated degradation of an ordered state – especially in the starved cell, i.e., in the absence of the external resources? How then the enzymatic activity can be involved in the stabilization of the ground state? One can phrase this problem in a slightly different way, by borrowing terminology from Leon Brillouin (Brillouin, 1949) – how, in the *in vivo* context, can enzymes be involved in the mechanism of so called 'negative catalysis'?

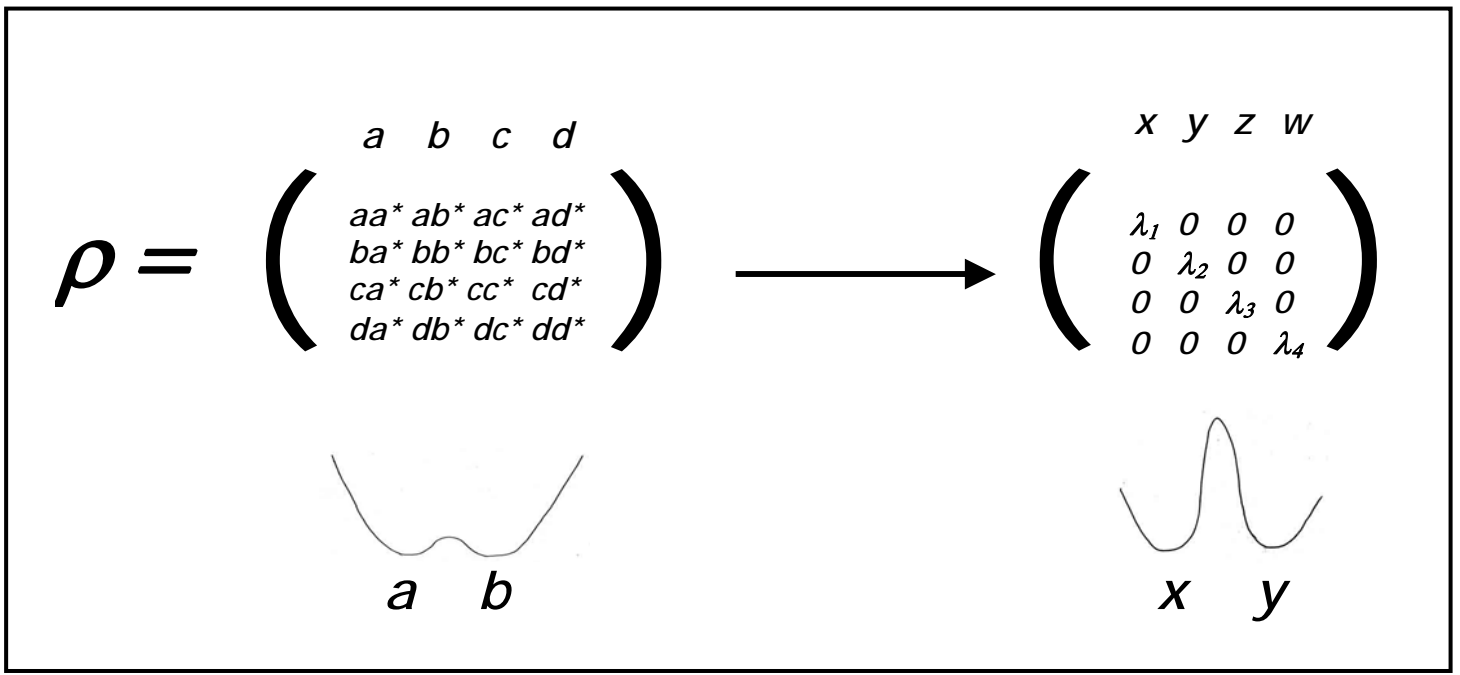

## 29. What kinetic barriers are from the 'first principles' point of view?

I summarize – how to reconcile the seeming contradiction between the established effect of enzymes *in vitro* (manifested in lowering kinetic barriers for their target reactions) and their opposite role *in vivo* (as establishing kinetic barriers protecting the ground state) proposed here?

The key is in recognizing that 'kinetic barriers' correspond to different things in these two cases. The density matrix formalism helps to see the kinetic barriers – a classical notion – for what they are in quantum terms – the off-diagonal elements of reduced density matrix describing system interacting with environment. Choosing a particular basis, a density matrix describing the state of a system will generally contain off-diagonal terms, corresponding to transitions between the basis states. Vanishing of an off-diagonal element corresponds to low transition rate between the respective states – i.e., to a high kinetic barrier in a more classical language.

Importantly, the same density matrix can be expanded in many different bases – and accordingly, the very definition of off-diagonal terms (and kinetic barriers) will depend on what basis we choose for the description of our system. Thus, by increasing off-diagonal terms in one basis ($|A\rangle, |B\rangle, |C\rangle, |D\rangle, ...$) – one can at the same time, and without any contradiction, have the off-diagonals decreased in another basis ($|X\rangle, |Y\rangle, |Z\rangle, |W\rangle,…$).

Now, back to the cells and the enzymes inside of them. To pay proper due to the presence of external environment, the state of a starving cell has to be described by a reduced density matrix. To start with, the obvious basis for this matrix (we will call it Molecular Biological basis, *MB* basis) would correspond to the configuration space of the cell – that specifies position, orientation and structure of every molecule in it. In this language, the effect of an enzyme in facilitating the transitions between different configurations of the cell (e.g., between $|A\rangle$ and $|B\rangle$ states) will correspond to increased values of the $|A\rangle\langle B|$ and $|B\rangle\langle A|$ off-diagonals of the density matrix expanded in the *MB* basis. However, speaking about the ground state of the starving cell, *we have a different basis in mind*. The configurations of the cell ($|A\rangle, |B\rangle, |C\rangle, |D\rangle, ...$ etc) contribute into elements of this basis (e.g., $|X\rangle = \alpha|A\rangle + \beta|B\rangle + \gamma|C\rangle +...$) – on the other hand, the off-diagonal terms (now between the $|X\rangle$ and $|Y\rangle$ states) are much weaker – i.e., this basis corresponds to the kinetically separated states. Thus, it is only if one keeps thinking about the cell organization in terms customary for molecular biology (in *MB* basis), the suggested role of enzymatic activity in stabilizing the ground state of cell will appear paradoxical.

Physicists are accustomed to performing the operation that we just depicted – they start from some physically obvious basis describing the system under question, and then 'canonically transform' it in order the find a more simple basis – not necessarily intuitive, but advantageous mathematically. Molecular biologists, on the other hand, have not advanced yet in this direction – and continue adhering to the original, *MB* basis for conceptualizing intracellular dynamics. More recently, systems biology approaches appeared that suggest more complicated descriptions (Palsson, 2006), however, as our discussion implies, to be consistent, quantum principles have to be taken into account (e.g., by allowing the use of complex numbers for the off-diagonal terms).

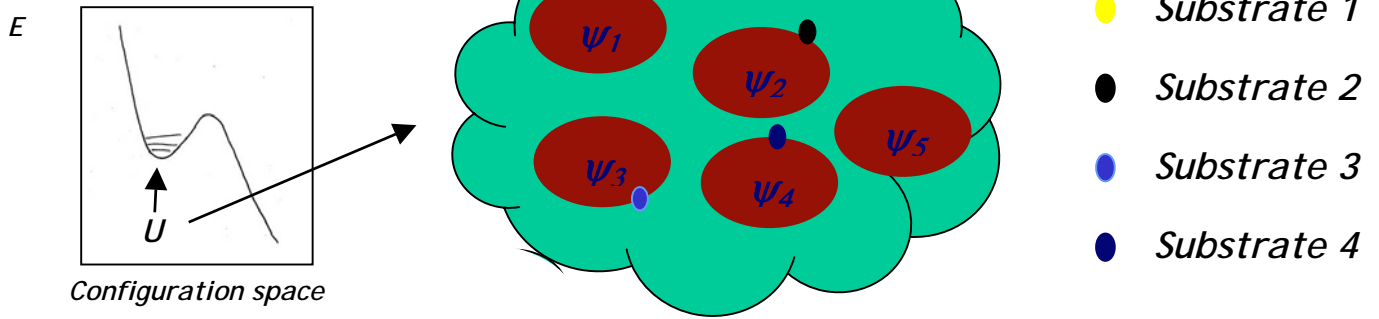

### 30. U state as a fluctuating state

In the next several slides, I will discuss how – under the assumption that we understand well the starving cell state – one can move on to the description of other, more difficult situations. In particular, we will be discussing how to describe cell growth – perhaps the first in the order of 'difficulty'. As we will see, the route that will be chosen is similar to the argument from the so called fluctuation-dissipation theorem from statistical physics. Namely, we will consider the fluctuation behavior of the starving cell state – and then take advantage of its similarity to growth.

The notion of fluctuation arises naturally in the description of the cell close to ground state. As just discussed, this state corresponds to a particular linear combination of molecular configurations of the cell, i.e. elements of the *MB* basis. One way to interpret this on a molecular biological language is to see the starving state having a certain probability to be observed in any of the given (supra)molecular configurations contributing to the ground state[20] – or as reversibly fluctuating between different (supra)molecular configurations.

Now I will point out another important consequence of taking all enzymatic activities into account for the description of the ground state. We will have to acknowledge that, as a part of this fluctuating picture of the starving state, there will be a *finite probability to observe any consumable substrate* in the cell. The significance of this conclusion for the euclidean approach will be illustrated on the further slides.

Here are some relevant definitions:

1. We will call a particular molecular structure *X* a *consumable substrate*, if the cell has ability (mainly dependent on the presence of genes encoding relevant enzymes) to convert *X* to other molecular configurations that can be further involved in cellular metabolism. For example, a functioning Lac operon in *E.coli* makes the sugar *lactose* a consumable substrate for this bacterium – by making lactose convertible to glucose, which is then broken down to elementary parts (via glycolysis, TCA cycle etc), further contributing to the molecular composition of the cell.

For a more formal description, we can consider an *MB* state |*A*⟩ of the cell characterized by the presence of a consumable substrate *X*. We define the presence of *X* in the cell as a value 1 of a projection operator *X*[21]. The property of a molecule *X* to be a consumable substrate implies that the description of the ground state has transitions (off-diagonals) that can transform the state |*A*⟩ into states |*B*⟩ that 1) also contribute to the ground state of the cell, however 2) have the value of the *X* observable 0 (i.e., in these |*B*⟩ states the substrate *X* was

---

[20] It is a separate issue of whether such measurement is practically possible. However, such a question would only underline the importance of recognizing the fundamental limitations on what can be observed on an individual cell level and the necessity of quantum-mechanical formalism to take these limitations into account
[21] This means that one can set up a measurement procedure, either breaking the cell apart and then applying a sensitive detection method, or performing some measurements on the single cell *in vivo*, that would allow to infer the presence of such molecular structure in a given individual cell

converted to another molecular form).

2. We will call two elements $|A\rangle$ and $|B\rangle$ of the *MB* basis *connected* ($|A\rangle \sim |B\rangle$), if state $|A\rangle$ can be reached from state $|B\rangle$ via a path that includes intermediate states $|C\rangle$, $|D\rangle$ and transitions (enzymatic acts, passive diffusion, binding events etc) between the states involved in the path. This property is obviously transitive (if $|A\rangle \sim |B\rangle$ and $|B\rangle \sim |C\rangle$, then $|A\rangle \sim |C\rangle$) and, due to the reversibility of unitary dynamics, reflective (if $|A\rangle \sim |B\rangle$, then $|B\rangle \sim |A\rangle$).

In the language of connectivity, a consumable substrate *X* could be defined as a molecular configuration present in a state $|A\rangle$ connected to the states of the ground state of the cell $|B\rangle$ that do no contain the molecular configuration *X*. On the other hand, since the ground state is a stationary solution of the dynamic equations, it should be naturally closed in respect to connectivity. This means that if the ground state includes an *MBB* state $|B\rangle$, then all *MBB* states $|C_i\rangle \sim |B\rangle$ must also be present with some amplitude in the ground state. It then follows that some states $|A\rangle$ that contain a consumable substrate *X* will have to be present in the ground state of the cell.

Although I argued for the unitary character of the intracellular dynamics in the starving cell, the idea that such a cell can spontaneously synthesize any substrate that it can ever consume might appear counterintuitive and even against the second law of thermodynamics. However, we should realize that we are working with an individual cell – i.e., at the nanoscale. For this kind of analysis, the so called *Fluctuation Theorem* is more appropriate (Evans and Searles, 2002; Wang et al., 2002). This is a generalization of the second law of thermodynamics, stating that on the nano-level the probability of physical processes to go, transiently, in the entropy decreasing direction becomes quite realistic. Thus, although counterintuitive in the eyes of a molecular biologist, raised on modeling the biochemical processes *in vitro* (or on growing millions of cells in logarithmic phase), the notion of a catalytic process (or a pathway) transiently going opposite to its 'usual biochemical' direction is perfectly consistent with physics and, moreover, realistically probable in the case of an individual starving cell.

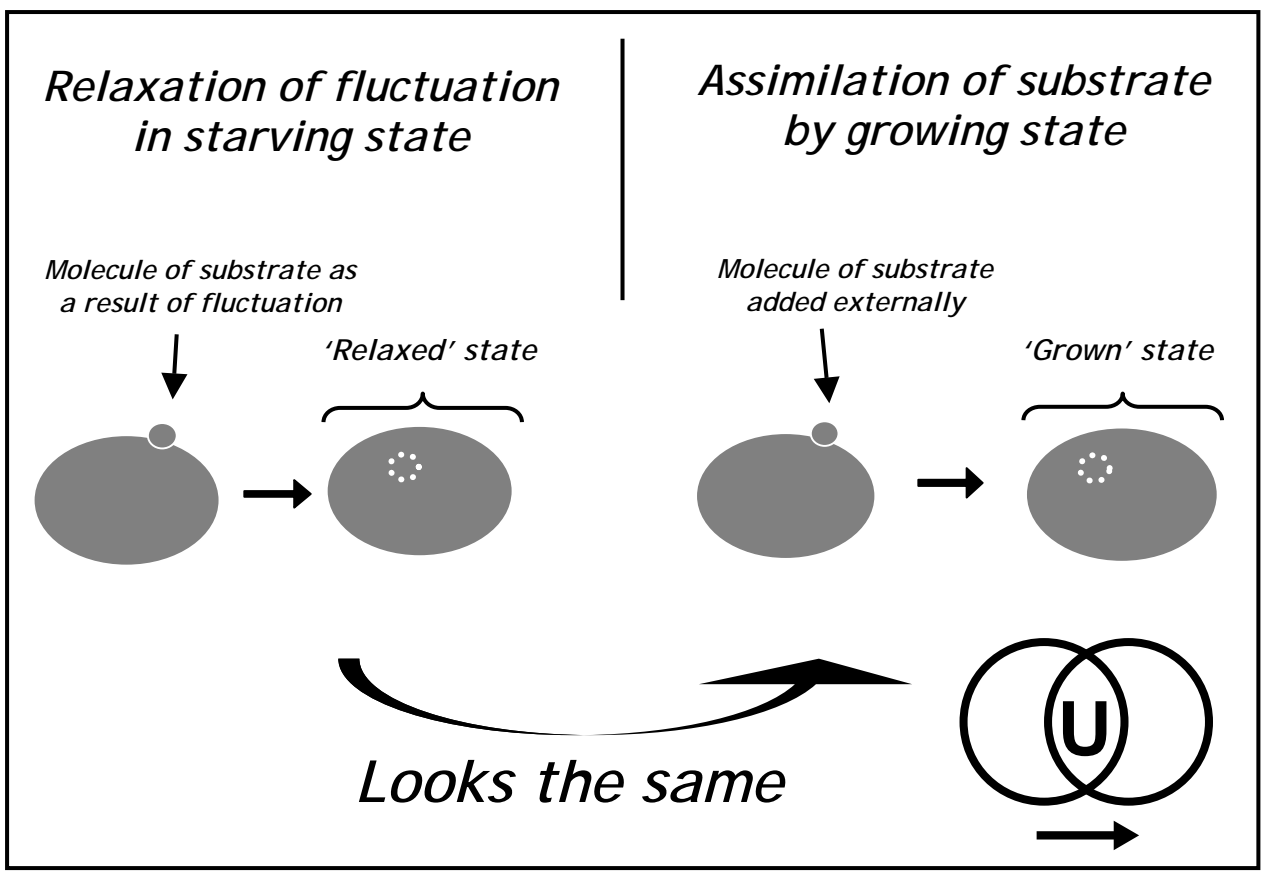

### 31. How to describe real growth?

Now, let us compare two situations, depicted on the following slide.

On the left part of the slide, we have our cell in an *MB* state $|A\rangle$ that contains a particular consumable substrate molecule *X*, which was generated by the reversible fluctuation process described on the previous slide – i.e., due to the fact that $|A\rangle$ is one of the components of the ground state. Because *X* is a consumable substrate subject to intracellular enzymatic interconversions, its appearance in the cell will be transient – i.e., will be quickly followed by relaxation of the fluctuation, due to transitions $|A\rangle \to |B_i\rangle$, such that $|B_i\rangle$ do not have the property *X*. $|A\rangle$ will evolve to a different state $U:|A\rangle \to \alpha|A\rangle + \Sigma\beta_i|B_i\rangle$, represented here by a superposition of many states of *MB* basis, resulting in *X* having vanishing probability to be detected (i.e. $\alpha \sim 0$). Given that we are dealing with a system close to a ground state evolving in a unitary way, let us assume here that we already know how to describe the process of fluctuation relaxation.

Now consider another situation, depicted on the right part of the slide. This time, the same molecule *X* is not generated spontaneously by a fluctuation mechanism, but is exogenously added to our cell[22]. From our experience, we know that, having encountered a consumable substrate, the cell will assimilate it by enzymatically converting it to other molecular configurations[23]. As a result, the external molecule *X* becomes a part of the cell (or, the 'stuff of the cell', see *slide 34*). As described before, all elements of *X* – the nuclei $n_k$ and electrons *e* – will become distributed over all possible states that can be reached via the processes constituting intracellular dynamics of the ground state. We will naturally identify this act of substrate assimilation with an elementary step of cell growth. As argued before, description of growth is a harder problem than that of the starving cell – and what we want to undertake next is to recover this description from the description of the cell in starving state.

To do this, we next note an obvious similarity between the left and right scenarios. Regardless of the history (either a reversible fluctuation or an exogenous addition by hand, respectively), in both cases we are having our molecule *X* as a particular molecular structure at a time $t_0$ and as broken to pieces and incorporated in the molecular composition of the cell at a later time $t > t_0$. What I would like to bring up now is a parallel between this comparison and yet another theoretical gadget of statistical physics – the so called *Fluctuation-Dissipation Theorem (FDT)*(Callen and Welton, 1951). This theorem is applied to derive properties of irreversible processes from the fluctuation behavior of the system in thermal equilibrium. It is based on a principal assumption that the response of a system in equilibrium to a small disturbance is the same as its response to a spontaneous fluctuation. Now, let's consider: 1) the starving cell in close to ground state as being in quasi-equilibrium with its environment, 2) the exogenous addition of a substrate *X* as a small disturbance and 3) the spontaneous appearance of the same substrate *X* in the starving state as a fluctuation. Then we will have all ingredients of the FDT in place – and thus should be able to apply its reasoning to obtain the description of growth from its formal analogy to the relaxation of a spontaneous fluctuation in the starving state.

---

[22] to be accurate, the recipient cell will have a slightly different composition, as it will be lacking the nuclei $n_k$ and electrons *e* constituting our molecule *X*: $C' = C-X$. However, as the cell states go, the $C'$ state will otherwise be an alive and well behaved state.

[23] the space of all available configurations will depend on the presence of particular enzymatic activities in the particular cell type, mostly encoded in genome.

This is how I see the logic of the proposed strategy – very much in the tradition of statistical physics, we first understand as much as possible the behavior of the (quasi)equilibrium state, the starving cell (taking advantage of its closeness to ground state), and afterwards we expand this understanding to the more difficult problems.

Previously, I implicated the mathematical procedure of analytic continuation in this expansion, which served as a reason to term the proposed approach 'Euclidean'. My motivation for this terminology will be explained on the next slide.

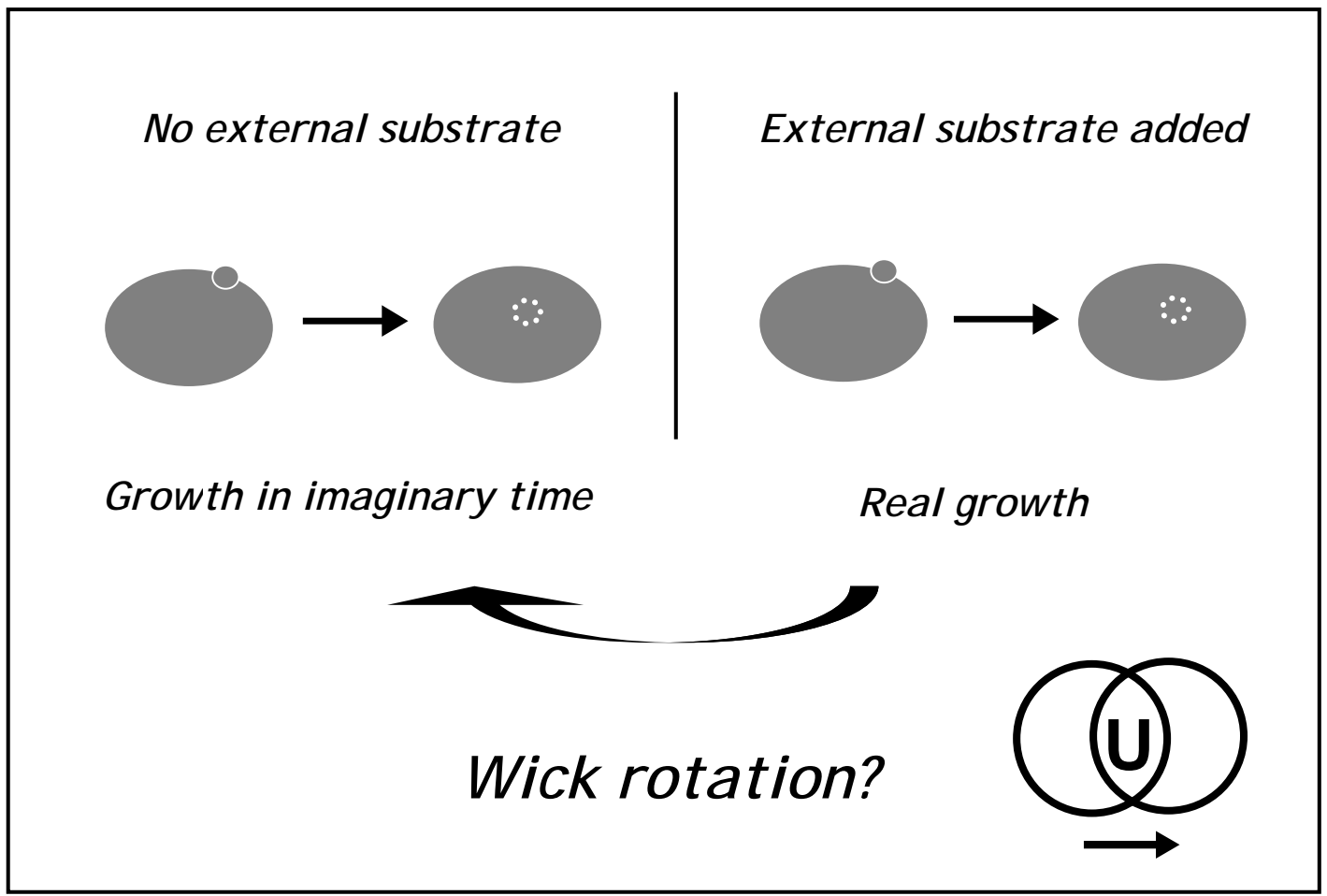

## 32. Why 'Euclidean'?

Before, I focused on the formal similarity between two processes: the elementary step of cell growth and the relaxation of a spontaneous fluctuation in the starving cell. In fact, there is also a crucial difference. In the first case (growth, *G-process*), we can control the process, by adding the substrate *X* to the cell one molecule after another. This is because of the very setup of the situation – the substrate and the cell are prepared independently and then put in contact with each other. In the second case (fluctuation relaxation, *FR-process*), the appearance of the substrate molecule *X* in the cell is out of our control – as it happens via a spontaneous fluctuation of a system left alone to its own devices. Here, the elementary components $n_k$ and *e* of the molecule *X* are always parts of the starving cell – whether before or after we could detect the molecular configuration *X* in it. More formally, we can say that at the right, the substrate molecule *X* and the 'rest of the cell' *R* form a product state: $|I\rangle|R_I\rangle$, so that *G-process* can be described with the quantum operation formalism. On the other hand, the *FR-process* (left-hand scenario) cannot be described with the quantum operation formalism, because these two subsystems interacted previously (Nielsen and Chuang, 2000).

Both the similarity between the *G-* and *FR-processes* and their differences should be reflected in a mathematical description. But how can we make their descriptions formally similar on one hand – but also different in some essential aspect, on the other hand? Here, we finally arrive at the motivation for the term 'euclidean'.

Previously, we used the similarity between these two processes to interpret the assimilation of substrate (an elementary step of growth) as a relaxation of a fluctuation. But an interpretation can also work the other way around – we can now, reciprocally, consider the starving state of the cell as a kind of growing state. But what kind of growth could this be, if we are not expecting to observe any actual changes? Notice that the difference between the left and right situations is in how they have been prepared – that is, in their *past*. This suggests that the time variable should be somehow involved in distinguishing between them. Given that no actual growth is experienced in the left scenario, I am tempted to suggest the following idea – the intracellular dynamics of the cell in starving state describes *growth in imaginary time*. Accordingly, the transition between the descriptions of these two situations will have to correspond to analytic continuation – i.e., to replacement of real time coordinate *t* by imaginary time coordinate *it*, the procedure known as Wick rotation. This is the first part of the motivation of the term 'euclidean' for the proposed approach.

Here is the second reason. There is already a long tradition to interpret Schroedinger equation as an analytic continuation of a real-time process – as a description of a diffusion process in imaginary time (Fenyes, 1952;

Nelson, 1966). It is based on its form, which can be made to look like a heat equation – with the only essential difference in that the real time coordinate *t* is replaced by an imaginary time coordinate *it*. Here we essentially suggest an alternative interpretation. As long as we are considering the dynamics of a system described by Schroedinger equation as an analytic continuation of some 'real-time' process (diffusion) – why cannot we choose an opposite kind of Wick rotation, implicating a different kind of real-time process? Since the negative exponent solution will have to be replaced by a positive exponent, it would be 'growth in imaginary time' instead of 'diffusion in imaginary time'.

Finally, a short philosophical intermission. Given that the Schroedinger equation is fundamental for the description of physical reality, the 'diffusion' interpretation supports the intuition of ancient atomists that randomness and stochasticity lie at the very core of the laws of our Universe. The alternative interpretation, proposed here, has its own fundamental implications for the general meaning of Schroedinger equation – consistent with a different school of thought. Regardless of whether we apply it to the description of living or inanimate systems, this new way to think about the Schroedinger equation suggests that *reproduction* (copying) – instead of a stochastic process – could be taken as a fundamental primitive process at the root of the physical laws. I consider it as an optimistic view – it suggests that Universe is more Life-friendly than one would expect from the standard mechanistic description of our world. Further discussion of this intriguing opportunity to justify Life principle as an integral aspect of physical reality is beyond the scope of this presentation (but see also section *35a*).

> *1. Assimilation force is the Cf force in a different disguise*
>
> ---
>
> *2. Assimilation converts substrate A into the state:*
> *'Stuff of cell' ($\rho_A$), whereas the rest of the elements in the cell are already in the same state $\rho_A$*
>
> ---
>
> *3. Similar to Bose condensation: if we have N particles in a state $|\psi_1\rangle$, the probability of another particle to enter the same state increases N+1 fold*
>
> ---
>
> *4. Observation by environment breaks symmetry of the ground state:*
>
> *Ground state is symmetrized $|1\rangle_A|0\rangle_B + |0\rangle_A|1\rangle_B$, so that $\rho_A = \rho_B$*
> *Environment: $|1\rangle_A|0\rangle_B + |0\rangle_A|1\rangle_B \to |1\rangle_A|0\rangle_B$ (or $|0\rangle_A|1\rangle_B$), so that $\rho_A \neq \rho_B$*

### *33. Relation to Bose condensation?*

The previous discussion might provide us with an alternative look at the nature of the *Cf* force – and accordingly, could lead to yet another way to estimate its energy scale. This time it can be based on the arguments from quantum statistics.

We start by clarifying the origin of the force that drives the assimilation of a substrate *X* upon its addition to the cell (i.e., the elementary step of cell growth). Given the relation of the process of substrate assimilation to the fluctuation relaxation, it is natural to expect that it should be the same *Cf* force responsible for the stability and low energy of the ground state. However, we need to deal with a potential complication first. Although the addition of a substrate *X* is formally a perturbation, it is not the same kind of perturbation that served us to derive the existence of the *Cf* force (sections *9* and *14*). Indeed, as introduced, *Cf* force does not act on the target molecule *X* – but rather on its complement, the rest of the cell $R_x$.

Are we then really talking about the same force? Yes – because these two kinds of perturbations look different only if we limit ourselves to one way of bipartitioning the cell into a target molecule and the rest of the cell (say, into *X* and $R_x$: $|C\rangle = |I\rangle_X|R_I\rangle_{RX} + |O\rangle_X|R_0\rangle_{RX}$). In fact, given that the cell is composed of many molecules, we are also free to divide the cell in many other ways – e.g., taking another molecule *Y* and its complement $R_Y$: $|C\rangle = |I\rangle_Y|R_I\rangle_{RY} + |O\rangle_Y|R_0\rangle_{RY}$. In this alternative way of representing the same state of the cell, our original molecule *X* will be a part of $R_Y$. Therefore, *X* in the substrate state $|I\rangle_X$, (which corresponds to a 'substrate addition' kind of perturbation), now will have to be represented as a perturbation of $R_Y$ (a 'microenvironment' perturbation). Accordingly, the adjustment of complement $R_Y$ as a manifestation of the *Cf* force can also include a catalytically induced change in the state of *X* – in particular, a substrate assimilation.

We can illustrate that we are talking about the same force by describing the effect of the *Cf* force in a toy scenario – with our cell containing only two elements: *A + B*, where *A* is the target molecule (substrate) and *B* is the microenvironment. The ground state is entangled $|C\rangle = |I\rangle_A|R_I\rangle_B + |O\rangle_A|R_0\rangle_B$.

If we perturb it, by taking the *B* in the state $|R_I\rangle_B$ so that the state of the cell is now:

$$|C'\rangle = |I\rangle_A|R_0\rangle_B + |O\rangle_A|R_0\rangle_B = (|I\rangle_A + |O\rangle_A)|R_0\rangle_B, \qquad [33.1]$$

the effect of *Cf* is to transform this state back to the ground state:

*Cf*: (|I⟩$_A$ + |O⟩$_A$)|R$_0$⟩$_B$ → |I⟩$_A$|R$_1$⟩$_B$ + |O⟩$_A$|R$_0$⟩$_B$ [33.2]

Notably, starting with a different state |R$_0$⟩$_B$ of *B*, we end up in the same final state

*Cf*: (|I⟩$_A$ + |O⟩$_A$)|R$_1$⟩$_B$ → |I⟩$_A$|R$_1$⟩$_B$ + |O⟩$_A$|R$_0$⟩$_B$ [33.3]

, i.e., given that we have an information loss, the effect of *Cf* is not described by a unitary operator.

Most importantly, we can see the same process from a different point of view. Instead of the adjustment of a microenvironment by a target molecule (as above), we can see it as a catalytic effect exerted by the microenvironment on the target molecule. For transparency, we now take a perturbation of the target *A*, say |I⟩$_A$ (the target in the input state) – and consider the effect of enzyme *B* on *A* as equilibration of the target between the input and output states:

*B*: |I⟩$_A$(|R$_0$⟩$_B$ + |R$_1$⟩$_B$) → |I⟩$_A$|R$_1$⟩$_B$ + |O⟩$_A$|R$_0$⟩$_B$ [33.4]

Similarly, taking |O⟩$_A$ instead of |I⟩$_A$, we again obtain:

*B*: |O⟩$_A$(|R$_0$⟩$_B$ + |R$_1$⟩$_B$) → |I⟩$_A$|R$_1$⟩$_B$ + |O⟩$_A$|R$_0$⟩$_B$ [33.5]

Again, this operation is irreversible, consistent with our consideration of substrate assimilation as a non-unitary process.

Given the complete symmetry in this description, we thus can consider the force responsible for substrate assimilation (i.e., growth-related relaxation of a perturbation) as the same *Cf* force – albeit in a different disguise.

Now, let us describe a more realistic situation. Instead of only two parts (*A* and *B*), we have *N* parts in our cell (where *N* typically a very large number), and also a more complex kinematic space (more than two basis states for *A*). As previously (Slide 28), we break down the *B* (the microenvironment) to many elementary parts. Let us assume for simplicity that *A* is an elementary particle and we have only one type of a particle in the cell – still not a completely realistic scenario, but sufficient to illustrate the main point. Then also *B = (A + A + A + …)*, and the Hilbert space of *B* is $H_b = \otimes H_{ak}$. What is important is that, in the ground state, all elements $A^k$ of *B* are expected to be in the same state, described by a density operator $\rho_A$ (corresponding to a mixed state, due to the entanglement between different parts of *B*). Moreover, although our substrate *A* was added to the cell originally in a state |I⟩$_A$ (as a perturbation) – upon its assimilation by the cell, it will also be forced by *Cf* into the same state $\rho_A$.

Now, what I would like to bring up is that, in this description, the effect of *Cf* becomes temptingly reminiscent of the *Bose condensation effect*. Quantum statistics tells us that given *N* bosons in a particular state, the probability for yet another boson to enter exactly the same state is increased by the factor *(N+1)* (Feynman et al., 1964). The situation is strikingly similar to our case – the ground state of the cell consists of a large number of particles in a state $\rho_A$, and another particle, when added, is forced into the same state $\rho_A$. Given that *N* is large (~$10^{11}$), the increase in probability can be quite significant. Assuming the connection between the two effects, we can thus roughly estimate the *Cf* energy contributed by one element from its assimilation by cell, using the formula for probability: *P ~ $e^{-\beta E}$*, where *β = 1/kT*:

$10^{11}$ ~ ($P_n/P_o$) ~ $e^{-\beta E_n}/e^{-\beta E_o}$  ⇒   $\Delta E$ ~ $kT ln(10^{11})$ ~ 25kT, or 50 times of thermal energy. [33.6]

Admittedly, these arguments are limited by the fact that we have more than one kind of particles in real cells. Moreover, not all particles behave like bosons (e.g., the electrons and protons are fermions and thus certainly do not obey the right statistics. See however (Law, 2005)). Nevertheless I find this similarity striking and deserving a notice, although its relevance remains an open question.

Finally, we need to make sense of the claim that all elements of the cell in ground state are in the same $\rho_A$ state. Can this notion be taken seriously, provided that cell is not a homogenous blob, but a highly differentiated structure – so that every element $A^k$ in the cell obviously has its own place and function (e.g., protons being parts of different molecules in the cell), thus clearly requiring different states for their description? To address this concern, we should keep in mind that we are talking about an ideal (ground) state – i.e., not perturbed by the 'observation' by environment. As soon as environment enters the picture, measurement of the cell takes an

effect – leading to the emergence of differences between its parts. These differences appear in the context of the entanglement of the parts between each other – so that the outcome of measurement of one element is not independent from the outcome of measurements of other elements.

Taking again a simple toy model for an illustration, a ground state is a symmetrized state:

$$|1\rangle_A|0\rangle_B + |0\rangle_A|1\rangle_B \text{ so that } \rho_A = \rho_B \qquad [33.7]$$

whereas measurement by environment breaks the symmetry, choosing, for example:

$$|1\rangle_A|0\rangle_B + |0\rangle_A|1\rangle_B \rightarrow |1\rangle_A|0\rangle_B \text{ so that } \rho'_A \neq \rho'_B. \qquad [33.8]$$

In this new state: 1) the *A* and *B* manifest each other differently. 2) Due to their entanglement, the choices of *A* and *B* states are correlated with each other. In a sense, despite the interaction with environment, the entanglement between the parts manifests a conservation of a certain invariant property of the system.

> # How to describe that Life utilizes enzymatically catalyzed molecular interconversions for their self-reproduction?
>
> | *Enzymatic catalyzis:* | *Self-reproduction:* |
> |:---:|:---:|
> | as | as |
> | Decoherence suppression | Substrate A becomes 'stuff of the cell' (B) |
> | or | or |
> | $\rho_A = Tr_B(\|C\rangle\langle C\|) = k\Sigma\|a_i\rangle\langle a_j\|\langle b_j\|b_i\rangle$ | $H_B = \otimes H_A$ |
> | For some $\langle a_i\|a_j\rangle = 0$, $\langle b_j\|b_i\rangle \sim 1$ | |
>
> ---
>
> ### If B = A
>
> $\|C\rangle = \|a_i\rangle_A\|a_j\rangle_B + \|a_j\rangle_A\|a_i\rangle_B$    cannot have both $\langle a_i\|a_j\rangle = 0$ and $\langle b_j\|b_i\rangle \sim 1$
>
> ---
>
> ### If B = (A+A+A+…)
>
> can have both $\langle a_i\|a_j\rangle = 0$ and $\langle b_j\|b_i\rangle \sim 1$
>
> i.e., if B is sufficiently large, a dramatic change in the substrate element $A^0$ can be accommodated by relatively small changes in every other element $A^k$

### 34. Back to large numbers.

Previously (*24*), I referred to the Schroedinger's argument that the law of large numbers is not applicable to the explanation of stability of biological order – leading him to propose that, instead, quantum principles should be involved. On the other hand, on the last slide (*33*) we were appealing to the Bose statistics factor **'N+1'** for the estimation of the *Cf* force strength. If this argument has any validity, it would immediately suggest that the stability of the ground state will, in fact, benefit from the large *N* – as the *Cf* force would increase accordingly.

Was Schroedinger wrong then – and does Life, in fact, need large numbers of elementary parts to ensure its stability? To clarify the issue, I will use now an argument independent from Bose statistics. I will argue that the cell indeed has to be sufficiently large, as compared to the molecules composing it. However, I will also argue that the large numbers are doing a more sophisticated job than usual – because Life is also taking advantage of quantum entanglement. Thus, Schroedinger was also right – even though cell does need many elements to work, quantum principles make these numbers considerably more affordable and realistic.

We start with the following task. Perhaps, the most important thing to know about Life is the fact that biological systems utilize enzymatically catalyzed molecular interconversions for their self-reproduction. We now ask how to formulate this property on the language of quantum mechanics. For simplicity, we will consider a scenario with only one kind of particle in the cell – hoping that generalization to the more realistic case will not dramatically devalue our argument. We will break our task down to two parts.

1) First, following previous discussion, we describe the enzymatic activity as decoherence suppression. In line with our general strategy, we consider a ground state of the cell. We divide the cell to two parts: an element *A* and its complement *B* – which we want to describe as a catalyst. For every such division, the state of the element *A* can be represented by a reduced density matrix as:

$$\rho_A = Tr_B(|C\rangle\langle C|) = k\Sigma|a_i\rangle\langle a_j|\langle b_j|b_i\rangle \qquad [34.1]$$

, where $k$ is a normalization coefficient, $|C\rangle$ is the ground state of the cell, $|a_i\rangle$ – the states of the element $A$ in the molecular conformation basis, $|b_i\rangle$ – the corresponding states of the rest of the cell $B$ (the complement, microenvironment of $A$). Note that by molecular conformation basis ($|a_i\rangle$) I mean all possible situations that the element $A$ can find itself in the cell, as a result of the enzymatic activities available to this particular cell. Importantly, it should not be confused with the **MB** basis, the basis for description of the state of the *whole* cell – here we are talking about the alternative states of an *individual element A* of the cell.

According to this description, for any two states $|a_i\rangle$ and $|a_j\rangle$, the low value of the $\langle b_j|b_i\rangle \sim 0$ corresponds to weak interference between these states, and thus to a high kinetic barrier – i.e., to a low transition rate between these states. On the other hand, if we want the microenvironment $B$ to protect the superposition between different states of $A$ from decoherence, we need strong interference between these states, or $\langle b_j|b_i\rangle \sim 1$. The implication is that if the states of microenvironment $B$, corresponding to the alternative states of $A$, are very close to each other, no measurement of the state of $B$ by the external environment will allow to tell whether $A$ is in $|a_i\rangle$ or $|a_j\rangle$ state. In this way, we essentially expressed the requirement of how intracellular microenvironment $B$ could serve to channel a state $|a_i\rangle$ of the element $A$ to the other orthogonal states $|a_j\rangle$, $|a_h\rangle$, … – in other words, how by suppressing decoherence between $|a_i\rangle$ and $|a_j\rangle$ (or $|a_h\rangle$, etc…), $B$ can act as a catalyst.

2). Second, we also need to describe the fact that this enzymatic activity is used for *self-reproduction,* – i.e., that as a result of this activity, a substrate $A$ is converted to the 'stuff of the cell'. In our case, it is particularly simple. According to the euclidean strategy, we limit ourselves to the ground state, describing self-reproduction in imaginary time. We demand that the microenvironment $B$ is composed of many copies of $A$ (we will label them, to distinguish from the substrate or 'target element', as $A^k$ – whereas the target element will be labeled as $A^0$), or : $H_b = \otimes H_{ak}$. Moreover, as discussed in the previous slide, the ground state is a symmetrized state:

$$|C\rangle = (|a_i\rangle_{A0}|a_j\rangle_{A1}|a_h\rangle_{A2}...) + (|a_j\rangle_{A0}|a_i\rangle_{A1}|a_h\rangle_{A2}...) + ... \qquad [34.2]$$

Now, we will show that these two parts of our description cannot be compatible with each other if the number of copies $A^k$ composing $B$ is not sufficiently large. Consider first an extreme case when $B$ is very small and consists of only one copy of $A^k$. We want from $B$ to serve as a catalyst – i.e., as formulated before, for some orthogonal states of the target element $A^0$ $\langle a_i|a_j\rangle = 0$, the corresponding states of microenvironment $B$ should be $\langle b_j|b_i\rangle \sim 1$. However, due to the obvious symmetry between $A$ and $B$ in this case (the ground state $|C\rangle$ is a symmetrized state, i.e., $|C\rangle = |a_i\rangle_A|a_j\rangle_B + |a_j\rangle_A|a_i\rangle_B + ...$) we certainly cannot have both $\langle a_i|a_j\rangle = 0$ and $\langle b_j|b_i\rangle \sim 1$.

However, if we now take $B$ composed of a large enough number of elements $A^k$, the condition $\langle b_j|b_i\rangle \sim 1$ would be more likely to satisfy. For, even if the target molecule $A^0$ changed dramatically (e.g., between the orthogonal states $|a_i\rangle$ and $|a_j\rangle$), each of the rest of the $A^k$ might need to change a little – and still contribute to a cumulative change in $B$ (which is composed of the $A^k$), sufficient to cover the change in $A^0$. For example, given $N \sim 10^{11}$, if the change in $A^0$ is a phase flip, it can be compensated by a close to infinitesimal phase shifts in the states of all remaining $A^k$ ($\sim \pi 10^{-11}$). In other words, when the number of elements $N$ is large, the states of the microenvironment $B$ $|b_i\rangle$ and $|b_j\rangle$ can be sufficiently close to each other and practically undistinguishable by the external environment – thus protecting the superposition between the corresponding orthogonal states of $A^0$ $|a_i\rangle$ and $|a_j\rangle$ from the environmentally induced decoherence.

| *Without entanglement* | *With entanglement* |
|---|---|
| *Tradeoff* | *No tradeoff* |
| *Either*     *Or* | *Both* |
| 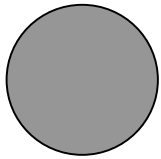   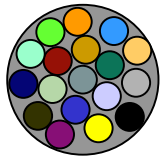 | 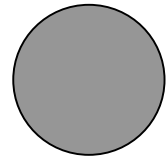   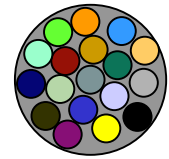 |
| Homogenous blob   Highly differentiated | Actual   Potential |
| $M_T \sim \sqrt{N}$   $M_T \sim \Sigma\sqrt{N} \sim N$ | *Symmetrized ground state:* |
|  | $\|C\rangle = (\|a_i\rangle_{A0}\|a_j\rangle_{A1}\|a_h\rangle_{A2}...) + (\|a_j\rangle_{A0}\|a_i\rangle_{A1}\|a_h\rangle_{A2}...) + ...$ |
| *Complexity* | *Complexity* |
| **Low**           **High** | **High (~ $N$)** |
| *Stability* | *Stability* |
| **High**          **Low** | **High (~ $\sqrt{N}$)** |

### 35. *The problem of tradeoff between complexity and stability.*

One might find it amusing that, after being dismissed as irrelevant (slide *24*), the law of large numbers reemerges as a necessary ingredient for quantum description of Life. Nevertheless, because of entanglement, Life does not need large numbers in such great excess – as compared to the classical picture. Here, I will use elementary arguments to illustrate this point.

In the classical description (left panel of the Figure), given a number *N* of elements in a system, there is a necessary tradeoff between the number of relevant variables and the stability of the system's dynamics. Taking a number $N \sim 10^{11}$, typical for a bacterial cell, consider two extreme cases:

1) The system is represented by $10^{11}$ copies of a single molecular species (left part of the left panel), i.e., we have one relevant variable only (it is a 'homogenous blob' – a drop of water, for example). Then, the fluctuation size $F \sim \sqrt{N}$ is very small relative to *N* (e.g., if $N \sim 10^{11}$, $\sqrt{N} \sim 10^6$, and $F/N \sim 0.001\%$) – thus, this sole variable is relatively stable.

2) In the other extreme case (right part of the left panel), each element $A^k$ of the system has a unique role. Now we have $10^{11}$ different variables in the cell – and the fluctuations in every one of them have to be dealt with individually. Hence the dramatic increase in the scale of 'total fluctuation' $F_T \sim \Sigma\sqrt{N^k} \sim N$ (due to $\sqrt{1} = 1$) – which can be measured, for example, by the total work of maintenance $M_T$ needed to support the functioning state of the system.

One can see that, given a fixed number of elements *N,* the more complex we wish our system to be (which amounts to the higher number of 'specialized parts', or relevant degrees of freedom in it), the more difficult it becomes to account for its stability (and more it has to rely on the external flow of resources to maintain its order). I term this *'the problem of tradeoff between complexity and stability'*, which is intrinsic for the classical description of complex dynamic systems.

Keeping with the same elementary argument, we will now see that if entanglement is taken into account, a system can enjoy both high number of relevant degrees of freedom and, at the same time, comparatively low total number of elements – thus significantly relieving the tradeoff problem (and the need for external resources). For simplicity, we consider again the extreme situation with every element in the cell $A^k$ having a unique role (Right panel). The key notion here is that of symmetrized ground state, discussed in the previous slide:

$$|C\rangle = |C_x\rangle + |C_y\rangle + \ldots = (|a_i\rangle_{A0}|a_j\rangle_{A1}|a_h\rangle_{A2}\ldots) + (|a_j\rangle_{A0}|a_i\rangle_{A1}|a_h\rangle_{A2}\ldots) + \ldots \qquad [35.1]$$

As evident from its form, each element $A^k$ of the system plays a different role in every different component $|C_i\rangle$ of the superposition (for example, $|a_j\rangle_{A1}$ could correspond to particular proton $A^1$ being a part of one molecule in the $|C_x\rangle$ component, and $|a_i\rangle_{A1}$ – to the same proton $A^1$ now as a part of another molecule in the $|C_y\rangle$ component, etc). Importantly, however, the differences between the elements $A^k$ in the ground state are only *potential* differences. It is because, by definition, the ground state is an ideal state, not perturbed via an observation by the environment – thus the *actual* differences inside the system could emerge only as a result of its interaction with environment. Since, in the ground state, the actual states of all parts of the system are identical, the fluctuation behavior of the system close to the ground state should scale as $\sqrt{N}$, instead of $N$. The reduced impact of fluctuations should be manifested in the fact that, upon the interaction of a system with environment, its different parts will respond in a correlated manner to the perturbations – due to their entanglement between each other.

Thus, using relatively simple arguments, we could illustrate an intriguing point about entanglement. Encouragingly, it can provide us with 'the best of both worlds' solution to the 'tradeoff problem' – although every part of the cell $A^k$ can potentially play a different role (i.e., the system can have highly differentiated structure), all these parts actually contribute into the same large number $N$, thus enabling the system to tap into the stabilizing effect of large numbers. I see this as another advantage of the euclidean approach – whereas classical description does not make a distinction between the actual and potential differences between the elements, leading to the ominous 'tradeoff problem', this distinction is neatly encoded in the notion of the symmetrized ground state. In a sense, the ground state is describing how every element of the cell can 'explore' all potential scenarios that it can be in the cell; but at the same time, the individual elements do not explore their possibilities independent from each other – i.e., the entanglement between them effectively reduces the number of dimensions in the system.

## 35a. Self-reproduction and condensed matter physics.

One can anticipate at least two objections to the toy model used in the argument in the end of the previous slide: 1). Only one kind of particle was considered, 2) No time hierarchy was taken into account. Certainly, dealing with these and other factors will add further layers of complexity in the description. For example, given the orders of magnitude difference in the rates of various intracellular processes, one might expect that the time scale would determine if two 'elements of the same kind' could be taken as undistinguishable or not (and thus whether they will be involved in the symmetrization), affecting our estimates of the *Cf* contributions.

This and other complications remain the open questions for the further research. Nevertheless, I hope that the generalization to the more realistic scenario will not dramatically devalue our conclusions. Similarly, despite its simplicity, the Ising model provides many useful insights into many aspects of condensed matter physics. Likewise, theoretical biology could benefit from an elementary model that captures, using the formalism of quantum theory, an essential aspect of Life – the fact that biological systems utilize enzymatically catalyzed molecular interconversions for their self-reproduction.

Finally, two features of our approach – the decoherence-suppression description of catalysis, and the reverse effects of the catalysis targets on their respective microenvironments – do not need to be limited solely to biological systems, and might also find applications in other domains of soft matter physics. The mesoscopic physics is in search of new fundamental principles (Laughlin et al., 2000). In this respect, it is tempting to ask whether the mean field theory, widely used in condensed matter physics, can be considered from the above self-reproduction perspective.

Take a ferromagnet for the illustration – a). the orientation of the magnetic dipole moment of every atom contributes to the mean field; b). and then the oriented state of every atom is maintained due to the presence of this very field. Now let us compare it with the *Cf* force principle acting in a living cell. Consider the information about all catalytic activities of the cell (including the location and timing of every enzymatic act) as describing a particular state of a 'mean field' – admittedly, much more complex than the common physical ones, but nevertheless corresponding to a *certain kind of order*. Just like a magnetic dipole in a ferromagnet, every element of the cell (i.e., every electron and nucleus) will be playing two complementary roles – a). it will contribute into the generation of the field (being a part of different catalytic molecules in its different 'symmetrized state incarnations', see **35.1**), but b). it will also be affected by the same very field (be the target of catalysis).

Thus, there is a similarity between the notion of a 'mean field' and 'self-reproducing order' that might be worth exploring further. An important difference from the more conventional mean field approach is that the 'value' of the field in the latter case cannot be obtained from simple *averaging* of some observable values of the individual elements – but will require more involved procedures. Consistent with the comment at the end of the section *32* of a potential fundamental role of reproduction (copying), as a primordial process – instead of a stochastic process – at the root of the physical laws, the notions of 'decoherence protection' and *Cf* force might provide a generalization of the notion of mean field, which deserves a separate discussion beyond the scope of this presentation.

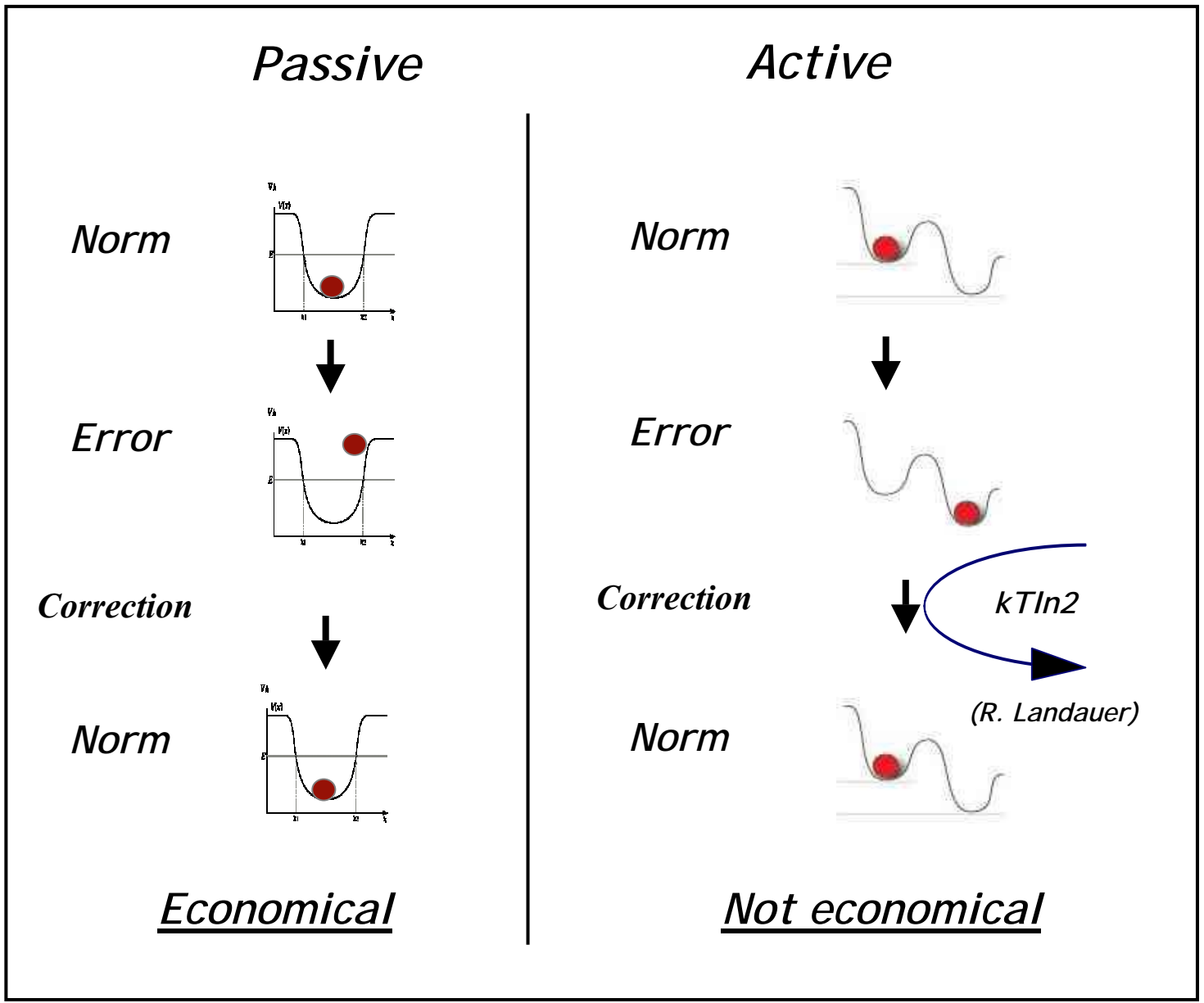

*36. Passive error correction*

The elementary arguments from the slide *35* show that entanglement can provide $\sqrt{N}$ gain in stability of a system close to ground state (as measured in the amount of work of maintenance needed to support its ordered state). Intriguingly, this gain is strikingly reminiscent of the quadratic speedup given by the Grover's search algorithm in the Quantum Information Theory (QIT). What relation the question of *stability* of intracellular dynamics could possibly have with the *efficiency* of a search algorithm? The notion of 'homeostasis' could provide the link. Intuitively, the more efficient (e.g., faster or less resource-heavy) the negative feedback response to a challenge is, the more stable the system is expected to be. Accordingly, we might see the $\sqrt{N}$ gain in stability (which can be measured as a decrease in the required work of maintenance $M_x$) as a result of an increased speed in finding how to respond to a particular environmental perturbation.

Here, using the language of QIT, I will contrast 'passive' versus 'active' mechanisms of error correction – and illustrate how quantum principles could justify more significant contribution of the 'passive error correction' mechanisms into biological organization than usually appreciated.

Calling the perturbation an 'error', we will be referring to the process of monitoring, detection and response to a perturbation as 'error correction' further on. From the work of Landauer (Landauer, 1961), it might appear at first that – regardless of a particular molecular mechanism – laws of physics set general lower bound on how much energy has to be dissipated to correct an error. Landauer had shown that erasure of one bit of information requires dissipation of at least *kTln2* of energy (so called Landauer limit) and later applied it to the issue of error correction – interpreting it as erasure of information about the state of the system generated by the error. This estimate, as discussed before, supports the view of an energy inflow as necessary to support the stability of any living state.

However, the Landauer bound applies only if we limit our notion of an 'error' to the perturbations that transfer our system into relatively stable alternative state – pictured here, on the right side of the slide, as located in a neighboring potential well. On the other hand, the potential well representation naturally allows us to consider also a different kind of perturbation – shown on the left side of the slide. The system responds to this kind of perturbation in a different manner, manifested in generation of a reactive force that does not allow the system to cross the barrier and land in an alternative energy minimum. This way to handle a perturbation is known as a 'passive error correction' mechanism (also called 'error prevention'). Like the previous, 'active error correction' mechanism, it is a response of the system to a perturbation. Moreover, in this case we can formally calculate the 'work of maintenance' by integrating the response force over the path of the deviation and obtaining the difference in energy between the perturbed and non-perturbed states. Clearly, however, the Landauer bound does not apply in this case, as our system can persist in the desired state long time without any energy dissipation and energy input from outside – i.e., with no maintenance cost whatsoever.

The main idea of my presentation was the use of quantum mechanical principles to justify a mechanism of stability of the organized state of a cell based on the notion of the ground state supported by the reactive enzymatically-driven **Cf** forces. As evident from the previous discussion, it is more in concordance with the 'passive error correction' scheme. Again, this is not to belittle the role of energy dissipation in the functioning of a living organism – but rather to limit its universality. In particular, it is natural to expect the ground state-based, 'passive error correction' mechanism being predominant in the case of dormant states – such as cryptobiosis, anabiosis, sporulation etc.

Now, for something truly interesting. Intriguingly, quantum mechanics allows us to see the 'active' and 'passive' error correction' mechanisms in biology from a common point of view – with the 'passive error correction' being a more advanced (i.e., energetically efficient, 'green') way to correct an error. In this view, both passive and active error correction mechanisms use energy derived from the outside environment to correct an error. However, the crucial difference of the 'passive error correction' is that all this energy is nothing but the energy introduced into the system by the perturbation itself! From this perspective, we can see the enzymatic mechanisms and their cooperated action in response to a particular error as becoming so efficient in the 'passive error correction' case that they can be represented simply as existence of kinetic barriers responsible for an 'elastic reaction of a rigid system' to the perturbation – without any additional $M_x$ coming from outside.

We can also consider this idea from the perspective of Maxwell's demon (Leff and Rex, 2002). The suggestion to see 'passive error correction' as an extreme case of 'active error correction' could be formulated in information terms. Namely, the 'passive error correction' corresponds to when the information obtained by the system from the measurement of the state of the error provides gain in free energy that exactly covers the cost of work $M_x$ to fix this very error. The potential implication of this idea is a different take on the issue of whether a nano-device could use information about the microstate of a system to extract free energy and perform useful work (à la Maxwell's demon). We can state that a nano-device (e.g., bacterial cell) can use thermal fluctuations as a resource, as long as one condition is met – that all this resource has to be spent to support the functioning of the device itself, with nothing left for an outside observer to use. I strongly emphasize that I do not propose a molecular perpetuum mobile here. Rather, I see a potential for *a new formulation of the 2nd law*, more useful for dealing with biological systems. Due to the space limitations, I will not explore this side of connection between biology and the fundamental issues of statistical thermodynamics any further.

## 37. Other quantum approaches in biology

How the proposed euclidean approach fares in comparison with the alternative quantum approaches to biology – most notably, the theory of Fröhlich (Fröhlich, 1968) and that of Penrose-Hameroff (Hameroff and Schneiker, 1987; Penrose, 1994)? Acknowledging that these approaches must have their own merits, I prefer to focus here on the advantages of my proposal. I have already discussed the technical benefits of starting the analysis from the starving cell state. Here I will emphasize additional strong points of the proposed approach.

1. Physically, Life is a kind of condensed matter. The art of condensed matter physics consists of writing effective field theories for the particular property of matter that we need to understand. The effective field theory approach is based on the principle of *universality* – i.e., that low energy physics (responsible for the phenomenon in question) is independent on underlying high energy (short distance, microscopic) physics (Altland and Simons, 2006; Laughlin and Pines, 2000). For example, the understanding of gyroscopic precession is oblivious to particular molecular details – and only requires the fact that we are dealing with a solid body. The seemingly innocent fact that the gyroscope is a solid puts so much constraint on the relative motions of its many parts that it is entirely sufficient to explain precession.

Regarding Life, the most distinct feature of the intracellular order is its functionality – i.e., the apparent 'meaningfulness' of its organization. In the spirit of the effective field theory, the task then becomes to find what aspects of Life as a physical system have to be taken into account for understanding this property (which is similar to knowing the fact that gyroscope is a solid to understand its precession) – and what aspects can be safely dispensed with (similar to the microscopic details of interactions holding the gyroscope together, irrelevant for the precession problem).

I believe that the advantage of the proposed approach is in doing exactly this – it combines two ingredients essential for understanding what shapes the intracellular order to make it meaningful. At the molecular level, Life is mostly about catalytic actions of enzymes – therefore the information about enzymatic acts (their location, efficiency, regulation etc) has to be involved in any description of intracellular dynamics and structure. On the other hand, we also have to learn how to integrate the tremendous amount of data accumulated by molecular biology about different enzymatic activities into one cogent picture, which can be called '*effective equation of motion of an individual living cell*'. As I argued previously, quantum entanglement is a new theoretical ingredient required for this task (Ogryzko, 2008a). Here, I suggested first steps towards accomplishing this combination.

2. The idea of ground state supported by reactive *Cf* forces is based on one nontrivial assumption only – *in ground state enzymes work*.

By itself, the idea of a starving cell in a ground state is trivial – as it is also consistent with a more mundane alternative of the enzymes being inactive in the cell or separated from their target molecules by compartmentalization (or related mechanisms), so that no enzymatically driven processes could occur. In this picture the notion of a ground state is still valid, but it would have a more static, fixed nature – as the stable states will correspond to particular supramolecular configurations of the cell. There still are kinetic barriers protecting the state from immediate disintegration, but the forces responsible would be only the regular binary interactions – such as van der Waals forces, hydrogene and covalent bonds, ion-ion interactions – without any interesting dynamics directly relevant to the biological meaning of intracellular structure.

To be sure, this possibility cannot be excluded. The nontrivial assumption that enzymes do work in the ground state is an attempt to develop a self-consistent alternative to this more traditional view. The advantage here is in treating two seemingly disparate problems – stability of the starved cell structure and stability of intracellular dynamics – from a unified perspective, which would allow for a common formal solution, related by the mathematical procedure of Wick rotation. An open question remains how the imaginary time description can accommodate the hierarchy of many different time scales present in a typical biological system.

3. Any new physical approach to intracellular dynamics and organization will have to *make sense for biologists* and provide new insights into biological problems. The 'meaningfulness' of biological organization is usually attributed to natural selection. As already discussed before, the proposed approach leads to a physical justification of the self-organization phenomenon – i.e., to the principle of 'optimization without natural selection of replicators'. The more conventional 'dissipative structure' approach to self-organization, due to its reliance on the law of large numbers, is hardly applicable to the intracellular dynamics (the 'tradeoff problem', sections *24 and 35*). On the other hand, a self-organization principle grounded in a solid physical foundation would be welcome in biology – as providing elegant answers to several open questions in evolutionary biology, exploited by the proponents of Intelligent Design.

4. Finally, any sensible theory should be experimentally testable. Life is the biologists' turf. Thus, any theory of intracellular dynamics and organization, no matter how deeply rooted in physical principles, will have to *be testable by the biologists*. But, ultimately, most of what experimental biologists know how to do is develop increasingly sophisticated ways to measure and manipulate the amounts, locations, conformations and activities of enzymes and other proteins (either directly or indirectly, via manipulation of the structure of genes). Then, it is encouraging that the level of abstraction chosen by this approach stops at the level of enzymatic actions and does not profess to go any deeper. The formulation of the 'effective equation of motion of the cell' in the language of catalytic activities (and other comparable processes) is as biologist-friendly as it can possibly be – and thus can be immediately translated into the verificative experimental schemes.

# Part 4. Experimental verification.

*Abstract 4*

I argue that biological adaptation could be a general situation to experimentally observe quantum entanglement in biological systems. Given that the most reliable and informative observable of an individual cell is the sequence of its genome, I propose that if entanglement does play any role in living cells, and we want to observe it – our best bet would be to use high throuput DNA sequencing. As an example, I consider my previously published work on the phenomenon of adaptive mutations.

## *38. Experimental verification: dissipation and adaptation*

I am coming to the last part of my presentation, concerned with the experimental verification of the proposed ideas. One can envision two general types of approaches.

The first approach is based on the notion of the cost of maintenance $M_X$, introduced previously (*18, 25, 35, 36*). This notion can be used to distinguish between the classical 'no entanglement' and the quantum 'entanglement' models of intracellular dynamics. Intuitively, because of the more important role of the passive 'error correction' (*36*), one would predict that the 'entanglement' model would rely less on energy dissipation – as compared to the classic model. Thus, the general approach to test this prediction could be based on independent estimation and comparison of two values: 1) experimental work of maintenance $M_E$, which can be estimated by measuring the *dissipation of energy* by the cell and 2) the theoretical work of maintenance $M_T$, which would correspond to the energy needed to be dissipated to preserve the cell order under the assumption of 'no entanglement' between the enzymatic events in the cell (i.e., according to the classical model). The first value $M_E$ could be relatively straightforward to measure – for example by calorimetry. The estimation of the second value $M_T$ will be more involved – and will require: a) a detailed knowledge of cellular metabolism and intracellular regulation (i.e., which variables are relevant and require maintenance), as well as b) measuring their 'deterioration rates', i.e., how fast the values of the variables cross the limits of acceptability. The classical model will be ruled out if we find that $M_E < M_T$. Given its reliance on evaluation of a thermodynamic quantity – entropy production – the described test bears resemblance to the notion of thermodynamic 'entanglement witness' (Hide et al., 2007).

The second type of approaches is based on the following general idea – *biological adaptation* represents a kind of experimental situation where the quantum entanglement could manifest itself very naturally. Accordingly, as will be argued later, those could be exactly the situations where one can envision entanglement to be harnessed for practical purposes.

To support the argument about the link between entanglement and adaptation, I will use the language of einselection as a convenient framework. First, consider a starving cell in a quasiequilibrium with a particular environment $E_1$. According to the einselection principle, we should have a certain set of preferred states, determined by their property of not becoming entangled with the environment $E_1$ with time. Now suppose that we change $E_1$ to some other environment $E_2$. We will describe, from three different perspectives, what happens with our cell after the $E_1 \rightarrow E_2$ transition.

1) Physically, we are having a situation when the previously stable states of the system become dynamically unstable and have to change in order to be in (quasi)equilibrium with the new environment $E_2$.

2) Mathematically, we describe it as a change from one preferred basis (einselected in the environment $E_1$) to another (corresponding to the environment $E_2$). This entails that we will need to represent the states of the old preferred basis as *superpositions* of the elements of the new preferred basis. Therefore, the density matrix describing the state of the system at the moment of the change in environment $E_1 \rightarrow E_2$ will have to contain non-zero off-diagonal terms in the new preferred basis.

3) Finally, from the biological perspective, we have nothing else but the process of adaptation of the cell to the new environment.

This comparison suggests that the change of the preferred basis could be a simple and economical way to describe biological adaptation (at least some instances of it). Importantly, this description naturally employs the notion of quantum superposition.

What all this has to do with entanglement? The connection becomes clear when we attempt to understand this description in terms of what is happening inside the cell. For an illustration, let's go back to the simplest possible presentation of the internal cell structure. We bipartition the cell to two parts: a molecule $T$ and the rest of the cell $R_T$. As argued before, the state of the starving cell is represented by a superposition $|C\rangle = \alpha_1|I\rangle|R_I\rangle + \alpha_2|O\rangle|R_O\rangle$, implying that the environment $E_1$ was not able to distinguish between the two components of the superposition (i.e., they are not the elements of the preferred states basis in the environment $E_1$). Now consider a change to a different environment $E_2$, where these two states will become the preferred states. Regardless of a specific outcome of the adaptation of our cell to the new environment, the choice of the state of $T$ ($|I\rangle$ or $|O\rangle$) will correlate with the choice of the state of $R_T$ ($|R_I\rangle$ or $|R_O\rangle$, correspondingly).

As is evident from the above description, these correlations between the states of different parts of the cell are due to their entanglement – first *prepared* via adaptation of the system to the environment $E_1$, and then *revealed* as a part of its adaptation to the new environment $E_2$. The existence of such correlations is a characteristic feature of our description. They cannot be expected from the classical molecular-biological picture of the cell – which always relies on a molecular mechanism (typically involving physical interactions (Local Operations) and diffusion (Classical Communication)) in order to account for the correlations between intracellular events. From this perspective, biological adaptation appears as a promising and rather general experimental setup where the quantum entanglement in the cell could be observed.

## 39. If we use cells as quantum computers, what could be the readout procedure?

Even if phenomenon of biological adaptation is the right place to look for entanglement, what could one measure in order to infer its existence?

One can notice an immediate problem – evident from the way how we introduced the place of entanglement in description of biological adaptation. We were talking about the so called *global entanglement* – i.e., not about a correlation between the states of individual elements of our system *X* and *Y* (e.g., particular molecules or atoms; this would correspond to *regional entanglement*), but rather between an element *T* and the rest of the system $R_T$. The notion of global entanglement is easier to study mathematically, since we can use the Schmidt decomposition theorem to describe a bipartition (and no equivalent to this theorem exists for a composite systems with more than two parts). However, as often happens, the conceptual simplicity comes with a cost. Global entanglement is more difficult to observe in practice, because it requires one to perform measurements of the state of $R_T$ – in our case a very complex system in itself, with many degrees of freedom. Although this is a valid concern in general, I hope that we can be helped by special instances of adaptation, which (based on my previous work on adaptive mutations phenomenon (Ogryzko, 2007; Ogryzko, 2008b)) involve entanglement manifested in a correlation between two individual localized events in the cell, i.e., closer in its spirit to a regional entanglement.

We can reformulate the problem of observing entanglement as the *readout problem* – i.e., as a question of what properties of the cell one could measure in order to infer an existence of entanglement in it? In fact, first we might want to ask a simpler question – in general, regardless of whether it can be an 'entanglement witness', what observable property of a *single cell* could be most robust, easy to measure and at the same time carry as much information as possible about the state of the cell as a physical object? Let's first find such a good readout observable and worry about entanglement later.

There are many reasons to consider sequence of the cellular DNA as such an observable. I will list three of them:

1. DNA is the most stable molecule in the cell.

2. Structurally, it is a linear aperiodic polymer – i.e., it is literally a molecular text, the main function of which is to be read and amplified. Thus, DNA sequence is much easier to unambiguously '*measure*' than anything else of a comparable complexity in the cell.

3. Last but not least, there is no need to develop special technology for the readout procedure to measure the state of DNA in order to test entanglement. We would be taking advantage of the dramatic progress in development of technology for high throughput DNA sequencing (Parkhomchuk et al.). In this field, the goal is to be able to determine complete sequence of human genome (of 3 billion DNA bases pairs) for the cost of about 1000 dollars or less. Most likely, such a goal will be reached within a decade. Accordingly, the cost of sequencing of a bacterial genome (of the 5 Mb size, containing about 10,000 bit of information) will be 1 dollar.

Now back to detecting entanglement in cells. Given that DNA sequence is such a convenient readout observable, one can ask – is it possible to arrange an experimental scheme based on the DNA sequencing that would allow us to infer an existence of entanglement in the cells? More specifically, since the promising circumstance to observe entanglement is a situation of adaptation – could we design an experiment based on adaptation of the cell to environment $E_1$, and then changing it to environment $E_2$, in such a way that the resulting adaptation would induce changes in the cell's DNA? Afterwards, we would sequence the DNA, determine what these changes are, and thus could infer that entanglement was taking place in the cell. Is such an experiment possible in principle?

Given that we are looking for a useful entanglement – i.e., for something that we could eventually take advantage of – let me reformulate the same question in a different, more 'business-oriented' way. Admittedly pushing the boundaries of imagination to the limits, we can ask – if cells use entanglement for their information processing needs, and we want to use DNA sequencing for a readout procedure, could we make the cells compute something for us and then record the results of their computation on DNA? Afterwards, we would be able to read these records by high throughput sequencing – in effect taking advantage of the convenience of genome sequence as a readout observable of the cell.

## *40. Problem: cells cannot directly change their genome*

I summarize. Given that the best proof of a theory is its practical application, the most convincing way to demonstrate that entanglement plays a role in the living cells would be if we could utilize cells as quantum computers. It appears then that to extract the results of this computation would be the easiest, if we could make the cells to record these results on their DNA – as we could read these records by an increasingly powerful and accurate high-throughput sequencing technology.

But that's too bad, because cells cannot write information directly on their DNA. This ability would amount to *Lamarckism* (Lamarck, 1809), a long discredited theory in biology. It is also explicitly forbidden by the 'Central Dogma of Molecular Biology' (Crick, 1970), the best known 'No-Go' statement in biology. According to this claim, the flow of information goes one way only – from genotype (DNA sequence) to phenotype (protein structure, organization of intracellular events in space and time). This dogma provides solid molecular support for the fundamental principle of Darwinian evolutionary theory that *evolution does not have foresight* – because the one-way information flow insures independence of heritable variations (that happen on the level of genome) from selection – which always happens afterwards (and on the level of phenotype).

Thus, regrettably, it appears that DNA sequence cannot be a good readout observable – either to detect entanglement nor to play any role in utilizing cells as quantum computers.

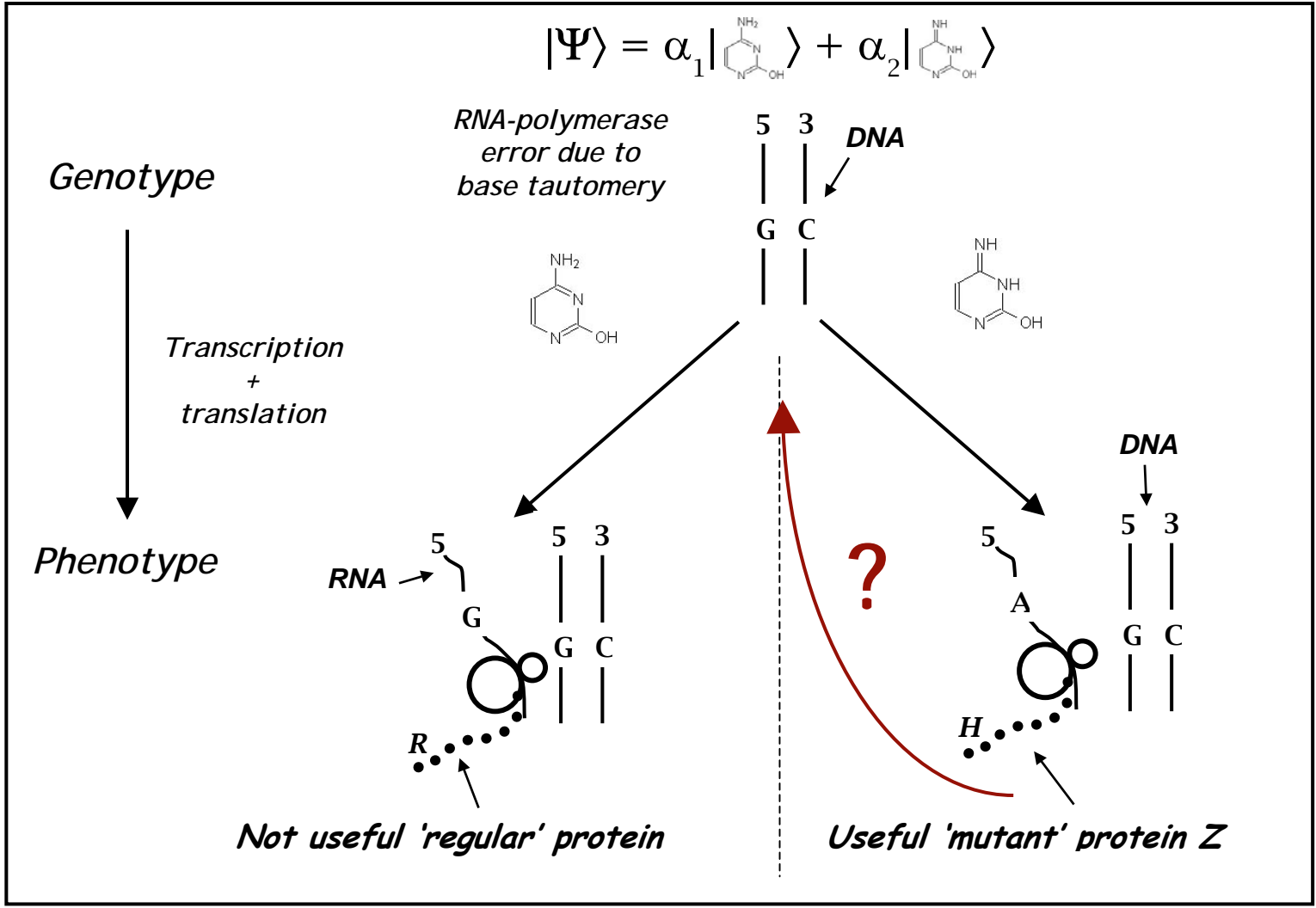

## 41. Central Dogma implies physical irreversibility

Our dreams, however, might not be all in vain. Let's take another look at the 'Central Dogma of Molecular Biology' – now from the physical perspective, along the lines of the Rolf Landauer's aphorism 'information is physical' (Landauer, 1992). After all, the Central Dogma is a statement about information processing on the molecular level – and the claim that information flow can go only in one direction is a statement about physics, namely it implies *physical irreversibility*. Could the science of quantum information and the euclidean approach provide a fresh take at the issue?

Let's illustrate first why Central Dogma implies physical irreversibility. Consider the standard molecular-biological account of a transcription error (a synthesis of a wrong mRNA sequence by RNA polymerase), leading to an appearance of a protein *Z* useful for our cell. Despite the fact that changing genome in this case would be beneficial – enabling the cell to express useful *Z* in future generations – the individual cell cannot accomplish such a 'feedback correction loop', because due to the irreversibility of the intracellular processes the cell does not keep the record about the cause of the appearance of the desired protein sequence. Roughly speaking, irreversibility implies that the same state of the cell with a useful protein *Z* could have been generated in many different ways – with no possibility to trace back what was the cause of its appearance.

For a specific example, such transcription error can result from base tautomery – i.e., transition of a proton from one position of a particular nucleotide base in DNA to another. At the moment when the useful protein *Z* resulting from this error has been tested for function, the redistribution of the proton position between the alternative states in the DNA base will have already erased the memory of how *Z* has emerged in the cell. Thus, the physical irreversibility of intracellular dynamics (more specifically, of the gene expression processes) precludes the cell from recovering the information necessary to fix a valuable variation – i.e., irreversibility is responsible for the asymmetric one-way information flow from genotype to phenotype.

Let's now turn to the euclidean approach. Consider a starving cell close to the ground state. The memory erasure argument is not valid in this case, because the intracellular dynamics is described by unitary transformations – and no information can be lost in the course of a unitary evolution (Pati and Braunstein, 2000). In a sense, the cell in ground state will keep memory about all gene expression errors that it could ever make. It is a separate question where and how this information is stored – the 'record' about the error does not have to be a particular molecular structure, but, in full accord with the notion of entanglement, could be encoded in *correlations* between the states of the parts of the cell. In any case, via consideration of starving cell close to ground state, the euclidean approach helps to clarify the main limitation of the Central Dogma of Molecular Biology. Whereas this statement is relatively safe to apply to a growing cell, where irreversible regime dominates – it becomes questionable in the starving cell case. Due to the unitary nature of the ground state dynamics, the molecular events in the cell, sufficiently close to the ground state, are expected to be significantly more correlated – and the information would be more difficult to loose, thus increasing chances for possible violation of Central Dogma on physical grounds.

*1. Letter to Nature 1990 (rejected)*
*Sent to J. Cairns, B. Hall, K. Matsuno and others*

*2. Biosemiotics school in Soushnevo 1990*
*http://home.comcast.net/~sharov/biosem/seminar.html*

*3. Semiotics congress in Berkeley 1994*
*(http://home.comcast.net/~sharov/biosem/txt/ogr3.html)*

*4. Biosystems 1997, 43(2):83-95*

*5. 'Quantum Mind-1' conference, Flagstaff, 1999*
*http://www.conferencerecording.com/newevents/qac99.htm*

*6. Mentioned in two popular books on science (McFadden, 2000; Staune, 2007)*

## 42. Quantum approach to adaptive mutations

So far, I was using purely theoretical arguments, following from the ideas of this presentation. Intriguingly, there are also empirical facts against the Central Dogma. The principal example is the phenomenon of adaptive mutations (Cairns et al., 1988; Foster, 2000; Hall, 1991; Ryan, 1955), which challenges the Darwinian notion of separation between variation and selection, and suggests that cell can directly change its own genetic sequence – more in accordance with the Lamarck's evolutionary view.

I will refer you to the original publications where you can find more details about my attempts to approach this phenomenon from the quantum theoretical perspective, based on these and related arguments (Ogryzko, 1997; Ogryzko, 2007; Ogryzko, 2008b; Ogryzko, 2008c).

## 43. Plating of bacteria as a measurement procedure

Here I will only briefly recapitulate the approach that I have chosen.

As argued before, the ability of the cell to grow in particular environment can be regarded as its bone fide quantum observable. Consider now the cell in starving state. The important point here is that the simplest two-part representation of the ground state is not only applicable to the description of a catalytic act: $|C\rangle = \lambda_1|I\rangle|R_I\rangle + \lambda_2|O\rangle|R_O\rangle$, as considered previously. In addition, it can also describe an *error*: $|C\rangle = \nu_1|W\rangle|R_W\rangle + \nu_2|M\rangle|R_M\rangle$, where $|W\rangle$ and $|M\rangle$ denote, in this toy example, a regular and tautomeric state of a nucleotide base, respectively. This formulation reflects two facts: 1) in the starving cell the nucleotide base tautomery happens, 2) given that the starving state is stable and is supported by the 'passive error correction' mechanism, the different alternative states of the microenvironment of the nucleotide ($|R_W\rangle$, $|R_M\rangle$) correlated with the nucleotide states ($|W\rangle$, $|M\rangle$) are not distinguishable in the environment of the starving cell – i.e., these differences cannot be amplified to become observable differences. In the density operator description, this situation is described by having non-zero off-diagonal terms between the $|W\rangle|R_W\rangle$ and $|M\rangle|R_M\rangle$ states.

Now we consider an addition of a generic substrate (e.g., glucose) that allows the cell to grow – irrespective of whether it was in a mutant $|M\rangle|R_M\rangle$ state or the wild type $|W\rangle|R_W\rangle$ state. We consider this procedure[24] as a measurement of the cell's capability to grow on this substrate. Given that now both components of the superposition can be amplified in this new environment, they become distinguishable from each other[25]. Formally, this newly acquired distinguishability should correspond to the disappearance of the off-diagonal terms that were reflecting interference between the $|W\rangle|R_W\rangle$ and $|M\rangle|R_M\rangle$ states.

Now consider a different substrate that allows only one of the components of the superposition to grow (e.g., lactose, if we had a classical *LacZ* selection system). In this case, we also will have disappearance of the off-diagonal terms – however, only the $|M\rangle|R_M\rangle$ state will amplify. Given that the starving cell can survive for several days, and due to the continuing tautomery process, there is going to be constant generation of new $|M\rangle|R_M\rangle$ states from the starving $|W\rangle|R_W\rangle$ states, and their subsequent amplification. This is exactly what is observed in the phenomenon of adaptive mutations. (See (Ogryzko, 2008b) for description of this process in terms of non-classical correlation between transcription and replication errors).

To summarize, the ability to mutate in adaptive manner appears naturally in the quantum-mechanical description of the cell, if we consider it as a physical system that utilizes enzymatically catalyzed molecular interconversions for self-reproduction. One can see this ability as a result of *non-commutativity* between two operators describing two observable properties of the cell: 1). its ability to reproduce and 2). its genome sequence. Importantly, although we used base tautomery as an example of genome variability, this scheme is equally applicable to other sources of variability. I consider this as a merit, given that the real phenomenon of adaptive mutations involves many kinds of genomic changes (adaptive transpositions, amplifications, suppressor tRNA mutations, frame shifts etc). This universal behavior, independent of a particular mechanism of genomic variability and gene expression, would be also consistent with the philosophy of the effective field theory, alluded to previously (*37*). In other words, regardless of the molecular (microscopic) details, quantum theory might be telling us that the ability of the cell to directly change its genome is a universal property of biological systems – an inevitable consequence of their self-reproductive capacity and genome variability.

---

[24] In the jargon of biologists, this procedure is called 'bacteria plating'.
[25] that they are distinguishable in the growth-permissive conditions is testified by the fact that we can take a part of the cell population generated as a result of the amplification, extract its DNA and sequence it – all without disturbing the state of the remaining part of the population. Note that this was not possible in the case of the starving cell, because no growth and amplification was possible.

## 44. Use of base analogs and other tricks to increase 'parallelism efficiency'.

Admittedly, we might be decades away from the practical use of quantum information processing capabilities of living cells. This fantastic possibility might not be possible for many independent reasons, however, the above discussion makes the prospect of using cells as quantum computers a little bit more plausible.

In the last slide of this part of my presentation, I would like to consider another problem that one would face using cells as quantum computers. If we will ever be able to come to the point of taking advantage of the quantum parallelism in the cell, we will be facing the following *problem of efficiency*. Consider base tautomery in DNA sequence as a source of quantum parallelism. Suppose that we have engineered a cell that can use the superposition of the DNA states as the input for a quantum algorithm implemented by the enzymatic mechanisms, installed in the cell by us specifically for this purpose. Suppose that, for a particular base, its $|W\rangle$ and $|M\rangle$ states (corresponding to the regular proton position on the base or the tautomeric one, respectively) encode $|0\rangle$ and $|1\rangle$ states of a particular qubit of the input[26]. The problem is that the contribution of the tautomeric $|M\rangle$ state is typically very small (about 4 orders of magnitude less compared to the $|W\rangle$ state) – on the other hand, we will be measuring the outcome of the computation in the ($|W\rangle$, $|M\rangle$) basis (this is the only basis that we could use when we amplify and sequence DNA). Therefore, one can immediately see the problem – most of the resources of the cell will be spent on exploring the $|0\rangle$ input. Considering that we will want to use combination of several bases in genome – and all of them will have only a small contribution of the $|1\rangle$ states – the parallelism could not be efficiently exploited – as most of the 'time' the cell will run the $|0,0,0,0,0,...\rangle$ component of the input.

This suggests that we might want to use other than regular base tautomery sources of variability. For example, transposition rearrangement of DNA is expected to give more equal contribution of the alternative states, that could encode $|0\rangle$ and $|1\rangle$ of a qubit. Alternatively, instead of regular nucleotides, we could use mutagenic base analogues – such as amino-purines or inosine, known to significantly increase the mismatch frequency. These and similar tricks could be used to generate input states much closer to the Hadamard transformed states (($|0\rangle$ + $|1\rangle$)/$\sqrt{2}$; ($|0\rangle$ - $|1\rangle$)/$\sqrt{2}$) – the most desirable input for exploiting quantum parallelism.

The main goal of this slide is to illustrate an important point – if quantum information has a role in Life, it will not be possible to neither explore nor exploit it without the help and expertise of biologists.

---

[26] Do not confuse the $|0\rangle$ and $|1\rangle$ states with the $|I\rangle$ and $|O\rangle$ states of cell – the input and output of an enzymatic act.

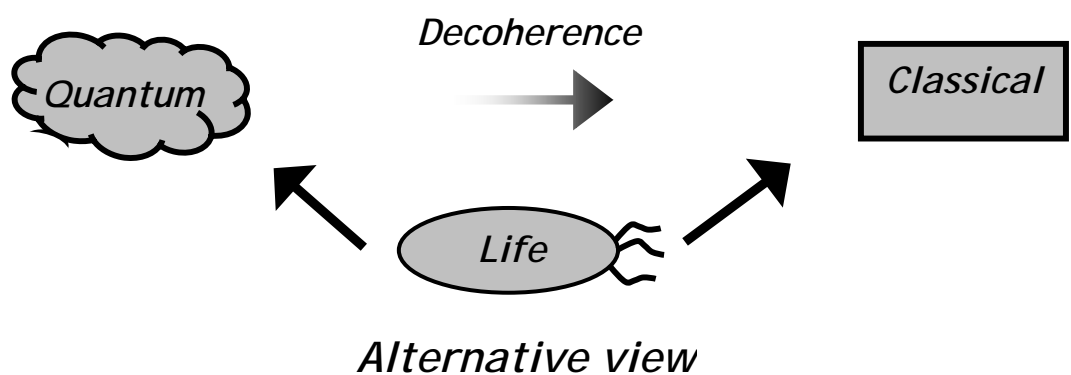

*Traditionally, physics of Life is considered as being beyond the 'quantum to classical' transition*

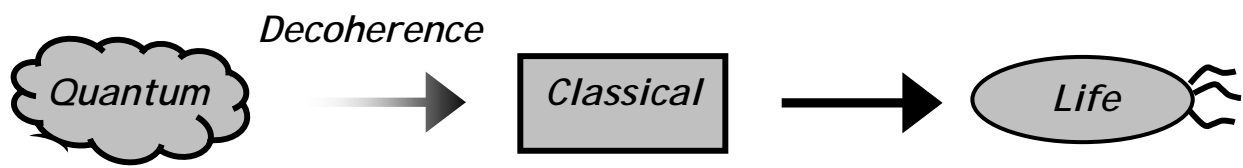

*Alternative view*

## 45. Philosophical conclusion. Environmentally induced decoherence in biology

In the comment to the last slide, I will indulge myself in a bit of more philosophy.

The narrative of the 'environmentally-induced decoherence' program, the 'new orthodoxy' in the physics' foundations, goes something like this. The success of quantum theory has been largely due to using a very important idealization – the notion of an *isolated system*. However, the more complex and big a studied system is, the more questionable and unrealistic this idealization becomes. The paradox of 'Schroedinger cat' illustrates the dire consequences of applying the notion of an 'isolated system' to a macroscopic object. The simplest way to spare the cat from the dishonor is to acknowledge that no physical system is isolated and to take its environment into account – the essence of the decoherence approach (Zeh, 1970; Zurek, 2003).

This simple recipe helps to recover the classical description from the quantum one. So far, so good. However, decoherence is also taken as an argument against nontrivial quantum effects in biology (Tegmark, 2000). As my presentation suggests, such conclusion is unwarranted. By allowing the environment to enter into our description of the system under study, we have, in fact, opened the door for biology to take the central stage. This is because the environment that we might need to consider in order to look for the preferred states of a particular system could be very varied and complex – but the 'physics proper' limits itself to very simple environments, represented by a thermal bath of some kind. For biologists, on the other hand, it is natural to study much richer and less trivial environments. In particular, the concept of '*adaptation*' of a system to its environment starts to play an important role. First, the most important part of understanding a biological object is in considering its relationship with its environment – the way how different life forms correspond (adapt) to their surroundings. Second, the biological organization is often manifested in the adaptation of different parts of the system to each other. Accordingly, biology can provide a more appropriate experimental and conceptual framework for general exploration of the phenomenon of decoherence – especially when one applies the decoherence scheme to the cases of environment other than a thermal bath. Along the way, it can help to understand some biological problems.

In my presentation, I discussed two aspects of how the role of environment in biology is different from its role in physics of inanimate matter. I also illustrated how could they be exploited in better understanding the physics of Life.

1) In biology the dichotomy between 'local' or 'micro-' environment (very ordered) and an 'outside' (more generic) environment plays an important role. I suggested to describe enzymatic activity as decoherence suppression by a specific microenvironment (See *5* and also *46)*. On the other hand, this microenvironment is not a 'bottomless pit' and can be reciprocally affected by the target catalyzed system – leading to intriguing self-adjustment effects (*Cf* force) in biological systems.

2) In biology environment often changes. Accordingly, some properties of a particular object that appear to be classical and subject to superselection rules in one environment, might exist in a superposition in another environment – manifesting itself in a nontrivial quantum behavior. I discussed interesting insights into biological adaptation offered by this general perspective (i.e., adaptive mutation).

The proposed new perspective on the decoherence program[27] also sheds new light on the relations between physics and biology. Usually, biology is considered as a science subordinate to physics. In this commonly shared view, the foundational problems of physics, such as the problem of 'quantum to classical transition', 'time arrow' etc, can be only dealt with by the methods and approaches of the 'physics proper'. Granted, biological problems cannot be of any use for quantum physics – if we assume that Life belongs squarely to the classical realm, represented by irreversible processes at a scale beyond the 'quantum to classical' transition. My view is different, as I expect that the progress of 'nano-' and '-omics' biology will lead to acknowledgment of the nontrivial role that quantum physics plays in Life. But if quantum principles could find a nontrivial manifestation in biological systems, the methods and ideas developed and tested on the terrain more familiar to biologists could significantly contribute to the progress in the foundations of physics. In a sense, the theoretical physics of the 21st century might as well turn out to be the theoretical biology.

---

[27] I would strongly discourage using the term 'Entanglement witness protection program'…

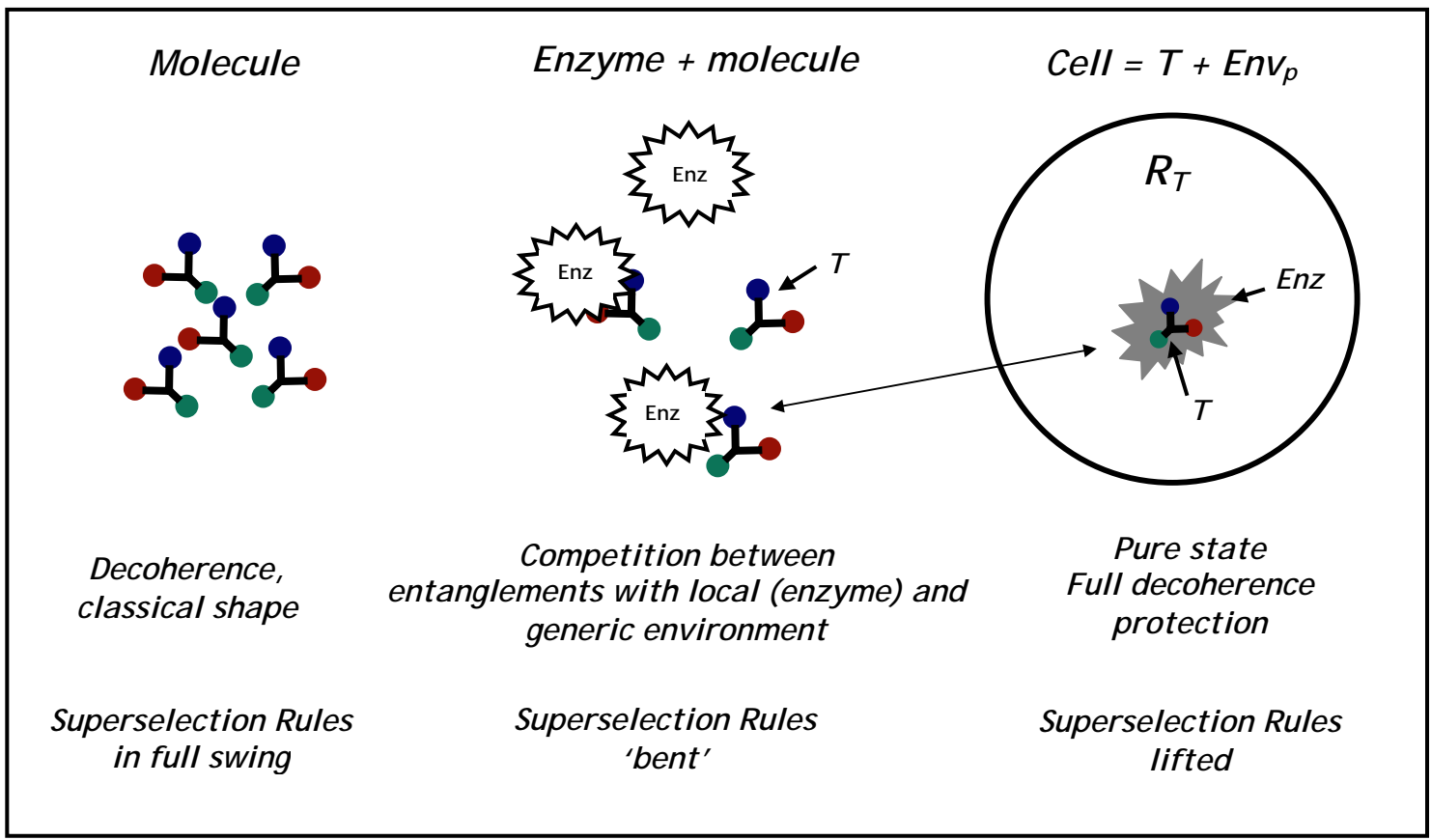

### 46. Appendix.1. Enzymatic activity in vivo and in vitro.

Previously (*5*), I alluded to the Schmidt decomposition theorem to support the general notion of enzymatic activity as decoherence suppression. This might create an erroneous impression that I consider the enzymatic molecule as an auxiliary system $Env_p$ which, when combined with the target molecule $T$, forms a composite system $(Env_p + T)$ undergoing a unitary evolution. Here I would like make a clarification – in order to avoid potential confusion between the notions of 'decoherence suppressing microenvironment' (which I propose to reflect the essence of enzymatic catalysis) and that of purifying microenvironment $Env_p$ (an abstract auxiliary system added to the system under consideration in order to obtain a unitary description).

We start with *in vivo* situation (right side of the slide). First consider the rest of the cell (the complement $R_T$ of a target molecule $T$) as the catalytic microenvironment. According to the Euclidean approach, there are benefits in considering the total system $(R_T + T)$ as evolving in a unitary way – thus, in this case, the $R_x$ could indeed be understood as a purifying system $Env_p$[28]. However, let us now change the focus and consider, as a catalytic microenvironment, an individual enzymatic molecule $Enz$ in this cell. In this case, we will have to trace out the degrees of freedom corresponding to the rest of the cell $(R_T - Enz)$ – thus obtaining a mixed state (or rather, improper mixture) of the subsystem: 'enzyme + target molecule' $(Enz + T)$. Accordingly, for many practical purposes, the description of an individual enzymatic act *in vivo* will be identical to its description *in vitro*, – i.e. *as if* the enzyme was interacting with a thermal environment – which is not described by a unitary evolution. Incidentally, this similarity between the descriptions of enzyme activity *in vitro* and *in vivo* illustrates the fact that the biggest novelty of our approach is not in suggesting how the enzymes work – but rather how their individual actions are correlated in the confines of the living cell – and this information is largely 'traced out', when we limit our description to a part of the cell.

Likewise, if we now consider enzymatic act *in vitro* (middle of the slide), it also does not make sense to consider the enzymatic molecule $Enz$ as a 'purifying microenvironment' $Env_p$ (i.e., the 'enzyme-substrate' complex evolving in a unitary way towards the 'enzyme-product' complex). First, it is misleading – as the external (thermal) environment plays an essential role in the *in vitro* description: 1) it usually provides activation energy (*modulo* tunneling effects), 2) it helps to dissociate the target molecule $T$ from the enzyme $Enz$, recycling the enzyme for the next round of activity, and 3) after the dissociation of the target $T$ from the enzyme, it recovers the classical molecular configuration of $T$ via decoherence. Second, complete purification by *Enz* is also unnecessary – because regardless of whether the dynamic of the *(Enz + T)* system is unitary or not, even a weak ability to revive the off-diagonal terms (in the molecular configuration basis of the reduced density matrix $\rho_T$ describing the target molecule *T*) will already qualify the macromolecule *Enz* as a catalytic microenvironment in experiments *in vitro*.

---

[28] Moreover, I argued (section 14) how the notion of a cell in ground state supported by the reactive $Cf$ forces helps to clarify the ontological status of this 'purified' state of $(R_T + T)$.

I would like to emphasize again that, discussing the role of decoherence suppression, it is not my intent to suggest a new mechanism of enzymatic activity based on some exotic quantum effect. It is rather an attempt to 'quantize' the description of enzymatic act – describe it in general terms 'from the first principles' – i.e., in the language of density matrix – in order to properly integrate the acts of different enzymes in the description of intracellular dynamics.

Now, after this caution, the good news. Even when considering the *in vitro* situation, the Schmidt decomposition might be useful in the task of 'quantization' of the description of enzymatic act. It tells us that it is not dramatically difficult to arrange for an environment *$Env_p$* that is permissive for the unitary transition of a molecule from one molecular configuration to another. According to this theorem, the size of the purifying system can be surprisingly small, and comparable to the system itself – namely, the Hilbert space for the auxiliary system can have the same number of dimensions as the Hilbert space of the target system.

Thus, given a density matrix describing a particular enzymatic transition (e.g., between *Left* and *Right* states of a chiral molecule), we do not need the whole Universe to be aligned in a special way in order to obtain a unitary process. Typically, a much smaller part of the Universe would suffice – although the exact effort required will depend on the situation (similar to the different amount of effort required to observe superposition of nano-objects fullerens compared to the macroscopic Schroedinger cat, or to the differences in the probability of entropy to decrease spontaneously in a nano-system versus a macrosystem, see *3*).

How does it all help us with the 'quantization' of the description of enzymatic act? All what we want from a microenvironment to qualify as an enzyme is to increase the probability of the unitary transition between the input $|I\rangle_T$ and output $|O\rangle_T$ states – which will correspond to an increase in the values of the off-diagonal terms in the reduced matrix $\rho_T$ describing the molecule *T* in the molecular conformation basis (i.e., increase interference between the $|I\rangle_T$ and $|O\rangle_T$ states). From this perspective, we can consider first an individual target molecule *T* (left side of the slide), without enzyme. Let us say that, depending on the desired density matrix $\rho_T$ (with a given values of the off-diagonals responsible for the catalytic transition), and the specific structure of our target molecule *T*, we can estimate the probability to procure a purifying microenvironment *$Env_p$*, so that the desired $\rho_T$ could be obtained from a pure state of the system *(T + $Env_p$)* after tracing out the information about *$Env_p$*. Since, typically, in the case of a molecule in solution, such transitions happen thanks to a thermal fluctuation in the environment of this molecule[29] – we would be, in fact, describing thermally activated barrier crossing in this 'the first principles' language. Accordingly, the general way to describe the effect of enzyme is to see it as facilitating the appearance of such purifying microenvironment *$Env_p$*.

The above discussion should clarify the difference between the notions of 'decoherence suppressing microenvironment' (enzyme *Enz*) and that of 'purifying microenvironment' *$Env_p$*. The purifying microenvironment has a finite probability to appear spontaneously in the absence of enzyme – whereas enzymes by themselves are not purifying systems, but increase the probability for a purifying microenvironment to appear.

---

[29] (modulo tunneling effects)

## *Hammerhead ribozyme*

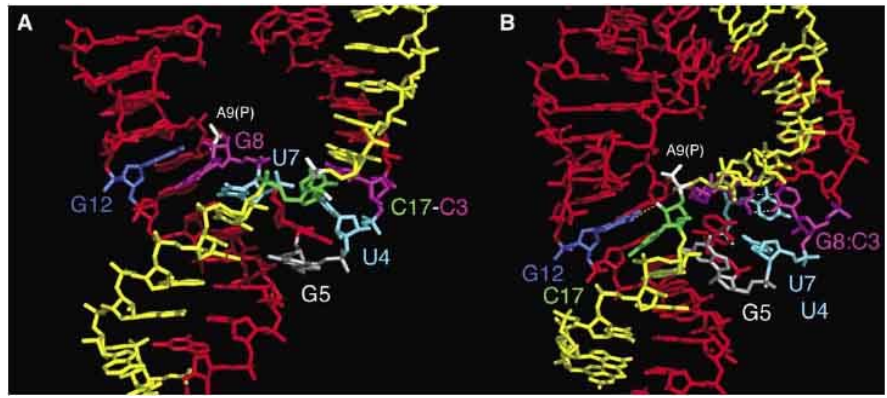

| *Version:* | *Minimal* | *Full-length* |
|---|:---:|:---:|
| • *Activity:* | *Low* | *High* |
| • *Evolutionary:* | *Primitive* | *Advanced* |
| • *Role of Cf:* | *More likely* | *Less likely* |
| • *Reductionism:* | *Not friendly* | *Friendly* |

### 47. Appendix 2. Is it possible to observe the *Cf* force in vitro? Evolutionary implications.

All through the talk, I was emphasizing that my main focus was on how quantum theory could justify the existence of non-classical correlations between individual catalytic events at the level of a living cell – and deliberately avoided the discussion of the particular mechanisms of enzymatic activity.

Here I want to point out that going one level down – to the analysis of individual enzymatic mechanisms – one can also benefit from the suggested ideas. More specifically, our principal focus was on the 'adjustment' effects that the target molecule $T$ exerts on the state of the catalytic microenvironment $E$ (i.e., the *Cf* force). Whereas, for convenience sake, we focused on the state of the cell $R_T$ as the subject of this effect, an individual enzyme would certainly qualify as a catalytic microenvironment as well. Thus, we cannot *a priori* exclude that an individual enzymatic act *in vitro* could also be a subject of the *Cf* force – although in this case the theory and experiment will be complicated by the need to include the external (generic) environment into the consideration.

How could one detect the *Cf* force *in vitro*? Using single molecule experiments, the enzyme activity was documented to vary significantly between molecules – because protein conformation is under constant fluctuation, and some conformations have better activity than others (Min et al., 2005; Xing and Kim, 2006). However, an experimental model that comes to mind first is not a protein, but the *hammerhead ribozyme*. There is evidence that the resting state of the so called minimal hammerhead ribozyme is a noncatalytic conformation – so that the active site (core) must assemble with each catalytic event (Martick and Scott, 2006; McKay, 1996; Wang et al., 1999). Given the existence of two alternative states of this molecule (facilitating and not facilitating the catalytic transition, correspondingly), one can use it as an experimental model to observe the effects of *Cf* on an individual enzyme molecule *in vitro* – by testing whether the probability to find the hammerhead ribozyme in the catalytic conformation will be increased in the presence of the target molecule.

Notably, the *Cf* force might be more difficult to observe on the full length hammerhead ribozyme – which, unlike the minimal version, comes stabilized in the catalytic conformation. On the other hand, being more robust, this version of the enzyme is significantly more efficient. Accordingly, the comparison between the 'minimal' and the 'full length' versions of the hammerhead ribozymes might serve as an illustration for a general trend in the evolution and origin of Life. We can expect that the action of primitive enzymatic molecules in the early days of Life was strongly dependent on the self-organizing effect of the *Cf* force – just like we expect it to play more noticeable role in the mechanism of the minimal hammerhead ribozyme. One could expect, however, that once their role in the primitive cell was established, the evolution would lead to changes in sequence that stabilize the active conformation of these enzymes – thus making their mechanism more robust and less dependent on the *Cf* effects. Literally, the function of the enzymes, first explored and established with the help of the *Cf* force, became codified in the aminoacid and nucleotide sequences – classical type of information.

Somewhat ironically, this vast 'digitalization project run by living Nature' has an unfortunate implication – it should make the modern Life more 'mechanical' and reductionism-friendly compared to the early Life. Thus, using modern Life as an experimental model, the role of self-organization becomes more challenging to demonstrate – although still possible – and here I remain the optimist.

## *Acknowledgements:*

I thank the participants of the workshop on 'Quantum Technology in Biological Systems' for the illuminating discussions. I thank D. Parkhomchuk and K. Augustin for helpful discussions and suggestions.